\documentclass[twocolumn]{aastex62}

\usepackage{amsmath}
\usepackage{color}
\usepackage{appendix}
\usepackage{multirow}
\usepackage[flushleft]{threeparttable}

\newcommand{\bse}{\texttt{BSE}~}
\newcommand{\mobse}{\texttt{MOBSE}~}

\newcommand{\Log}{{\rm Log}}
\newcommand{\rem}{{\rm rem}}

\newcommand{\bbh}{{\rm BBH}}
\newcommand{\Ms}{{\rm M}_\odot}
\newcommand{\Zs}{{\rm Z}_\odot}

\newcommand{\Min}{{\rm min}}

\newcommand{\iso}{{\rm iso}}
\newcommand{\dyn}{{\rm dyn}}

\newcommand{\gc}{{\rm GC}}
\newcommand{\yc}{{\rm YC}}
\newcommand{\nc}{{\rm NC}}

\graphicspath{{./}{./}}

\received{6 January 2020}
\accepted{\today}

\submitjournal{ApJ}

\shorttitle{Fingerprints of binary black holes formation channels}
\shortauthors{Arca Sedda M. et al}

\begin{document}

\title{Fingerprints of binary black hole formation channels encoded in the mass and spin of merger remnants}

\correspondingauthor{Manuel Arca Sedda}
\email{m.arcasedda@gmail.com}

\author[0000-0002-3987-0519]{Manuel Arca Sedda}
\affil{Zentrum f\"{u}r Astronomie der Universit\"{a}t  Heidelberg, Astronomisches Rechen-Institut,M\"onchhofstrasse 12-14,Heidelberg, D-69120, DE}

\author[0000-0001-8799-2548]{Michela Mapelli}
\affil{Physics and Astronomy Department Galileo Galilei, University of Padova, Vicolo dell'Osservatorio 3, I--35122, Padova, Italy}
\affil{INFN-Padova, Via Marzolo 8, I--35131 Padova, Italy}
\affil{INAF-Osservatorio Astronomico di Padova, Vicolo dell'Osservatorio 5, I--35122, Padova, Italy}

\author[0000-0003-0930-6930]{Mario Spera}
\affil{Physics and Astronomy Department Galileo Galilei, University of Padova, Vicolo dell'Osservatorio 3, I--35122, Padova, Italy}
\affil{INFN-Padova, Via Marzolo 8, I--35131 Padova, Italy}
\affil{Department of Physics and Astronomy, Northwestern University, Evanston, IL 60208, USA}
\affil{Center for Interdisciplinary Exploration and Research in Astrophysics (CIERA), Northwestern University, Evanston, IL 60208, USA}

\author{Matthew Benacquista}
\affil{Center for Gravitational Wave Astronomy, University of Texas Rio Grande Valley, One University Blvd, Brownsville TX 78520, USA}
\affil{Division of Astronomy, National Science Foundation, 2415 Eisenhower Ave, Alexandria, VA 22314, USA}

\author[0000-0002-8339-0889]{Nicola Giacobbo}
\affil{Physics and Astronomy Department Galileo Galilei, University of Padova, Vicolo dell'Osservatorio 3, I--35122, Padova, Italy}
\affil{INFN-Padova, Via Marzolo 8, I--35131 Padova, Italy}
\affil{INAF-Osservatorio Astronomico di Padova, Vicolo dell'Osservatorio 5, I--35122, Padova, Italy}

\nocollaboration
\begin{abstract}
Binary black holes (BBHs) are thought to form in different environments, including the galactic field and (globular, nuclear, young and open) star clusters. Here, we propose a method to estimate the fingerprints of the main BBH formation channels associated with these different environments. We show that the metallicity distribution of galaxies in the local Universe along with the relative amount of mergers forming in the field or in star clusters determine the main properties of the BBH population. Our fiducial model predicts that the heaviest merger to date, GW170729, originated from a progenitor that underwent 2--3 merger events in a dense star cluster, possibly a galactic nucleus. The model predicts that at least one merger remnant out of 100 BBH mergers in the local Universe has mass $90 < M_\rem/ {\rm ~M}_\odot \leq{} 110$, and one in a thousand can reach a mass as large as $M_\rem \gtrsim 250\Ms$. Such massive black holes would bridge the gap between stellar-mass and intermediate-mass black holes. The relative number of low- and high-mass BBHs can help us unravelling the fingerprints of different formation channels. Based on the assumptions of our model, we expect that isolated binaries  
are the main channel of BBH merger formation if $\sim 70\%$ of the whole BBH population has remnants masses $<50\Ms$, whereas $\gtrsim{}6$\% of remnants with masses $>75\Ms$ point to a significant sub-population of dynamically formed BBH binaries.
\end{abstract}


\section{Introduction} 
\label{sec:intro}

The direct detection of gravitational waves (GWs) by the LIGO -- Virgo  collaboration (LVC) \citep{abbott16a,abbott16b,abbott16c,abbott17a,abbott17b,abbott17c,LIGO19} has marked the dawn of gravitational-wave astronomy. 

During the first two observational runs \citep{LIGO19}, the LVC detected ten binary black hole (BBH) mergers \citep{abbott16a,abbott16b,abbott16c,abbott17a,abbott17b,abbott17c,LIGO19} and one double neutron star merger \citep{abbott17d}. The black holes (BHs) detected thus far are consistent with a power-law mass distribution with index $1.3^{+1.4}_{-1.7}$ (at 90~\% confidence level) and no more than $\sim{}1$~\%{} BHs  with mass $>45$ M$_\odot$ \citep{abbott19}.

One of the intriguing puzzles related to LVC detections is the large mass of the observed mergers. Indeed, 8 out of 10 detected BBHs have components with masses above $20\Ms$. This seeming {\it overabundance} of heavy stellar BHs contrasts with the 22 BHs observed in X-ray binaries that have masses in the range $1.6-18\Ms$ \citep{xbinrev,casares17}. 

This difference between the mass range of LVC BHs and BHs in X-ray binaries might be ascribed to the detectors sensitivity, which is much higher for larger BH masses \citep[see for instance][]{fishbach17a}, to other observational biases (e.g. X-ray binaries for which we have a dynamical mass measurement are within few Mpc in a predominantly metal-rich environment), to a predominantly different formation channel \citep{perna19}, or to gravitational lensing \citep{broadhurst18}.
One of the critical parameters affecting the natal mass of BHs is the metallicity of their progenitors, $Z$, as metal-poor stars are expected to produce heavier BHs \citep[]{mapelli09,mapelli10,mapelli13,belczynski10,spera15} and to have a higher merger efficiency than metal-rich stars  \citep{dominik13,giacobbo18b,askar17}. Merging BBHs form either from isolated binary stellar evolution in galactic fields \citep[][and references therein]{tutukov73,zwart00b,hurley02,belczynski02,belczynski10,mapelli13,marchant16,belczynski16a,giacobbo18,spera18,ASBEN19}, or through dynamical interactions in dense young massive clusters \citep{zwart00b,banerjee10,ziosi14,mapelli16,banerjee16,banerjee18,dicarlo19,rastello19}, globular clusters  \citep{sigurdsson93,lee95,miller02,wen03,oleary09,downing10,rodriguez15,antonini16,rodriguez16,askar17,samsing18,ASKLI18,zevin19,hong18,rodriguez2018} or nuclear clusters and galactic nuclei \citep[][Arca Sedda, in prep.]{miller09,vanL16,stephan16,bartos16,antonini16b,OLeary16,hoang18,ASCD17b,ASG17,antonini18c,fragione19}.

The main properties of merging BBHs --- component masses, semimajor axis, and eccentricity --- depend primarily on their birth-site. Mergers taking place in galactic fields are usually characterised by low eccentricities unless they are part of a hierarchical triple \citep{antonini17}.
In star clusters, instead, the zoology of BBH mergers is quite vast. Dynamical scatterings can drive the shrinkage of a BBH down to a point where gravitational waves (GWs) dominate the evolution \citep{rodriguez18,samsing18,ASKLI18,zevin19}, or can trigger the formation of triples that can efficiently affect the BBH end phase for both stable \citep{antonini17,rastello19} or unstable systems \citep{ASKLI18}. 

Dense stellar systems, like globular or nuclear clusters, can potentially retain merger products and favour multiple mergers, thus allowing BH mass buildup \citep{fishbach17a,gerosa18,antonini18c,qin18,ASBEN19,rodriguez2019,doctor19}. These {\it second generation} BHs can significantly affect the BH mass spectrum. In galactic nuclei, BBH evolution and coalescence is even more complex due to the possible presence of a quiescent SMBH \citep{ASG17,ASCD17b,hoang18,hoang19,antonini12,rasskazov19,fernandez19} or an AGN in the galactic centre \citep{bartos16,yang19}.

Placing constraints on the population of merger products is also important to improve our knowledge of BH formation. For instance, chirp masses can be used to constrain the global BH natal kick distribution \citep{zevin17,barrett18}, while their spin distribution can carry information on BH natal spins and BBH spin alignment in isolated \citep{gerosa18} and dynamical environments \citep{morawski18}. As LIGO and Virgo reach full sensitivity and the number of detections increases, it will be possible to determine what is the most likely BH spin amplitude and the BBH spin orientation \citep{stevenson17,talbot17,farr17,ASBEN19,bouffanais2019,bavera20}. 
Dissecting the formation history of BBHs from GW observations is a many-faceted problem that requires simultaneous accounting for single and binary stellar evolution, stellar dynamics, general relativity and cosmology. Addressing this problem by means of direct N-body simulations combined with population-synthesis simulations is a computational challenge (see e.g. \citealt{wang16,banerjee16,dicarlo19,rastello19}). If we want to probe a large portion of the parameter space, we need a much faster approach than full N-body simulations.  Recently, \cite{ASBEN19} proposed a way to take into account these different aspects with a fast and self-consistent approach.
Following a similar technique, in this paper we provide an astrophysical framework to characterize the formation channels of BBHs. 

We combine state-of-the-art stellar evolution recipes, theoretical models for BBH merger processes, observational constraints on the local Universe metallicity distribution, and numerical relativity fitting formulae to calculate post-merger BHs final mass and spin. We explore how theoretical uncertainties can affect the results, and discuss what we can learn from potential differences between observations and our model.

The paper is organized as follows: in Section \ref{sec:method} we discuss the method and the underlying assumptions beneath our fiducial model; Section \ref{sec:results} presents the main results of our fiducial model, providing a comparison to the known population of GW sources (Section \ref{sec:o12}), showing how different formation channels impact the percentage of massive mergers in BBH populations (Section \ref{sec:massBBH}), and discussing how such a model can be used to constrain the formation pathway of massive BBH mergers like GW170729 (Section~\ref{sec:gw170729}); Section \ref{sec:discus} is devoted to investigate the impact of theoretical uncertainties on our results; in Section \ref{sec:concl} we draw the conclusions of this work.

\section{Method} \label{sec:method}

Tracking the evolutionary path of merging BBHs requires taking into account several parameters simultaneously: the metallicity distribution of galaxies and star clusters in the local Universe; the possibility that GW observations are biased toward heavy mergers; the probability for a merger to take place in metal-poor or metal-rich environments, in the field or in a star cluster. 

\begin{figure*}
    \centering
    \includegraphics[width=13cm]{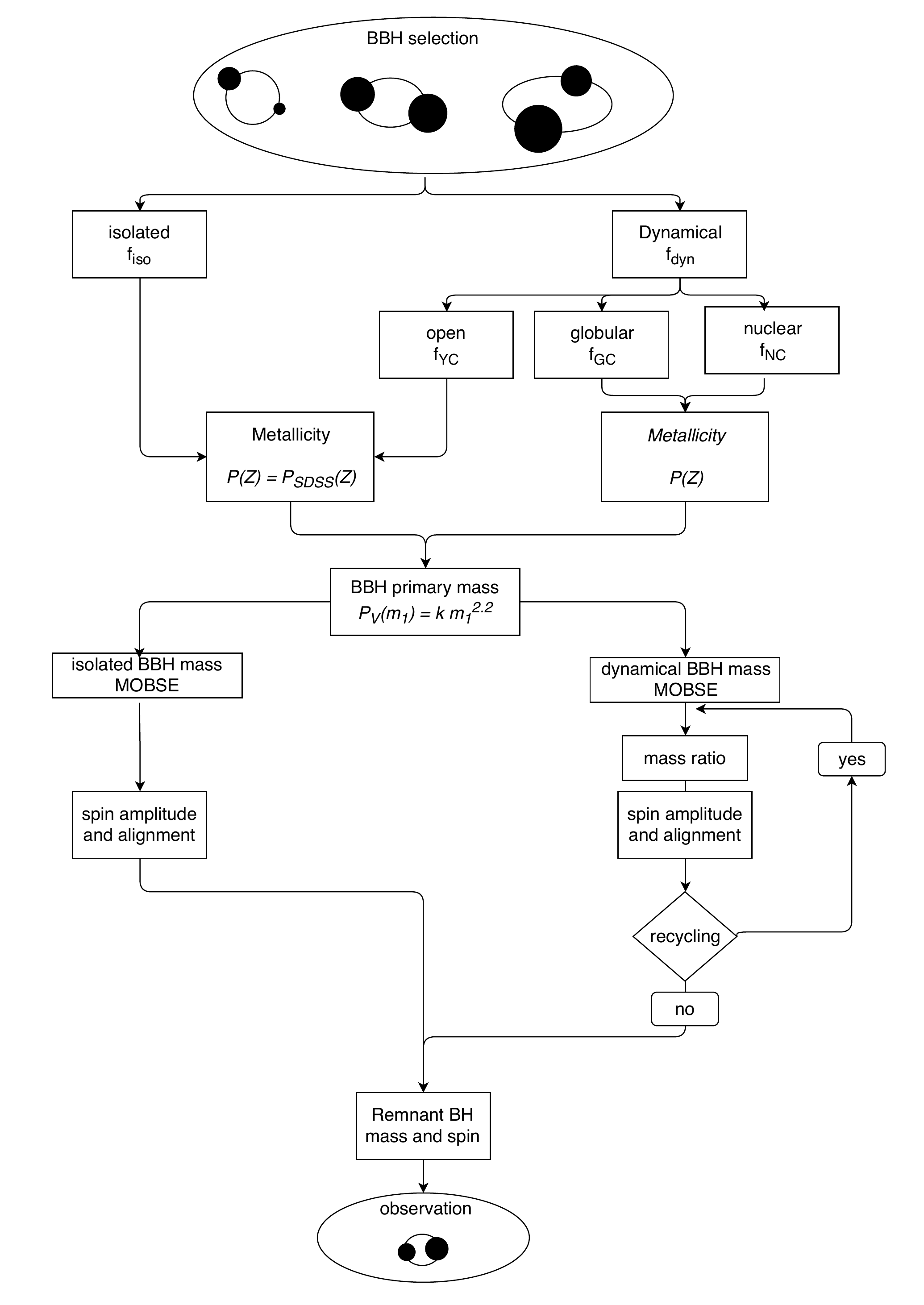}
    \caption{Sketch of the procedure adopted to create the BBH catalogue}
    \label{fig:sketch}
\end{figure*}

Our multi-step procedure can be outlined as follows: for each BBH that coalesces we
\begin{enumerate}
    \item select its birth-place metallicity;
    \item select the BBH formation channel assuming different probability thresholds for isolated and dynamical channels;
    \item use single/binary stellar evolution to calculate the natal mass of the components, taking into account an observational selection function to select the BBH primary mass;
    \item calculate the natal spin amplitude of the components;
    \item calculate the orientation of the spins according to a given distribution;
    \item calculate the merged BH final mass and spin via numerical relativity fitting formulae.
\end{enumerate}
The procedure is sketched in Figure \ref{fig:sketch}.

Table~\ref{tab:1} summarizes the features of our fiducial model and of the other models we consider to estimate the impact of our assumptions on the final results.
For each model, we create a sample of $10^5$ BBH mergers.
\begin{table*}
    \begin{center}
    \caption{Parameters of the models investigated}
    \begin{tabular}{c|cc|cccccc|ccc|c|cc}
         \hline
         \hline
         {\bf ID} & \multicolumn{2}{c|}{{\bf Formation channel}} & \multicolumn{6}{c|}{{\bf Dynamical channel}} & \multicolumn{3}{c|}{{\bf Metallicity}} & {\bf OBS} &\multicolumn{2}{c}{{\bf Spins}} \\
          \hline
         & $f_\iso$ & $f_\dyn$ & \multicolumn{2}{c}{$f_\gc$} & \multicolumn{2}{c}{$f_\nc$} & \multicolumn{2}{c|}{$f_\yc$} & \multicolumn{2}{c}{$P(Z)$} & $f(Z)$ & $\alpha_{m_1}$ & $P(a_1)$ & $n_{\theta}$ \\
             &          &          & $q_\Min$ & $v_{\rm max}$ & $q_\Min$ & $v_{\rm max}$ &$q_\Min$  & $v_{\rm max}$ & iso+yc & gc+nc & & & & \\
             &          &          &  & \footnotesize{[km/s]} &  & \footnotesize{[km/s]} &  & \footnotesize{[km/s]} & &  & & & & \\
         \hline         
         \multicolumn{15}{c}{{\bf Fiducial model}}\\
         \hline
         \multirow{2}{*}{1} & \multirow{2}{*}{0.67} & \multirow{2}{*}{0.33} & \multicolumn{2}{c}{0.6} & \multicolumn{2}{c}{0.2} & \multicolumn{2}{c|}{0.2} & \multirow{2}{*}{SDSS} & \multirow{2}{*}{LOG} & \multirow{2}{*}{$Z^\beta$}& \multirow{2}{*}{$2.2$} & \multirow{2}{*}{uniform} & \multirow{2}{*}{$0$} \\
         & & & 0.2 & 15 & 0.2 & 100 & 0 & 3 &  &  &  & &    \\
         \hline
         \multicolumn{15}{c}{{\bf Metallicity distribution choice}}\\
         \hline
         \multirow{2}{*}{2a} & \multirow{2}{*}{0.67 } & \multirow{2}{*}{0.33 } & \multicolumn{2}{c}{0.6} & \multicolumn{2}{c}{0.2} & \multicolumn{2}{c|}{0.2} & \multirow{2}{*}{SDSS} & \multirow{2}{*}{SDSS} & \multirow{2}{*}{1} & \multirow{2}{*}{2.2} & \multirow{2}{*}{ uniform} & \multirow{2}{*}{$0$} \\
         & & & 0.2 & 15 & 0.2 & 100 & 0 & 3 &  &  &  & &    \\
         \hline
         \multirow{2}{*}{2b} & \multirow{2}{*}{0.67 } & \multirow{2}{*}{0.33 } & \multicolumn{2}{c}{0.6} & \multicolumn{2}{c}{0.2} & \multicolumn{2}{c|}{0.2} & \multirow{2}{*}{SDSS} & \multirow{2}{*}{LOG} & \multirow{2}{*}{1} & \multirow{2}{*}{2.2} & \multirow{2}{*}{ uniform} & \multirow{2}{*}{$0$} \\
         & & & 0.2 & 15 & 0.2 & 100 & 0 & 3 &  &  &  & &    \\
         \hline
         \multirow{2}{*}{2c} & \multirow{2}{*}{0.67 } & \multirow{2}{*}{0.33 } & \multicolumn{2}{c}{0.6} & \multicolumn{2}{c}{0.2} & \multicolumn{2}{c|}{0.2} & \multirow{2}{*}{LOG} & \multirow{2}{*}{LOG} & \multirow{2}{*}{1} & \multirow{2}{*}{2.2} & \multirow{2}{*}{ uniform} & \multirow{2}{*}{$0$} \\
         & & & 0.2 & 15 & 0.2 & 100 & 0 & 3 &  &  &  & &    \\
         \hline
         \multicolumn{15}{c}{{\bf Natal environments choice}}\\
         \hline
         \multirow{2}{*}{3a} & \multirow{2}{*}{1} & \multirow{2}{*}{0} & \multicolumn{2}{c}{-} & \multicolumn{2}{c}{-} & \multicolumn{2}{c|}{-} & \multirow{2}{*}{SDSS} & \multirow{2}{*}{-} & \multirow{2}{*}{$Z^\beta$}& \multirow{2}{*}{2.2} & \multirow{2}{*}{uniform} & \multirow{2}{*}{$0$} \\
         & & & - & - & - & - & - & - & & & & &    \\
         \hline
         \multirow{2}{*}{3b} & \multirow{2}{*}{1} & \multirow{2}{*}{0} & \multicolumn{2}{c}{-} & \multicolumn{2}{c}{-} & \multicolumn{2}{c|}{-} & \multirow{2}{*}{SDSS$/30$} & \multirow{2}{*}{-} & \multirow{2}{*}{$Z^\beta$}& \multirow{2}{*}{2.2} & \multirow{2}{*}{uniform} & \multirow{2}{*}{$0$} \\
         & & & - & - & - & - & - & - & & & & &    \\
          \hline
         \multirow{2}{*}{3c} & \multirow{2}{*}{1} & \multirow{2}{*}{0} & \multicolumn{2}{c}{-} & \multicolumn{2}{c}{-} & \multicolumn{2}{c|}{-} & \multirow{2}{*}{LOG} & \multirow{2}{*}{-} & \multirow{2}{*}{$Z^\beta$}& \multirow{2}{*}{2.2} & \multirow{2}{*}{uniform} & \multirow{2}{*}{$0$} \\
         & & & - & - & - & - & - & - & & & & &    \\
          \hline         \multirow{2}{*}{4a} & \multirow{2}{*}{0 } & \multirow{2}{*}{1 } & \multicolumn{2}{c}{0.6} & \multicolumn{2}{c}{0.2} & \multicolumn{2}{c|}{0.2} & \multirow{2}{*}{SDSS} & \multirow{2}{*}{LOG} & \multirow{2}{*}{$Z^\beta$}& \multirow{2}{*}{2.2} & \multirow{2}{*}{ uniform} & \multirow{2}{*}{-} \\
         & & & 0.2 & 15 & 0.2 & 100 & 0 & 3 &  &  &  & &    \\
         \hline
         \multirow{2}{*}{4b} & \multirow{2}{*}{0 } & \multirow{2}{*}{1 } & \multicolumn{2}{c}{1.0} & \multicolumn{2}{c}{0.0} & \multicolumn{2}{c|}{0.0} & \multirow{2}{*}{-} & \multirow{2}{*}{LOG} & \multirow{2}{*}{$Z^\beta$}& \multirow{2}{*}{2.2} & \multirow{2}{*}{ uniform} & \multirow{2}{*}{-} \\
         & & & 0.2 & 15 & - & - & - & - &  &  &  & &    \\
         \hline
         \multirow{2}{*}{4c} & \multirow{2}{*}{0 } & \multirow{2}{*}{1 } & \multicolumn{2}{c}{0.0} & \multicolumn{2}{c}{0.0} & \multicolumn{2}{c|}{1.0} & \multirow{2}{*}{SDSS} & \multirow{2}{*}{-} & \multirow{2}{*}{$Z^\beta$}& \multirow{2}{*}{2.2} & \multirow{2}{*}{ uniform} & \multirow{2}{*}{-} \\
         & & & - & - &  - & - & 0 & 3 & &  &  & &    \\
         \hline
         \multirow{2}{*}{4d} & \multirow{2}{*}{0 } & \multirow{2}{*}{1 } & \multicolumn{2}{c}{0.0} & \multicolumn{2}{c}{1.0} & \multicolumn{2}{c|}{0.0} & \multirow{2}{*}{-} & \multirow{2}{*}{LOG} & \multirow{2}{*}{$Z^\beta$}& \multirow{2}{*}{2.2} & \multirow{2}{*}{ uniform} & \multirow{2}{*}{-} \\
         & & & - & - & 0.2 & 100 &- &  - &  &  &  & &    \\
         \hline
         \multirow{2}{*}{4d$\dagger$} & \multirow{2}{*}{0 } & \multirow{2}{*}{1 } & \multicolumn{2}{c}{0.0} & \multicolumn{2}{c}{1.0} & \multicolumn{2}{c|}{0.0} & \multirow{2}{*}{-} & \multirow{2}{*}{LOG} & \multirow{2}{*}{$Z^\beta$}& \multirow{2}{*}{2.2} & \multirow{2}{*}{ uniform} & \multirow{2}{*}{-} \\
         & & & - & - & 0.2 & 100 & - &  - &  &  &  & &    \\
         \hline
         \multirow{2}{*}{4e} & \multirow{2}{*}{0 } & \multirow{2}{*}{1 } & \multicolumn{2}{c}{0.0} & \multicolumn{2}{c}{1.0} & \multicolumn{2}{c|}{0.0} & \multirow{2}{*}{-} & \multirow{2}{*}{LOG} & \multirow{2}{*}{$Z^\beta$}& \multirow{2}{*}{2.2} & \multirow{2}{*}{ uniform} & \multirow{2}{*}{-} \\
         & & & - & - & 0.2 & 0 & - &  - &  &  &  & &    \\
         \hline
         \multirow{2}{*}{5} & \multirow{2}{*}{0.50} & \multirow{2}{*}{0.50} & \multicolumn{2}{c}{0.33} & \multicolumn{2}{c}{0.33} & \multicolumn{2}{c|}{0.33} & \multirow{2}{*}{SDSS} & \multirow{2}{*}{LOG} & \multirow{2}{*}{$Z^\beta$}& \multirow{2}{*}{$2.2$} & \multirow{2}{*}{uniform} & \multirow{2}{*}{$0$} \\
         & & & 0.2 & 15 & 0.2 & 100 & 0 & 3 &  &  &  & &    \\
         \hline
         \hline
    \end{tabular}
    \end{center}
    \begin{tablenotes}
     \item Col. 1: model ID number. Col. 2-3: fractional number of isolated or dynamical mergers, respectively. Col. 4-6: fractional number of sources forming in globular, nuclear, and young clusters. Sub-rows indicate, for each cluster type, the minimum mass ratio allowed and maximum velocity ($v_{\rm max}$) allowed for a merger to be retained an undergo a further merger. Col. 7-8: metallicity distribution adopted, either the one inferred from observations (SDSS) or flat in logarithmic values (LOG), and the weighting function used to account for the dependence between metallicity and merger probability. Col. 9: Slope of the observational mass selection function. Col. 10: natal spin distribution. Col. 11: slope of the distribution function adopted to model spins alignment: isotropic ($n_\theta = 0$), mildly aligned ($n_\theta = 2$), or fully aligned ($n_\theta\rightarrow\infty$) distribution.  
	\item Note: In model 4d$\dagger$ we assume the same values of set 4d, but the maximum mass allowed for single BHs is set to $40\Ms$.
    \end{tablenotes}
    \label{tab:1}
\end{table*}

The range of assumptions featured by our fiducial model is detailed in the following subsections. 

\subsection{Metallicity distribution}

In order to obtain a reliable distribution for the metallicity of BBH host galaxies, we use the analysis performed by \cite{Gallazzi05}, based on 44254 galaxies drawn from the Sloan Digital Sky Survey Data Release Two (SDSS DR2). Note that the galaxy sample considered here spans the redshift range $0.005 < z \leq 0.22 $, thus it provides a reliable representation of the volume scanned by the LVC during the O1, and partly O2, runs\footnote{The instrumental horizon of LIGO and Virgo  was $\sim 1.3$ Gpc \citep{martynov16} (redshift $z\simeq 0.25$) during O1 and will grow up to $4$ Gpc ($z\sim{}1$) at design sensitivity \citep{LIGO19}.}. 

\begin{figure}
    \centering
    \includegraphics[width=\columnwidth]{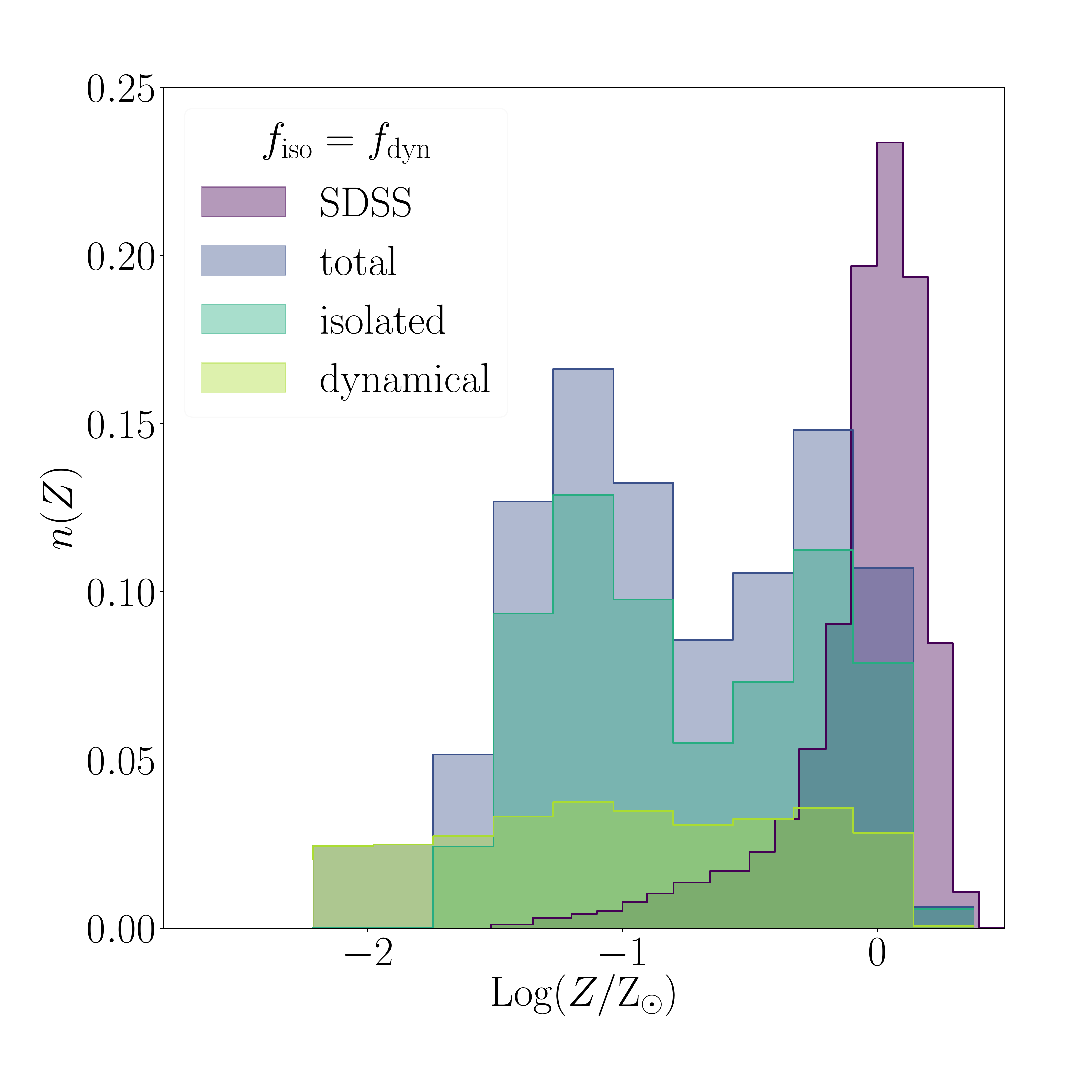}
    \caption{Metallicity distribution for isolated (dark green filled steps) and dynamical mergers (light green filled steps), overlaid on the metallicity distribution from the SDSS (purple filled steps). The total population of dynamical and isolated mergers is also shown (dark blue filled steps). The histogram is normalized to the total number of sources. We assume that the dynamical mergers are half of the total population, corresponding to model ID 5. We assume that all dynamical environments (globular, young, and nuclear clusters) contribute equally, being their fraction $f_\gc = f_\yc = f_\nc$.}
    \label{fig:Zdist}
\end{figure}

As shown in Figure \ref{fig:Zdist}, the metallicity distribution shows a clear peak toward solar values, while the population of metal-poor galaxies ($Z<0.1\Zs$) accounts for less than a few percent of all the galaxies in the sample. 

The preponderance of metal-rich galaxies might have a major impact on the mass of BBH mergers, as metal-rich stars are expected to produce lower mass BHs  \citep{mapelli09,mapelli10,mapelli13,belczynski10,spera15,belczynski16, giacobbo18} and to lead to smaller merger efficiency \citep{dominik13,dominik15,rodriguez16,askar17,giacobbo18,belczynski16}.
However, there are at least two aspects that should be considered here. On the one hand, a galaxy can be characterised by an intrinsic metallicity spread of up to 0.3 dex \citep[see for instance][]{Pilyugin14}. On the other hand, star clusters do not necessarily feature their host galaxy metallicity. This is clearly seen in our Galaxy. Indeed, while open clusters trace the Milky Way metallicity gradient pretty well \citep{netopil15}, the metallicity of globular clusters is significantly lower than that of disc stars \citep{harris14}. The Milky Way nuclear cluster consists of stars with large spread in age and metallicity \citep{do15}, is characterized by a complex star formation history, similar to its extra-galactic counterparts \citep{rossa}.

On top of this, population synthesis and N-body simulations suggest that the number of mergers strongly decreases at metallicity $Z>10^{-3}$ for both isolated  (see, for instance, \citealt{giacobbo18b}) and dynamical \citep{askar17} BBH mergers. 

In order to include all these features in our model, we create a two-layer procedure to select the BBH birth-site metallicity. 

First, we assume that galaxies and open clusters are characterized by a metallicity distribution, $P(Z)$, equal to the one inferred from SDSS DR2 observations $P(Z) \equiv P_{\rm SDSS}(Z)$. Note that this selection procedure allows us to naturally take into account the observed mass-metallicity relation. 
For globular and nuclear clusters, instead, we allow two possible choices: i) same as for galaxies and open clusters; or ii) logarithmically flat distribution, $P(\Log Z)=$ const. 
Observations of globular clusters indicate a bimodal metallicity distribution whose properties, e.g. peak amplitudes, broadening, or limiting values, vary noticeably from one galaxy to another \citep[see for instance]{lamers17}, whereas nuclear clusters feature metallicities broadly distributed from sub-solar to solar values \citep{rossa,paudel11,neumayer20}. Our choice for the metallicity distribution of globular clusters and nuclear star clusters matches the main features of the complex metallicity distribution observed in globular and nuclear clusters in different environments, while keeping our model as simple as possible. 
Second, we weight the metallicity distribution with the probability for a merger to take place in an environment with a given $Z$. For simplicity, we assume that such probability has a power-law form, $f(Z)\propto Z^\beta$, with $\beta = -1.5$, so to be consistent with isolated binaries \citep{giacobbo18} and star clusters \citep{askar17} results. The quantity $f(Z)P(Z)$ represents the probability for a merger to take place in galaxies at different metallicity. In all the models we consider metallicity values in the range $0.0002 \leq Z \leq 0.03$.

Figure \ref{fig:Zdist} compares this quantity assuming that the metallicity-merger dependence is either absent, thus the probability to detect a BBH with progenitor metallicity $Z$ depends solely on the observed metallicity distribution, or is a power-law. In the latter case, we also dissect the distribution into star clusters and galactic field, assuming that the population is equally divided between dynamical and isolated mergers. The plot shows how effective the contribution of metal-poor galaxies can be to the overall BBH merger population if the $f(Z)$ dependence is taken into account.
In our fiducial model, the metallicity selection for BBH progenitors is weighted with a power-law $f(Z)\propto Z^{-1.5}$ and following the SDSS distribution ($P(Z)=P_{\rm SDSS}$) for isolated and open clusters BBHs, or a logarithmically flat distribution ($P(\Log Z) = $const) for globular and nuclear clusters BBHs.

We explore the effects of different choices in Section \ref{sec:met}.

\subsection{Observational selection bias}
\label{sec:bias}

The actual size of the volume, $V$, visible to LIGO and Virgo depends in a non-trivial way on different parameters. More massive mergers emit GWs with higher strain amplitudes, thus being observable from greater distances. On the other hand, the GW frequency at merger is lower for higher mass systems, leading the signal-to-noise ratio to be smaller, as the signal spends less time in the detector sensitivity band. The exact relation connecting $V$ and the merger properties involves also sky location, angle of inclination and component spins. However, in the BBH mass range\footnote{In the following, we use $M_\bbh$ to refer to the sum of the BBH components mass, and $M_\rem$ to refer to the mass of the merger remnant.} $10 < M_\bbh/\Ms < 100 $, the mass dependence of the detection volume scales with a power-law of the primary mass $m_1$, namely $V \propto m_1^{\delta}$ \citep{fishbach17b}, with $\delta = 2.2$. This relation is valid at fixed mass ratio and under the assumption that the spin does not affect the BBH detectability \cite{fishbach17b, abbott19}. Note that at fixed BBH mass, lower mass ratios correspond to smaller volumes. The spin dependence can increase the volume up to $30\%$, depending on the binary components masses \citep{capano16}, although it is less trivial to show how spins affect the volume - binary total mass dependence. In our earlier paper, we explored how different choices for this relation affect the global population of observed BBH mergers \citep{ASBEN19}. 

Unless otherwise stated, throughout the paper we assume the power-law dependence $V \propto m_1^{2.2}$.

\subsection{Formation channel probability}

In this work, we consider either an isolated formation channel, namely the BBHs resulting from isolated evolution of a stellar binary, or a dynamical formation channel, according to which a BBH forms in star clusters from repeated scatterings among BHs originating via single stellar evolution.

The probability for a BBH merger to have originated via one mechanism or the other depends on our knowledge of the processes that regulate the BBH formation itself.
A possible way to quantify such a probability is by comparing the theoretical merger rates, namely the number of events taking place per time unit and volume unit, obtained for both channels, and compare this to observational limits, namely $\Gamma = 9.7 - 101$ yr$^{-1}$ Gpc$^{-3}$ based on the 10 current detections \citep{LIGO19,abbott19}. 
The most recent calculations suggest for the isolated channel a merger rate in the range $\Gamma \sim 10-250$ yr$^{-1}$ Gpc$^{-3}$  \citep{dominik13,belczynski16,mapelli17,eldridge17,mapelli18,kruckow18,spera18,ASBEN19, giacobbo2019,neijssel2019}.
For the dynamical channel, instead, the merger rate depends on the type of hosting cluster, being $\Gamma \sim 5-50$ yr$^{-1}$ Gpc$^{-3}$ for globular clusters \citep{rodriguez16,askar17,rodriguez18,ASBEN19}, $\Gamma \sim 0.1-5$ yr$^{-1}$ Gpc$^{-3}$ for open star clusters \citep{banerjee16,kumamoto19,rastello19}, $\Gamma \sim 0.1-100$ for young star clusters (\citealt{ziosi14},\citealt{mapelli16},\citealt{dicarlo19} and Di Carlo et al., in prep.), and $\Gamma \sim 0.5-10$~yr$^{-1}$ Gpc$^{-3}$ for nuclear clusters  \citep{antonini16b,ASG17,hoang18,hong18,vanL16,ASCD17b,rasskazov19}.

Alternative theories for BBH formation, like primordial BHs, lead to merger rates similar to those investigated here \citep[see for instance][]{bird16}. 

For clarity's sake, we build-up our mock sample assuming that isolated mergers have a probability $f_\iso$, with a complementary probability $f_\dyn = 1-f_\iso$ for dynamical mergers. According to this definition, in a sample of $N$ sources we will have, on average, $f_\iso N$ isolated mergers and $f_\dyn N$ dynamical mergers. For each BBH, we draw a number $n$ between 0 and 1 assuming a flat distribution. In the case $n<f_\iso$, the BBH is assumed to be isolated, otherwise it is dynamical. Thus, the actual number of BBHs associated with a channel or the other will be affected by the statistical fluctuations inherent in the selection process. 

In the fiducial model, we assume $f_\iso = 3f_\dyn$. 
Moreover, a further layer of complexity needs to be added to properly model dynamical mergers, for three main reasons. 
The first is connected to the evidence that different cluster types are characterised by different merger rates, although the amplitude of such difference is highly uncertain. To account for this effect, we associate different probabilities to different cluster types, namely  $f_\gc$ for globular, $f_\yc$ for young and open clusters, and $f_\nc$ for nuclear clusters. These quantities are defined in such a way that $f_\gc+f_\yc+f_\nc = 1$. This choice implies, for example, that a given sample of dynamical mergers will contain a fraction $f_\gc$ of mergers originated in globular clusters. 

To initialize  $f_\gc$,  $f_\yc$ and  $f_\nc$, we take advantage of the most recent results connected with dynamical BBH mergers. 

As discussed above, the most recent models suggest that mergers developing in globular clusters can outnumber those forming in  nuclear cluster by a factor up to 3--5.
Our knowledge of the merger rate from young star clusters is more uncertain, because massive stars form preferentially in young star clusters \citep{portegieszwart2010}. Hence, young star clusters can provide a large fraction of the BBH mergers that occur in the field (see e.g. \citealt{dicarlo19} and \citealt{bouffanais2019}).

Based on these speculations, we assume $(f_\gc,f_\yc,f_\nc) = (0.6,0.2,0.2)$ as fiducial value. Nevertheless, it must be noted that these numbers rely upon a number of unknown parameters, like the number of young and globular clusters in a given cluster, the merger efficiency (i.e. the number of mergers per unit of cluster mass), the cluster metallicity distribution. We investigate how these quantities affect the results in Section \ref{sec:discus}.

Another intriguing feature of dynamical mergers is the mass ratio. The most massive BHs quickly segregate to the host cluster centre and tend to pair together. This can lead to the preferential formation of BBHs with high mass ratios, regardless of the cluster type  \citep{downing10,rodriguez16,seoane16,ASBEN19,dicarlo19}. In order to take into account this effect, we assume that dynamical mergers have mass ratios above a minimum value $q_{\rm min}$. 

\subsection{Single and binary black hole natal mass and spin}

In order to calculate the mass of BBH components (for isolated binaries) and of single BHs (for dynamical binaries), we take advantage of the \mobse \citep{giacobbo18} population synthesis code. The code is an upgraded version of the \bse \citep{hurley02} stellar evolution package, which allows the user to follow the evolution of binary and single stars from the birth down to the final evolutionary stages.  
The main distinctive feature of \mobse{} with respect to other population-synthesis codes descending from \bse{} is that mass loss by stellar winds in \mobse{} depends on both the metallicity and the stellar luminosity of a massive star: the closer the stellar luminosity gets to the Eddington ratio, the higher the mass loss, regardless of its metallicity. In addition, \mobse includes a treatment of pair instability and pulsational pair instability supernovae \citep{woosley17,spera17}.
In the following, we make use of model $\mathrm{CC15\alpha1}$, presented in \cite{giacobbo18}, which assumes low natal kicks for both core-collapse and electron-capture supernovae.

In order to cover the span of metallicity typical of metal-rich and metal-poor systems, we create 12 different single and BBH populations, characterized by $Z$ values between $Z = 0.0002$ and $Z=0.02\equiv\Zs$.

To create the sample of isolated BBHs we first generate the binary stars following \cite{giacobbo18}. The primary star mass is selected according to a \cite{kroupa01} mass function truncated between $5 - 150 \Ms$, the mass ratio is thus extracted according to $P(q) \propto q^{-0.1}$ to obtain the secondary star mass. The binary period is assigned according to $P(\tau)\propto \tau^{-0.5}$ assuming limiting values of $\tau \equiv {\rm Log} (P/\mathrm{day}) = 0.15-5.5$, whereas the eccentricity is drawn between 0 and 1 according to $P(e) \propto e^{-0.42}$. Note that the assumptions above are motivated by observations of Galactic O-type stars \citep{sana12}. From the whole sample of binaries modelled with MOBSE, we retain only those whose product is a BBH for which the sum of the time needed for the two stars to become a BBH and the merger time is smaller than 14 Gyr \citep[for a description of the resulting BBH mass spectrum see][]{giacobbo18}.

For dynamical BBHs, we draw the mass of each  progenitor star according to a \cite{kroupa01} mass function within the same range of values assumed for isolated binaries. The natal mass of BHs is calculated via \mobse{}, the two BHs in  dynamical BBHs are randomly paired following a uniform distribution between  the minimum mass ratio $q_{\rm min}$ (which depends on the considered model, as described in the previous section) and the maximum mass ratio $q=1$. The probability to randomly draw a BBH from the isolated or the dynamical samples is then weighted with the assumed observational bias (see Section \ref{sec:bias}).

Determining BH natal spins represents a still largely debated issue in stellar evolution community. Some recent work proposed a relation between the spin amplitude and the mass of the stellar carbon-oxygen core \citep[see for instance][]{Belczynski17}. According to this prescription, BHs with natal masses $\leq 40\Ms$ have natal spins above 0.8, with a little dependence on the progenitor metallicity, while the spin decreases at increasing the BH mass. As opposed to this, other studies propose that massive stellar progenitors undergo an efficient angular momentum loss that leads the BH to have a spin $\sim 0.075-0.04$ times the inverse of the mass, at least for BHs heavier than $30\Ms$ \citep{seoane16}. The situation is even more complex if BHs form in a binary. 
In order to cope with our ignorance about the processes that regulate BHs natal spin amplitude, we assume a uniform distribution of spins between 0 and 1 in both isolated and dynamical binaries. 

Another crucial point is related to spin orientation. Spin alignment directly affects the remnant BH final spin amplitude. In the case of dynamical BBHs, the spin orientation is expected to be isotropically distributed, $P(\theta)=$~const.
On the other hand, predicting the alignment of isolated BBHs is more complex. In principle, one can expect that the mutual tidal field exerted from one component to the other would somehow maintain the spins aligned. However, during the stages that lead a star to turn into BH several processes can drive the spin re-orientation, like supernova explosion. Following \cite{ASBEN19}, we control the level of spin misalignment assuming that the angle between the two spins, $\theta$, is characterised by a distribution $P(\cos\theta) = k(n_\theta) (\cos\theta + 1)^{n_\theta}$. This choice implies that increasing $n_\theta$ values correspond to more aligned distribution, with $n_\theta = 0$ ($\infty$) corresponding to a isotropic (fully aligned) distribution. 

In our fiducial model, we assume $n_\theta = 0$, which corresponds to isotropically oriented spins, i.e. the same as for dynamical binaries.

\subsection{Black hole remnant final mass and spin}

During the merger, part of the binary mass is radiated away in the form of GWs, thus the final BH mass $M_\rem$ will be a fraction of the progenitor BBH mass $M_\bbh$. Using the LVC data catalogue \citep[GWTC-1,][]{LIGO19}, we find that the $M_\rem-M_\bbh$ is excellently described by a linear relation of the form
\begin{equation}
    M_\rem = AM_\bbh,
\end{equation}
where $A = 0.954\pm0.002$. Although tantalizing, calculating the remnant BH mass via this relation can neglect some important dependencies. 
A rigorous approach would require numerical relativity, which allows us to follow the last stages of BBH evolution and infer crucial information on the GWs produced during the BBH inspiral, merger and ringdown \citep{pretorius05,campanelli06,baker06,sperhake2015}. 
Therefore, we calculate $M_\rem$ taking advantage of the fitting procedure described in \cite{jimenez17}, although this is tailored to aligned-spin binaries based on numerical relativity simulations. We use the same approach to calculate the remnant spin $a_\rem$, making use of the so-called ``{\it augmentation}'' technique \citep{rezzolla08,hughes03}, which allows us to include the in-plane spin components in $a_\rem$ calculation. As a cross-check, we calculate $a_\rem$ also using the fitting formulae provided by \cite{hofmann16}, following our previous paper \citep{ASBEN19}. We note that the $M_\bbh-M_\rem$ relation calculated for a sample of over $10^5$ models is exquisitely described by a linear relation with slope $A=0.934\pm 0.001$, regardless of the spin orientation, BBH mass ratio or total mass. 

\subsection{Gravitational Wave recoil}
Promptly after the merger, the remnant receives a kick due to anisotropic GW emission, whose amplitude can exceed $10^3$ km s$^{-1}$.
Nonetheless, the potential well of the heaviest clusters (globular and nuclear clusters) might be sufficiently deep to retain some of the post-merger BHs. 
This allows the remnant BHs to possibly undergo multiple mergers \citep{miller02,fishbach17b,gerosa17,rodriguez18,ASBEN19,antonini18c,kimball2019,rodriguez2019,doctor19}, leading to higher and higher BH masses. The probability for BHs to undergo at least two mergers can rise up to $\sim 40\%$ for both globular \citep{rodriguez18} and nuclear clusters \citep{antonini16}. To account for multiple mergers, in our model we calculate the GW recoil kick as \citep{campanelli07,lousto08,lousto12}
\begin{align}
\vec{v}_k   =& v_m\hat{e}_{\bot,1} + v_\bot(\cos \xi \hat{e}_{\bot,1} + \sin \xi \hat{e}_{\bot,2}) + v_\parallel \hat{e}_\parallel, \label{eqKick1}\\
v_m         =& A\eta^2 \sqrt{1-4\eta} (1+B\eta), \\
v_\bot      =& \frac{H\eta^2}{1+q_\bbh}\left(a_{2,\parallel} - q_\bbh a_{1,\parallel} \right), \\
v_\parallel =& \frac{16\eta^2}{1+q_\bbh}\left[ V_{11} + V_A \Xi_\parallel + V_B \Xi_\parallel^2 + V_C \Xi_\parallel^3 \right] \times \nonumber \\
             & \times \left| \vec{a}_{2,\bot} - q_\bbh\vec{a}_{1,\bot} \right| \cos(\phi_\Delta - \phi_1).  \label{eqKick2}
\end{align}
Here, $\eta \equiv q_\bbh/(1+q_\bbh)^2$ is the symmetric mass ratio, while $\vec{\Xi} \equiv 2(\vec{a}_2 + q_\bbh^2 \vec{a}_1) / (1 + q_\bbh)^2$. The subscripts $\bot$ and $\parallel$ mark the perpendicular and parallel direction of the BH spin vector with respect to the direction of the BBH angular momentum. The unit vectors ($\hat{e}_\parallel, \hat{e}_{\bot,1}, \hat{e}_{\bot,2}$) constitute an orthonormal basis with one component directed perpendicular to ($\hat{e}_\parallel$) and two components lying in the BBH orbital plane. We set $A = 1.2 \times 10^4$ km s$^{-1}$, $B = -0.93$, $H = 6.9\times 10^3$ km s$^{-1}$, and $\xi = 145^\circ$ \citep[see][]{gonzalez07,lousto08}, and $V_{A,B,C} = (2.481, 1.793, 1.507)\times 10^3$ km s$^{-1}$ \citep{lousto12}. 
$\phi_\Delta$ represents the angle between the direction of the infall at merger (which we randomly draw in the BBH orbital plane) and the in-plane component of $\vec{\Delta} \equiv (m_1+m_2)^2 (\vec{a}_2 - q_\bbh \vec{a}_1)/(1+q_\bbh)$, while $\phi_1 = 0-2\pi$ is the phase of the BBH, extracted randomly between the two limiting values.

According to the above equations, the GW recoil kick imparted to the merger remnant can vary between $\sim 10 - 3000$ km s$^{-1}$, higher than the typical velocity dispersion of both open ($\sigma \simeq 1-5$ km s$^{-1}$) and globular clusters ($\sigma \simeq 10-15$ km s$^{-1}$). Note that we refer to the value of $\sigma$ calculated at the cluster half-mass radius, though it can be quite higher in the inner core, especially if the cluster hosts a central massive black hole or a stellar black hole cusp. In the case of nuclear clusters, whose velocity dispersion can have escape velocities $\sigma \geq (1-3)\times 10^2$ km s$^{-1}$ \citep[see for instance][]{georgiev09}, the chance for a post-merger BH to be retained in the host cluster and undergo a further merger is not negligible. 
For each BBH, we calculate the GW recoil $|\vec{v}_k|$ via Equations \ref{eqKick1}-\ref{eqKick2} and we allow the remnant to undergo another merger if
$ |\vec{v}_k| < v_{\rm max}$, where $v_{\rm max} = (3,~15,~100)$ km s$^{-1}$ for young, globular, and nuclear clusters, respectively.

\begin{figure}
\centering
\includegraphics[width=\columnwidth]{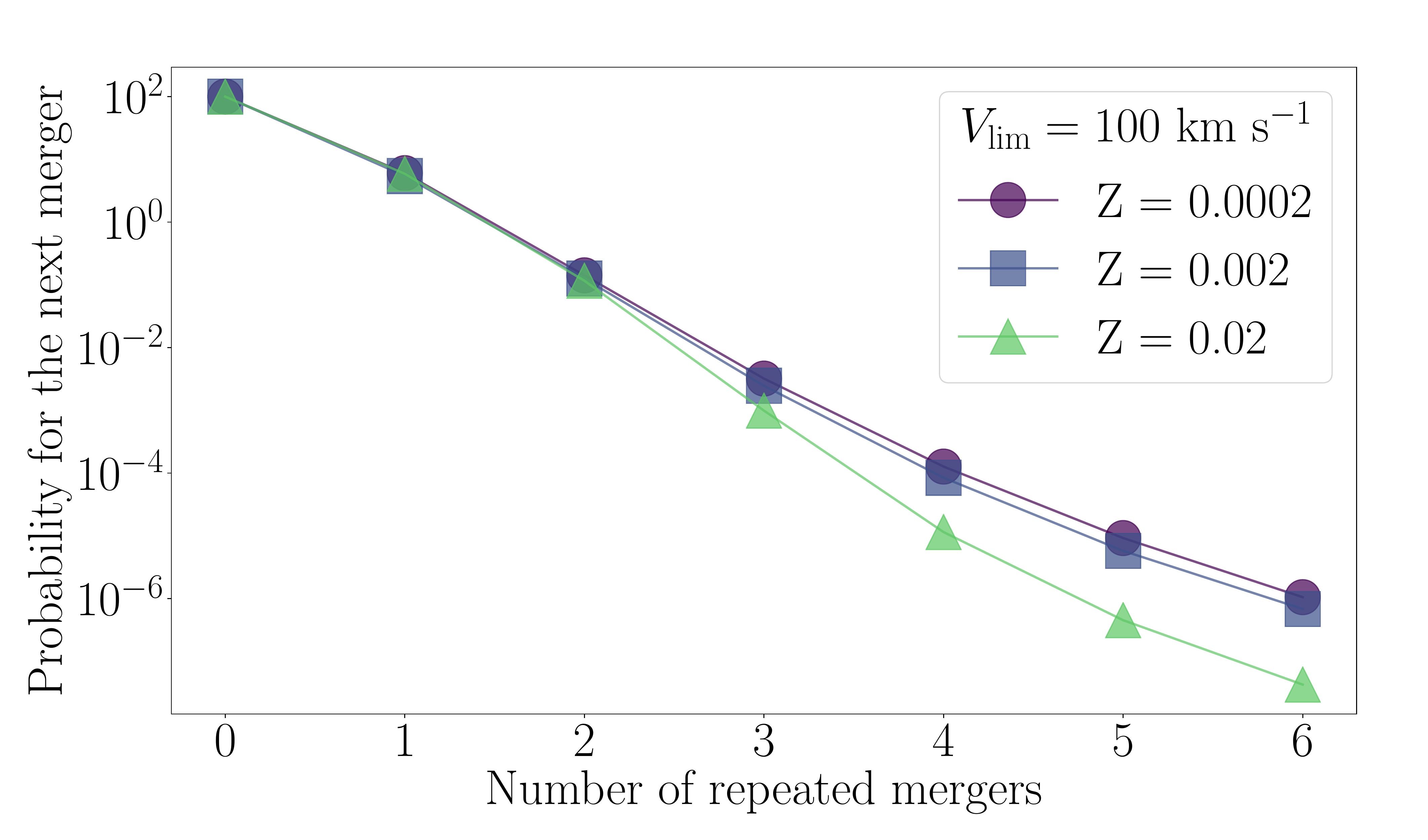}
\caption{Probability for a BH merger to undergo another merger as a function of the number of mergers that the BH has already experienced. We assume that the remnant is ejected from the parent cluster if the GW velocity exceeds $V_{\rm lim} = 100$ km s$^{-1}$. We consider metallicity $Z = 0.0002$ (purple cyrcles), $0.002$ (blue squares), and $0.02$ (green triangles).}
\label{fig:kick}
\end{figure}

Figure \ref{fig:kick} shows the combined probability, for a total of 10,000 BBHs harboured in a nuclear cluster, to receive a GW kick below 100 km s$^{-1}$ and undergo a further merger, assuming  a Kroupa initial mass function \citep{kroupa01} for the progenitor stars and calculating the  BH natal mass according to the single BH mass spectrum described in the previous section \citep{giacobbo18}. After the first merger, the remnant BH has a chance of $\sim 5-10\%$ to be retained. At each stage, the probability is calculated as the product of all the previous ones, and decreases by roughly one order of magnitude at any successive stage.

\section{Results}
\label{sec:results}

In this section, we use the fiducial model to infer a population of merging BBHs. We discuss how different assumptions impact the remnant mass and spin distributions. 

\subsection{Fiducial model}

\begin{figure*}
    \centering
    \includegraphics[width=0.6\textwidth]{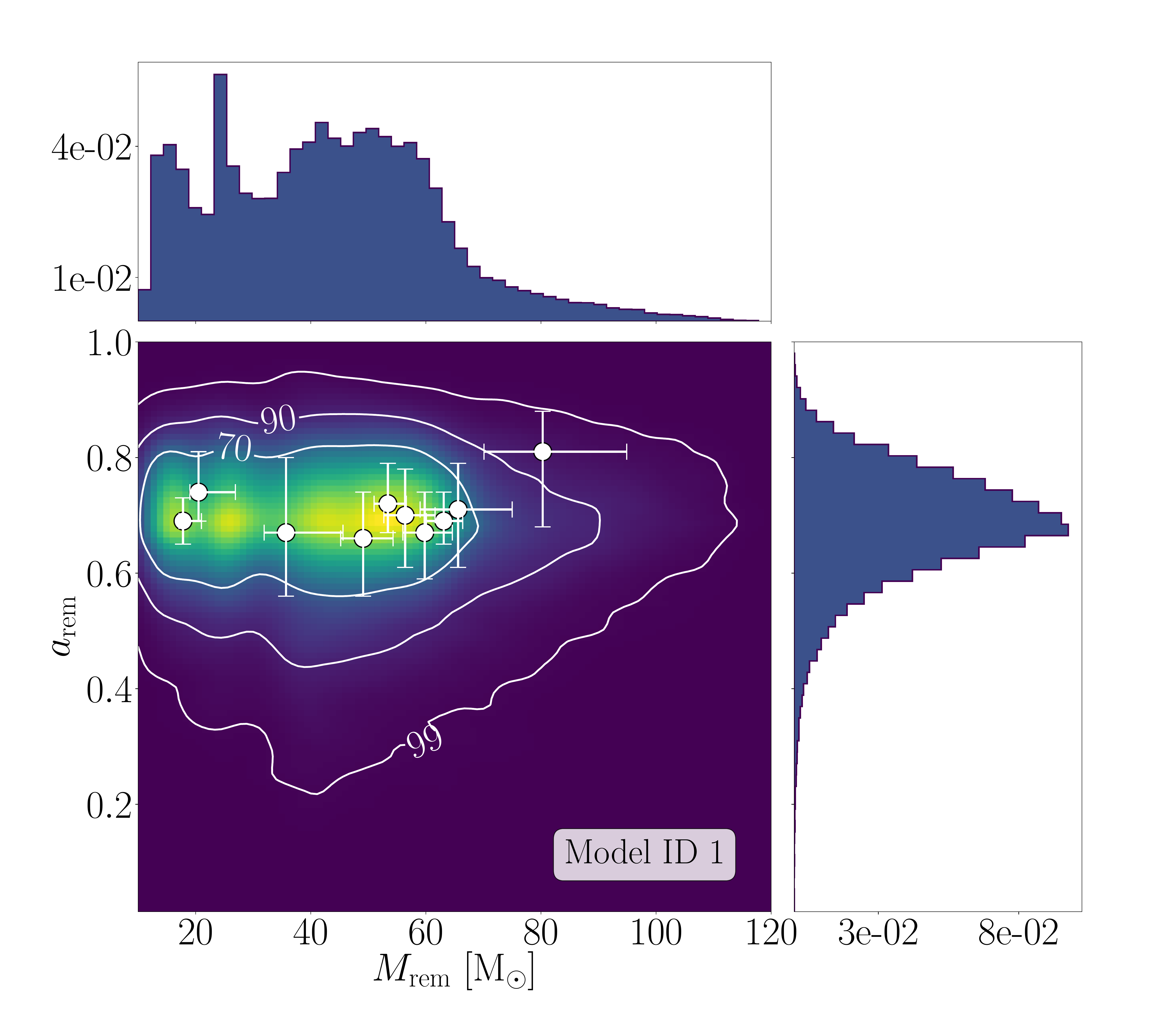}\\
    \includegraphics[width=0.45\textwidth]{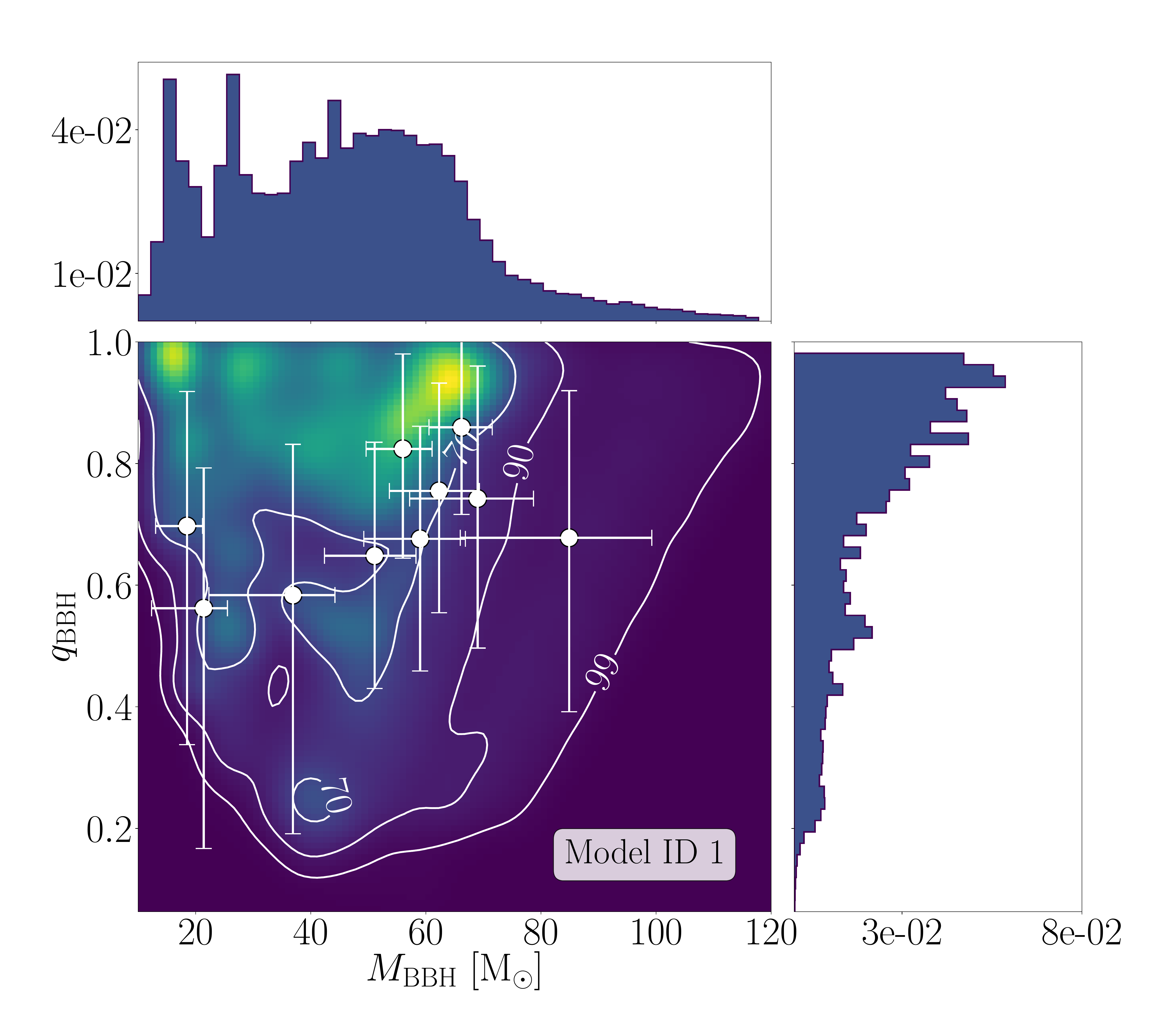}
    \includegraphics[width=0.45\textwidth]{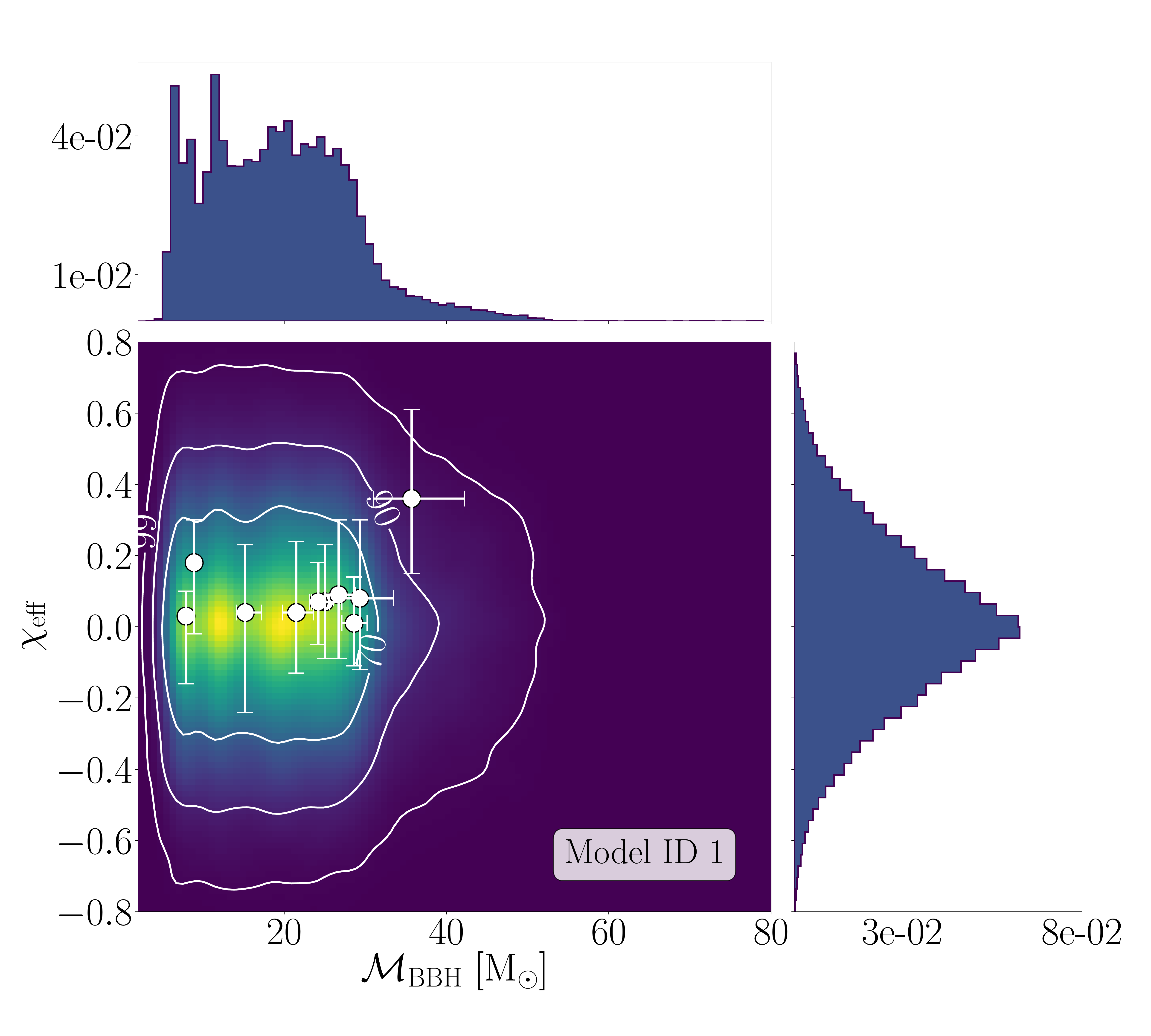}
    \caption{Top panel: remnant mass and spin. Bottom panels: BBH total mass $M_\bbh$ and mass ratio $q_\bbh$ (left panel) and BBH chirp mass $\mathcal{M}_\bbh$ and effective spin parameter $\chi_{\rm eff}$ (right panel). All panels refer to the fiducial model (SET ID 1). The coloured map represents the normalised density in each bin, smoothed with a Gaussian kernel density estimation. Contour lines encompass $70\%,~90\%,~99\%$ of the total sample, respectively. }
    \label{fig:fiducial}
\end{figure*}

The top panel of Figure~\ref{fig:fiducial} shows the final mass and spin distribution of $10^5$ mergers calculated upon the assumptions listed above. For comparison's sake, we overlap the ten BBH mergers detected by the LVC in the first and second observing runs \citep[see Table III in][]{LIGO19}.  

The remnant mass is characterised by a complex distribution that shows two peaks at roughly $\sim 18 \Ms$ and $\sim 25 \Ms$, along with a broader component peaking at $M_\rem \sim 55\Ms$. The spin distribution peaks at $a_\rem\sim 0.7$, with a FHWM $\simeq 0.1$. 
Comparing our $M_\rem-a_\rem$ plane with actual detections, it is apparent how BHs with both low-mass, i.e. $\lesssim 20\Ms$, and high-mass fall in the maximum likelihood of our distribution. Only the most massive BH detected so far, GW170729, seems to lie out of the main distribution, showing both larger mass and spin compared to the overall distribution. This might suggest a peculiar formation history. In Section \ref{sec:gw170729}, we discuss one possible route to the formation of GW170729-like sources as the result of multiple mergers in dense star clusters.
The central panel in Figure~\ref{fig:fiducial} compares the total mass ($M_{\rm BBH}$) and mass ratio ($q_{\rm BBH}$) of our  BBHs with O1 and O2 LVC detections. 
We also compare our model with observed chirp mass ($\mathcal{M}_{\rm BBH}$) and effective spin parameter ($\chi_{\rm eff}$), defined as
\begin{align}
\mathcal{M}_{\rm BBH} &= (m_1m_2)^{3/5}/(m_1+m_2)^{1/5}, \\
\chi_{\rm eff}   &= (a_1\cos\theta_1 + q_\bbh a_2\cos\theta_2)/ (1+q_\bbh),
\end{align}
where $\theta_i$ is the angle between the BBH angular momentum vector and the direction of the spin of the $i$-th BH. As shown in the bottom panel of Figure~\ref{fig:fiducial}, all the LVC sources fall inside the region enclosing $70\%$ of our models with the only exception of GW170729.

We note that single BHs can have mass up to $\sim{}65$ M$_\odot$ in the simulations by \cite{giacobbo18b}, while BHs in isolated binaries that reach coalescence within a Hubble time have a maximum mass of $\sim{}45$ M$_\odot$. Hence, the seeming dearth of remnants with a final mass above $80 \Ms$ is due to a combination of factors: i) the population is dominated by isolated binaries, which constitute the $67\%$ of mergers in our fiducial model, and only BHs with mass $<45$ M$_\odot$ coalesce in the isolated binary sample; ii) as shown in Figure \ref{fig:Zdist}, the assumption of a cluster metallicity distribution flat in the logarithm implies that only roughly half of dynamical mergers have a metallicity below $\sim 0.1\Zs$, i.e. smaller enough to trigger the formation of BHs heavier than $40-50\Ms$.

\begin{figure*}
    \centering
    \includegraphics[width=\textwidth]{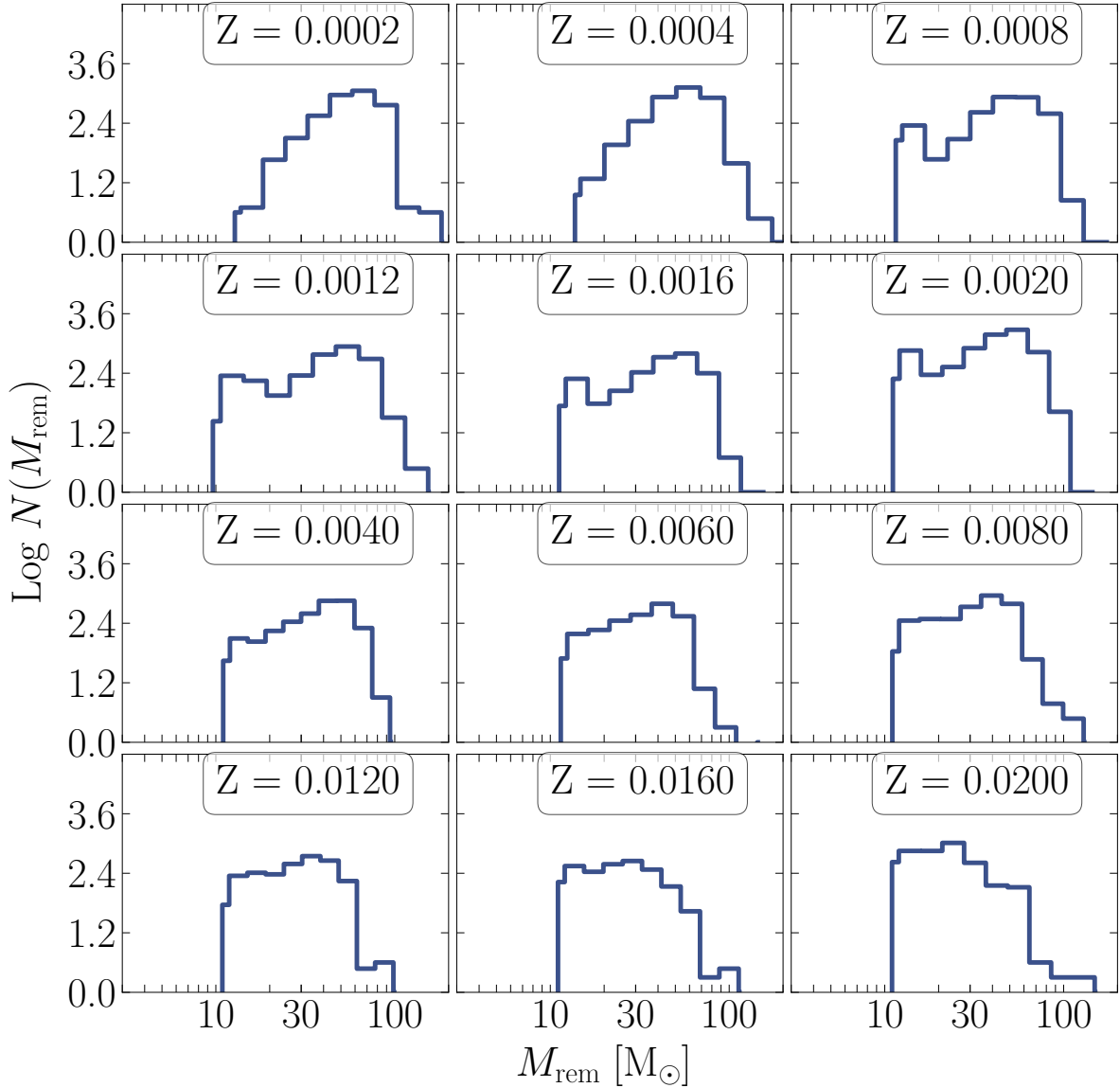}
    \caption{Remnant final mass for different metallicity bins and for the fiducial model. Metallicity increases from top to bottom and from left to right.}
    \label{fig:MZ1}
\end{figure*}

Metallicity is one of the most important ingredients in determining the remnant BH mass. Figure \ref{fig:MZ1} shows the $M_\rem$ distribution for different metallicity bins. The mass distribution at the lowest metallicities shows an evident peak toward values $M_\rem \sim 50-80\Ms$, and an interesting tail to values exceeding $100\Ms$. The high-end of this distribution becomes fainter and fainter at increasing $Z$ values, while at the same time the overall distribution becomes bimodal, acquiring a sizeable population of BHs with masses in the $10-30\Ms$ mass range. At metallicities above $Z\geq 0.008$ the whole distribution shifts toward lower $M_\rem$ values and the high-end tail truncation value of the distribution progressively decreases, reaching  $M_{\rem, {\rm max}}\sim 80 \Ms$ at solar metallicity. 

Unlike the remnant mass, the remnant spin distribution does not show any appreciable dependence on the metallicity, due to the assumption that BH natal spin amplitude is independent on the metallicity or the BBH formation channel.

\subsection{Matching O1+O2}\label{sec:o12}

Using our fiducial model, we now quantify the probability to obtain the currently known population of GW sources with our method. 
We create a sample of 100000 mergers for which we store total mass and mass ratio. To quantify the matching between observations and modelled binaries, we define two different comparison strategies. In the first, for each LVC source\footnote{Data are taken from \url{https://www.gw-openscience.org/}, see also \cite{LIGO19}.} we calculate the fraction of modelled BBHs having a total mass within $30\%$ the observed value. For instance, in the case of GW170104, we find that nearly $57.1\%$ of the modelled mergers have a total mass within $30\%$ the observed value, i.e. $\sim 50\Ms$. In the second, we calculate the fraction of modelled BBHs that have both a total mass and a mass ratio within $30\%$ the observed value\footnote{Note that the error associated to the observed quantities in some cases exceeds $30\%$, especially for the observed mass ratio.}, i.e. $35< M_\bbh/\Ms < 65 $ and $0.45 < q_\bbh < 0.85$ in the case of GW170104.

\begin{figure}
    \centering
    \includegraphics[width=\columnwidth]{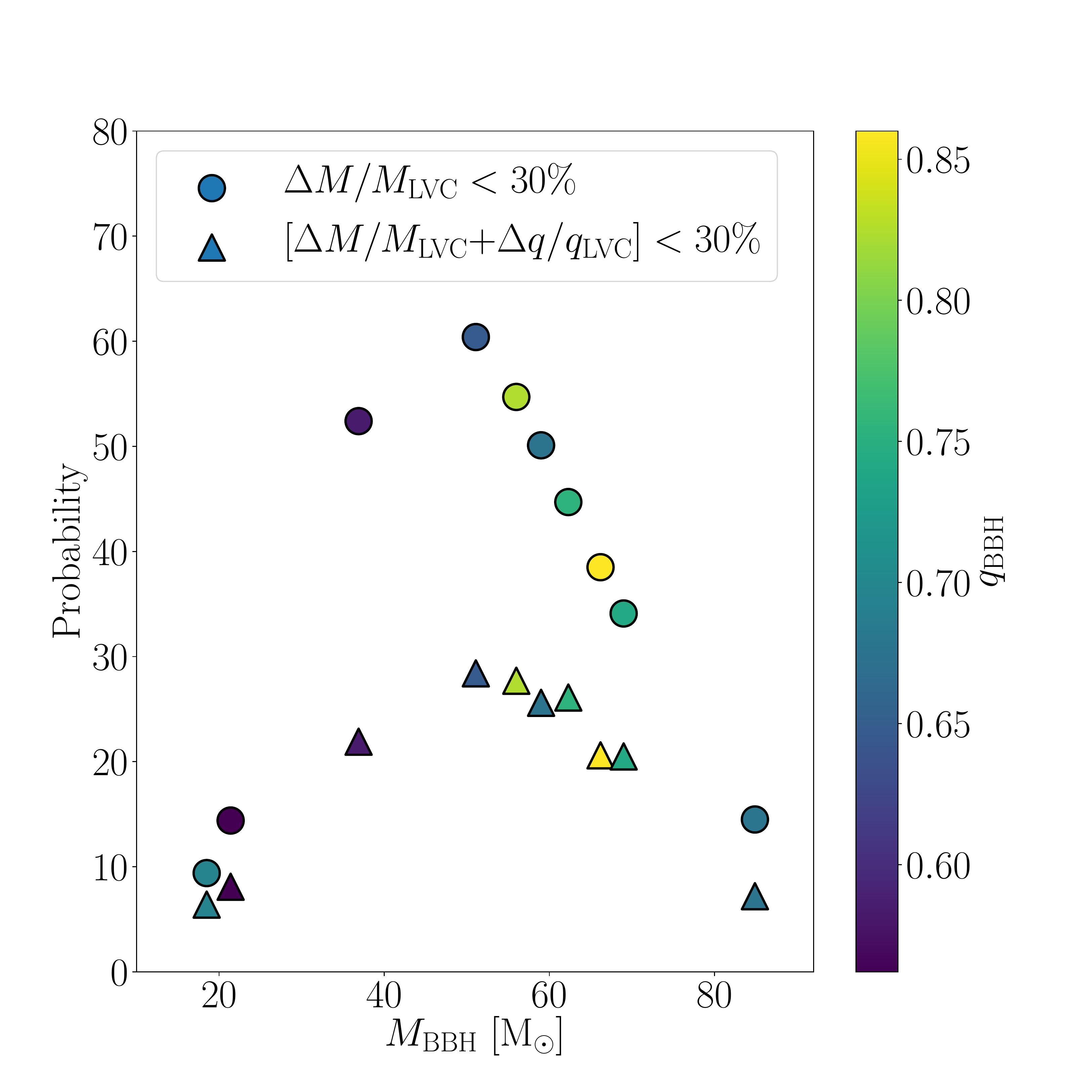} 
    \caption{Probability to select a BBH compatible with one of the 10 confirmed LVC BBHs from O1 and O2 \citep{abbott19}. Each circle or triangle corresponds to one LVC BBH. The color coding marks LVC BBH mass ratios. Triangles (circles) represent the probability to draw -- from the fiducial model sample -- a merger with mass ratio and total mass (total mass only) within $30\%$ from the corresponding observed value.}
    \label{fig:further}
\end{figure}

Figure \ref{fig:further} shows these probabilities for the current population of 10 LVC sources. We find that our fiducial model can match the mass of all O1+O2 BBHs. For instance, mergers with mass and mass ratio similar to the first observed source, GW150914, have a $\sim{}20\%$ probability to be selected in our fiducial model. The probability raises up to $\sim{}40\%$ if we limit the comparison to the BBH mass only. The matching probability is even larger for sources with masses in the range $50-65\Ms$ (GW170104, GW170809, GW170814), while it drops to $\sim 5-10\%$ percent when applied to the lightest mergers, $M_\bbh \lesssim 20\Ms$. 

\subsection{Massive BH remnants}\label{sec:massBBH}
\begin{figure*}
\center
\includegraphics[width=\columnwidth]{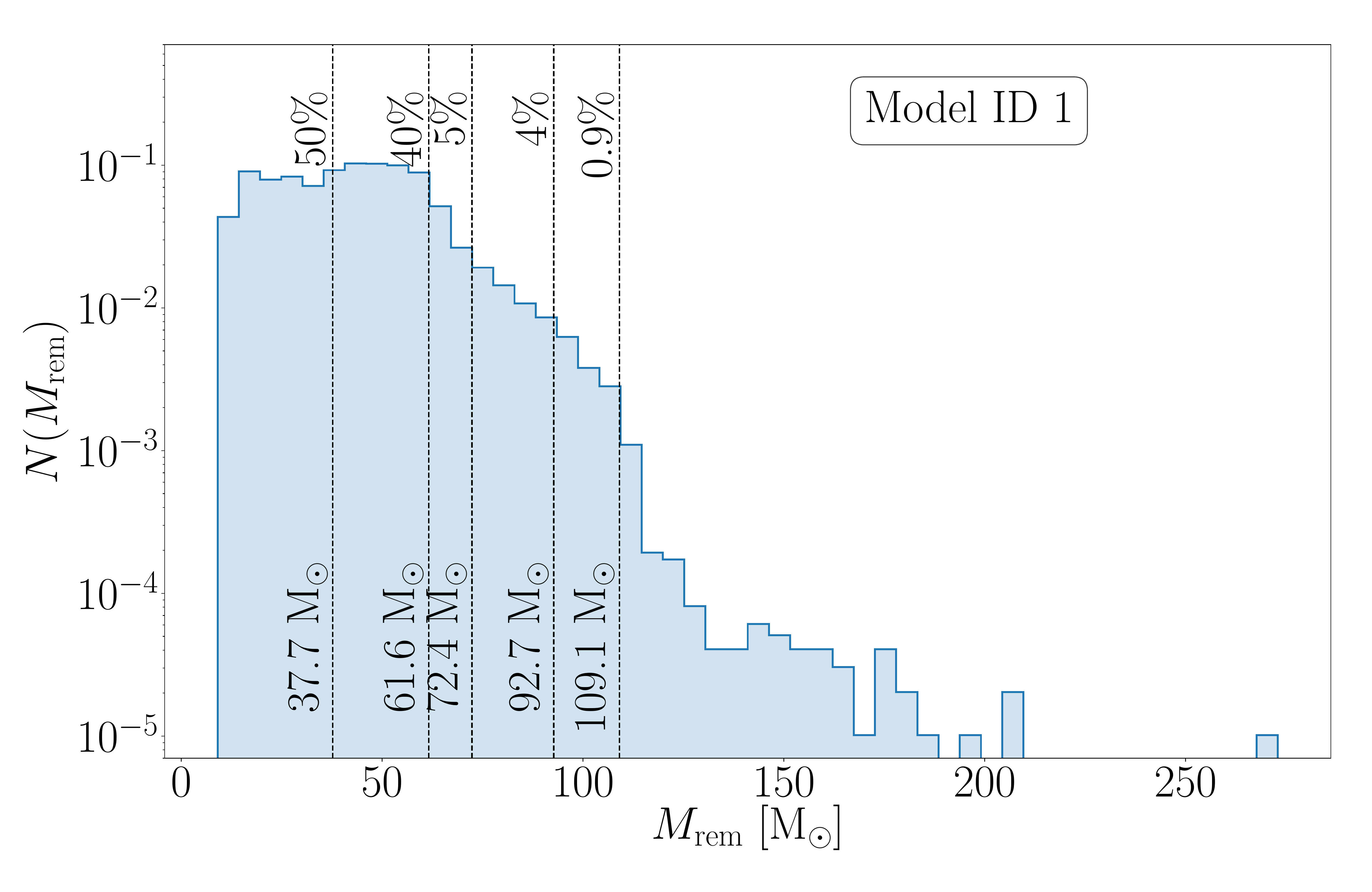}
\includegraphics[width=\columnwidth]{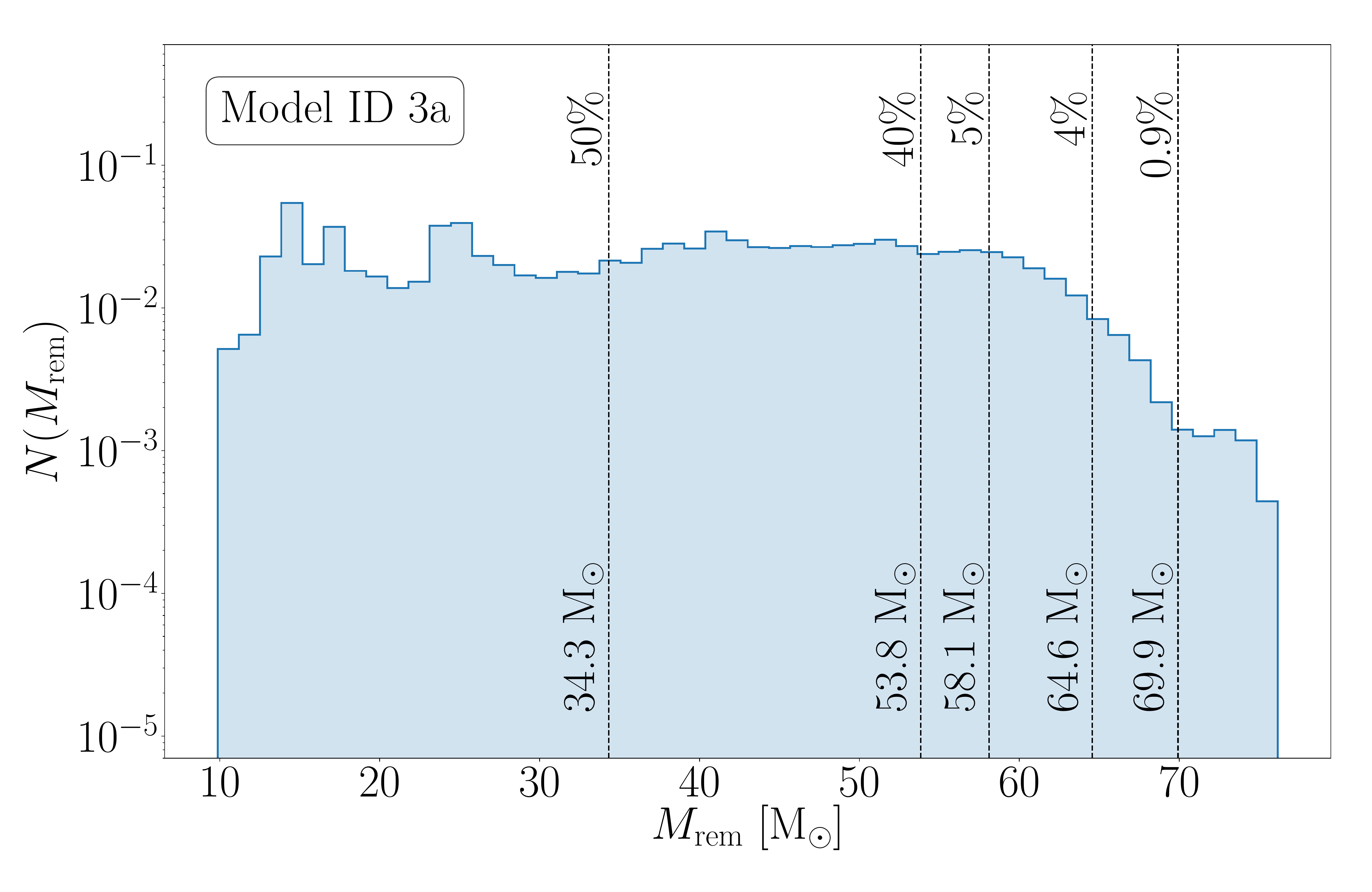}
\includegraphics[width=\columnwidth]{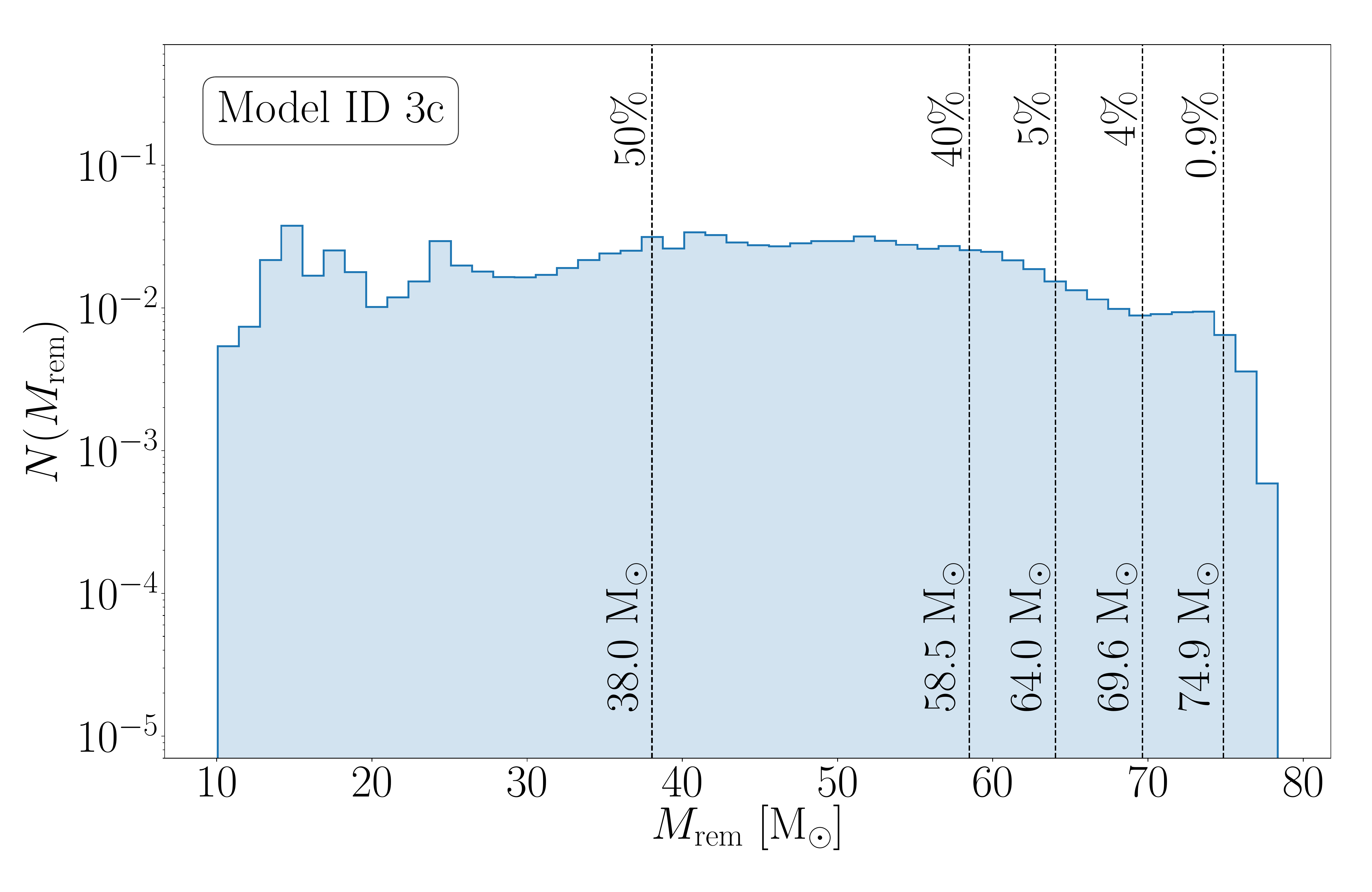}
\includegraphics[width=\columnwidth]{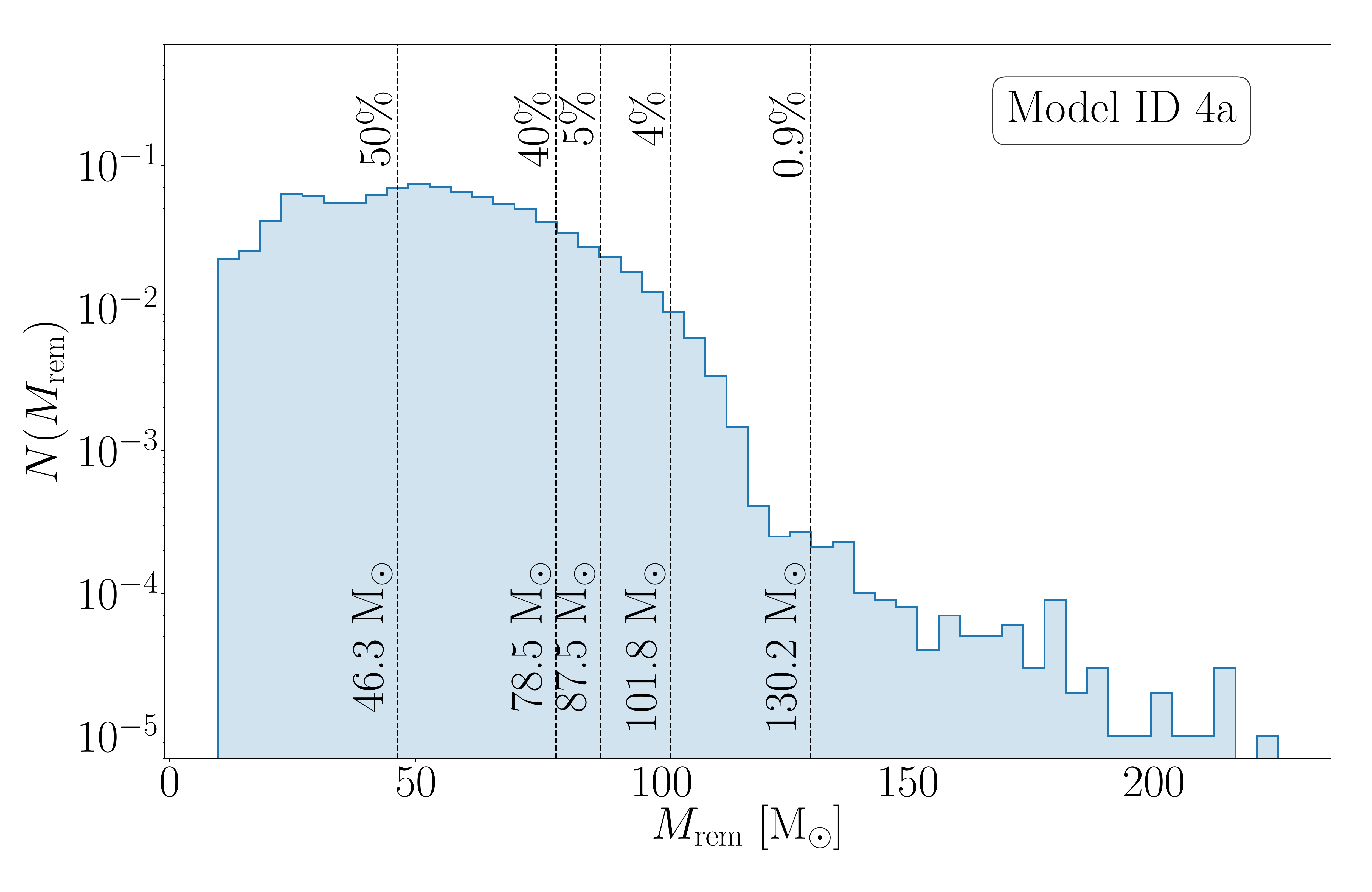}
\caption{Remnant mass distribution for $10^5$ BBH mergers in the fiducial model (ID 1, top left panel), model 3a (top right panel), model 3c (bottom left panel), and model 4a (bottom right panel). The labels indicate the percentage of binaries with remnant mass in a given mass range. Out of 100 BBHs, the fiducial model predicts $5$ mergers with remnant mass $M_\rem > 72\Ms$ and $\sim 1$ with mass above $\sim 93\Ms$. }
\label{fig:O3}
\end{figure*}

One of the most interesting features of our BBH merger products is the possible formation of BHs with masses in the IMBH mass range, i.e. $\sim 10^2\Ms$.
In order to understand the frequency of the formation of such massive merger products in our fiducial model, in Figure \ref{fig:O3} we show  the distribution of remnant masses for $10^5$ BBHs created according to models ID 1 (fiducial), 3a (isolated with $P(Z) = $ SDSS), 3c (isolated with $P({\rm Log} Z) =$ const), and  4a (dynamical). 

Our fiducial model predicts up to 5 merger products heavier than $\sim 93\Ms$ and at least 1 with mass above $\sim 114\Ms$ out of 100 BBH mergers in the local Universe. Observing such a massive BH would provide us with a window on the lowest metallicities, and could represent an exquisite signature of dynamical formation.
Indeed, according to our model, remnant BHs with masses above $80-90\Ms$ come from BBHs formed in metal-poor environments\footnote{It must be noted that such massive BHs can also form as the product of massive main sequence stars collisions \citep{spera18}. In dense clusters, BHs forming via this channel can capture a companion and lead to the formation of even heavier BHs through coalescence, as shown by \cite{dicarlo19}.}, where the contribution of isolated binaries is relatively small, due to the metallicity distribution in the local Universe (see also Figure \ref{fig:MZ1}).

Compared to the fiducial model, the $M_\rem$ distribution for dynamical models (4a) shows a tail that extends to larger values of the remnant mass. If dynamical mergers dominate the global population of BBH mergers, this model predicts at least 1 BH remnant with mass beyond $140\Ms$ out of 100 mergers. In contrast, isolated models (ID 3a and 3c) produce a narrower distribution of $M_\rem$, limited to roughly $75-80\Ms$, a limit set by the choice of the binary stellar evolution recipes implemented in \mobse{}.
Given the evident differences in the merger mass distribution among dynamical and isolated binaries, we calculate the percentage of BBHs with remnant mass in the range $\leq 35\Ms$, $35-50\Ms$, $50-75\Ms$, $\geq 75\Ms$ for models ID 1, 3a, 3c, and 4a.
\begin{table}
    \centering
    \caption{Percentage of mergers with a remnant mass in different mass ranges}
    \begin{tabular}{cccccc}
        \hline
        \hline
        ID & $N_{obs}$ & $P_{<35}$ & $P_{35-50}$ & $P_{50-75}$ & $P_{\geq75}$  \\
           &           & ($\%$) & ($\%$) & ($\%$) & ($\%$) \\
		\hline
		1  & 100 & 36.1 & 28.0 & 30.1 & 5.8\\
		3a & 100 & 41.9 & 30.8 & 27.2 & 0.0\\
		3c & 100 & 33.2 & 31.5 & 34.7 & 0.6\\
		4a & 100 & 26.1 & 21.9 & 35.2 & 16.8\\
		LVC& 10  & 20.0 & 20.0 & 50.0 & 10.0\\
		\hline
        \hline
    \end{tabular}
    \label{tab:mremo3}
\end{table}
Table \ref{tab:mremo3} summarizes the percentage of BBHs with a remnant mass in different mass ranges and for different models. Comparing the fiducial (ID~1), isolated only (ID 3a and 3c) and dynamical only (ID~4a) models makes apparent a striking difference between the predicted percentage of binaries with either low ($<30\Ms$) or large mass ($\geq 70\Ms$). The models in which we assume that the  merger population is mostly composed of isolated binaries (ID 3a and 3c) predict $\sim 65-75\%$ of mergers with $M_{\rem}<50\Ms$, and almost no binaries with $M_{\rem}\geq 75\Ms$. The percentage of mergers with masses falling in the low-end and high-end tail of the mass distribution can be extremely useful to place constraints on the contribution of dynamical mergers to the overall population. In the fiducial model, for instance, we find that the percentage of mergers lying in the high-end tail of the mass distribution ($P_{\geq 75} = 5.8\%$) is $\sim{}1/6$ of the percentage of mergers lying in the low-end tail of the distribution ($P_{\geq 35} = 36.1\%$). 

In model 4a, where dynamical mergers dominate the overall population, the percentage of mergers with masses $>50\Ms$ ($52\%$) and $<50\Ms$ ($48\%$) is very similar, and the heaviest mergers ($P_{\geq 75} = 16.8\%$) are $64\%$ of the lowest mass BBHs ($P_{\leq 35} = 26.1\%$).
Therefore, we expect that 
\begin{itemize}
    \item if more than $60\%$ of BBH mergers have mass $<50\Ms$, the isolated channel outweighs the dynamical one;
    \item the absence of remnants with masses above $75\Ms$ would imply a negligible contribution of dynamical mergers to the overall population;
    \item a comparable number of mergers with $M_{\rem}<35\Ms$ and $M_{\rem}\geq 75\Ms$ suggests that dynamical binaries  are the majority of the overall population.
\end{itemize}

As shown in Table \ref{tab:mremo3}, 60\% of the 10 confirmed BBHs in O1 and O2 have merger remnants heavier than $50\Ms$  \citep{abbott19}, showing an interesting similarity with our model 4a (dynamical mergers only). 

\subsection{A multi-merger route to the formation of GW170729--like sources} \label{sec:gw170729}

One of the most interesting sources detected by the LVC is GW170729, a BBH merger that left behind a highly spinning ($a_\rem = 0.81^{+0.07}_{-0.13}$) and massive ($M_\rem = 80.3^{+14.6}_{-10.2}\Ms$) BH.
Compared to the global distribution of mergers shown in Figure \ref{fig:fiducial}, sources of this kind have a relatively low probability to form. Indeed, less than $10\%$ of the simulated sources in the fiducial model have $M_\rem>80\Ms$.

\begin{figure}
    \centering
    \includegraphics[width=0.85\columnwidth]{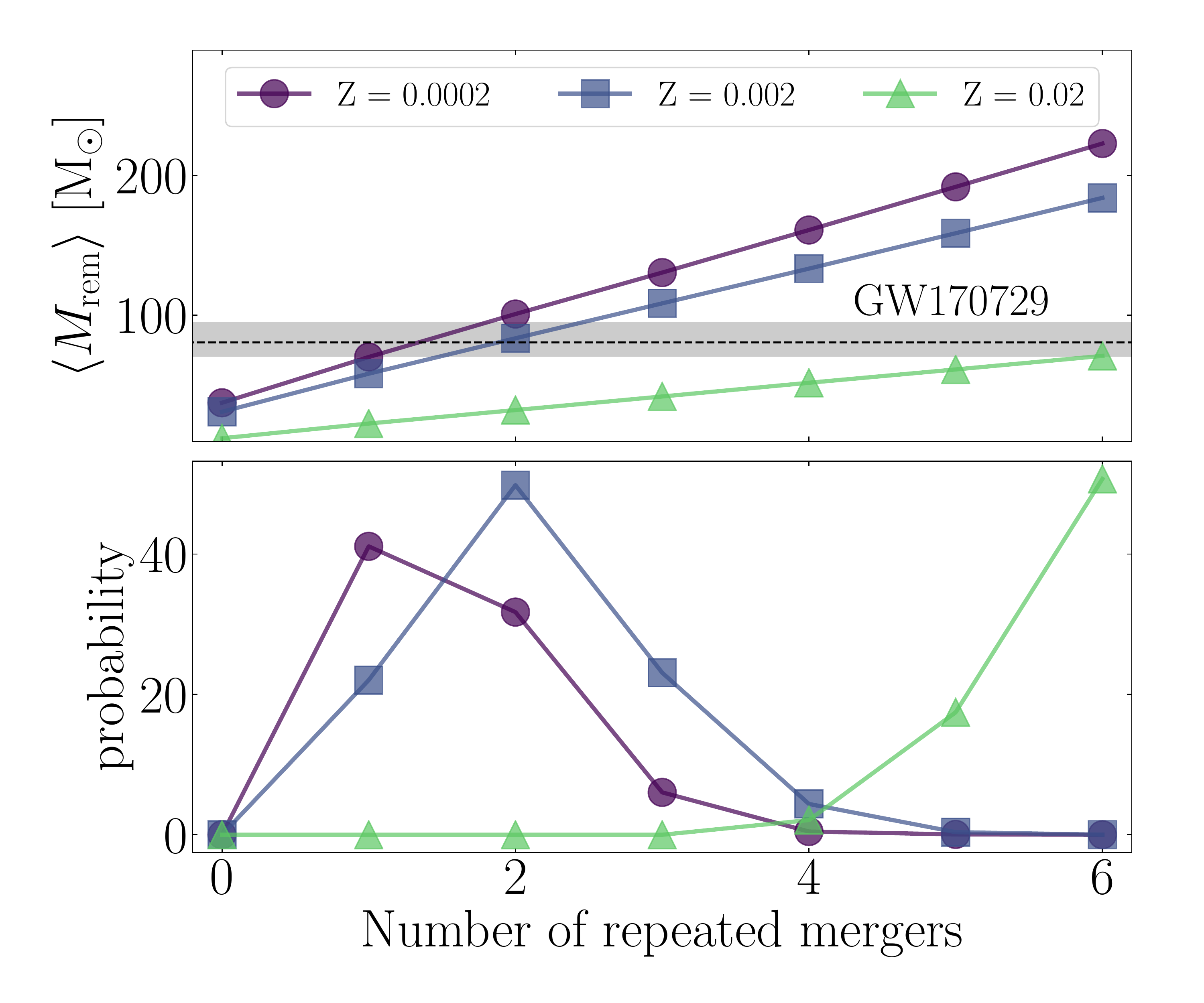}
    \includegraphics[width=0.85\columnwidth]{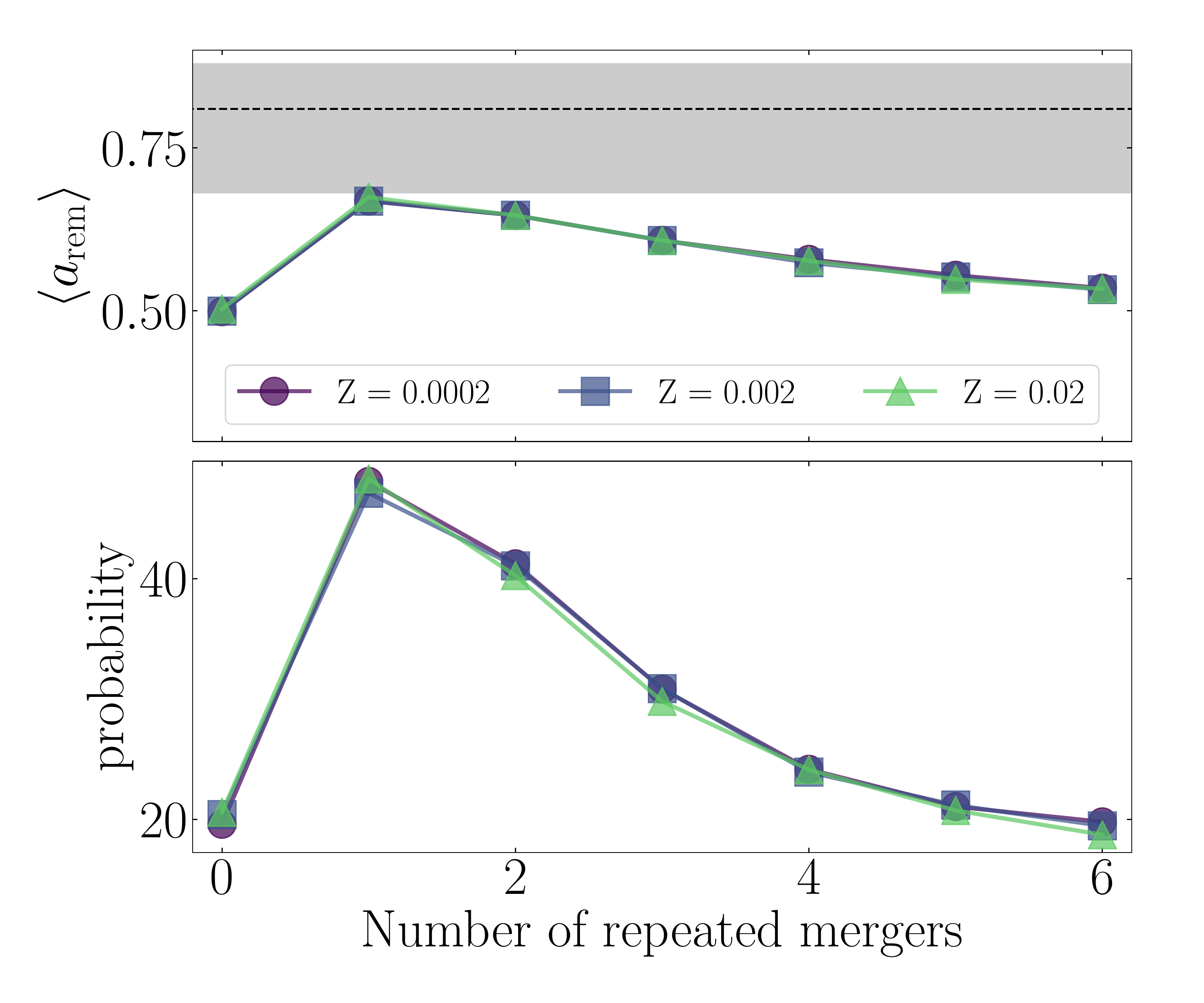}
    \includegraphics[width=0.85\columnwidth]{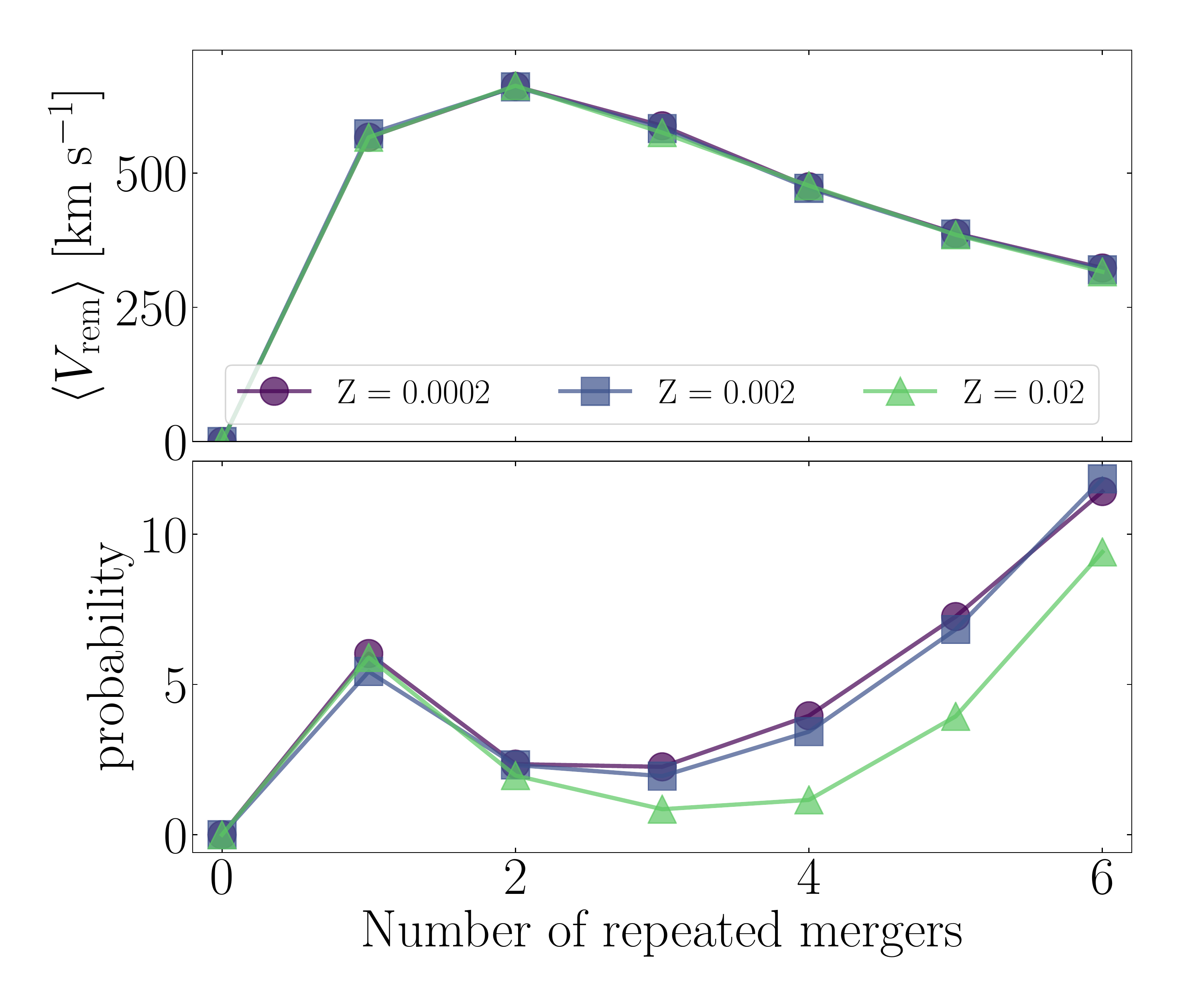}
    \caption{Top panel: Remnant BH mass (upper box) and matching probability for GW170729 (lower box) as a function of the number of mergers. The shaded area encompasses the $90\%$ credible interval of the GW170729 mass. Central panel: same as above, but for the remnant BH spin. Bottom panel: average GW recoil velocity for all modelled mergers (upper box) and probability that the kick remains below 100 km s$^{-1}$ (lower box), thus allowing a next-generation merger. }
    \label{fig:gw170729}
\end{figure}

One possibility is that GW170729 formed dynamically, possibly from a progenitor that underwent multiple mergers. In the following, we use our machinery to test this hypothesis.

We use the BH mass spectrum from \mobse{} to select the initial mass of two BHs, assuming that they merge inside a star cluster. We select the primary BH mass in the range $10-65\Ms$, while the companion mass is extracted assuming a flat mass ratio distribution. Spins for both BHs are drawn with uniform amplitude between 0 and 1 and with isotropically distributed orientations. 
We calculate the remnant mass and spin, and associate a new companion from the same mass spectrum to mimic a second merger. To mimic a sequence of mergers, we repeat this procedure $n_{\rm mer}$ times, assuming three values for the metallicity ($Z ~=~ 0.0002,~0.002,~0.02$). For each $Z$ value, we create 10,000 merger trees and we calculate for each $n_{\rm mer}$ value the mean mass ($\langle M_{\rm rem} \rangle$) and spin ($\langle a_{\rm rem} \rangle$) of the remnant BH. Moreover, we calculate the probability to form a remnant BH with a mass, or spin, within $30\%$ from the observed value for GW170729. These quantities are shown in Figure \ref{fig:gw170729}. 
The upper box in the top panel highlights that $\langle M_{\rm rem} \rangle$ values compatible with GW170729 can be achieved with 1--2 mergers in the metallicity range $0.0002-0.002$, while at least 5--6 repeated mergers are needed to explain such large mass in an environment characterized by a solar metallicity. Indeed, as shown in the lower box of the top panel, the probability to form a BH with remnant mass close to GW170729 is $\sim 40-60\%$ if the number of mergers is $n_{\rm mer} = 1-2$ and the metallicity is low ($Z<0.002$), while for solar metallicities the probability ranges between $20\%$ and $60\%$ assuming $n_{\rm mer} = 5$ or $6$, respectively. The average remnant spin, shown in the lower panel of the Figure, exhibits a peak at $n_{\rm mer} = 1-2$, regardless of the metallicity, where $\langle a_{\rm rem} \rangle \sim 0.65$, i.e. slightly off the observational error. This quantity reduces as we increase the number of mergers, thus limiting the possibility for GW170729 to have originated through more than 2 subsequent mergers. 

Furthermore, the post-merger recoil kick can eject the remnant BH outside the star cluster. Assuming that the progenitors of GW170729 formed in a nuclear cluster, we have calculated the recoil velocity $v_k$ and the probability that $v_k < v_\mathrm{max} \equiv 100$ km s$^{-1}$ for each BBH and for each merger. The upper box of the bottom panel in Figure \ref{fig:gw170729} shows the average recoil kick received at the last merger, while the lower box indicates the probability that such velocity falls below 100 km s$^{-1}$. Looking at the bottom box, we see that there is a $5\%$ probability for a BH formed via a ``first generation'' merger to get a kick below 100 km s$^{-1}$, whereas this probability falls to $\sim 2\%$ if the BH is originated via two subsequent mergers. To determine the combined probability for a BH to undergo a series of mergers we need to multiply the probability to be retained at every step. This implies that, in a typical NC, the probability for a BH to be retained after one merger event is around $\sim 5\%$, whereas the probability to be retained after two successive mergers is around $5\% \times 2\% = 0.1\%$.

Figure \ref{fig:gw170729} suggests that GW170729 likely formed in a metal-poor environment ($Z<0.002$), such as a dense globular cluster, via either a single merger or 2 subsequent mergers.

\section{Discussion: quantifying the uncertainties} \label{sec:discus} 
In this section, we discuss how  BBH formation channels, distribution of galaxy metallicity, and merger probability-metallicity correlation affect our results.

\subsection{Impact of metallicity} \label{sec:met}

Metallicity is one of the parameters that most influences the merger remnant mass and spin distribution. As discussed in previous sections, the Ansatz behind our fiducial model is that the metallicity distribution of merging BBHs depends on the formation channel, and that a larger merger probability corresponds to a lower metallicity.

In order to quantify the role of metallicity in shaping the $M_\rem-a_\rem$ plane, let us assume that the merger probability does not depend on the metallicity, namely that the average number of mergers in different metallicity bins is nearly constant, $f(Z)=$const. This implies that the metallicity distribution of galaxies in the local Universe provides a one-to-one match to the metallicity of merging BBHs\footnote{Note that we refer to the metallicity of the stellar progenitor.}. 

\begin{figure}
    \centering
    \includegraphics[width=7cm]{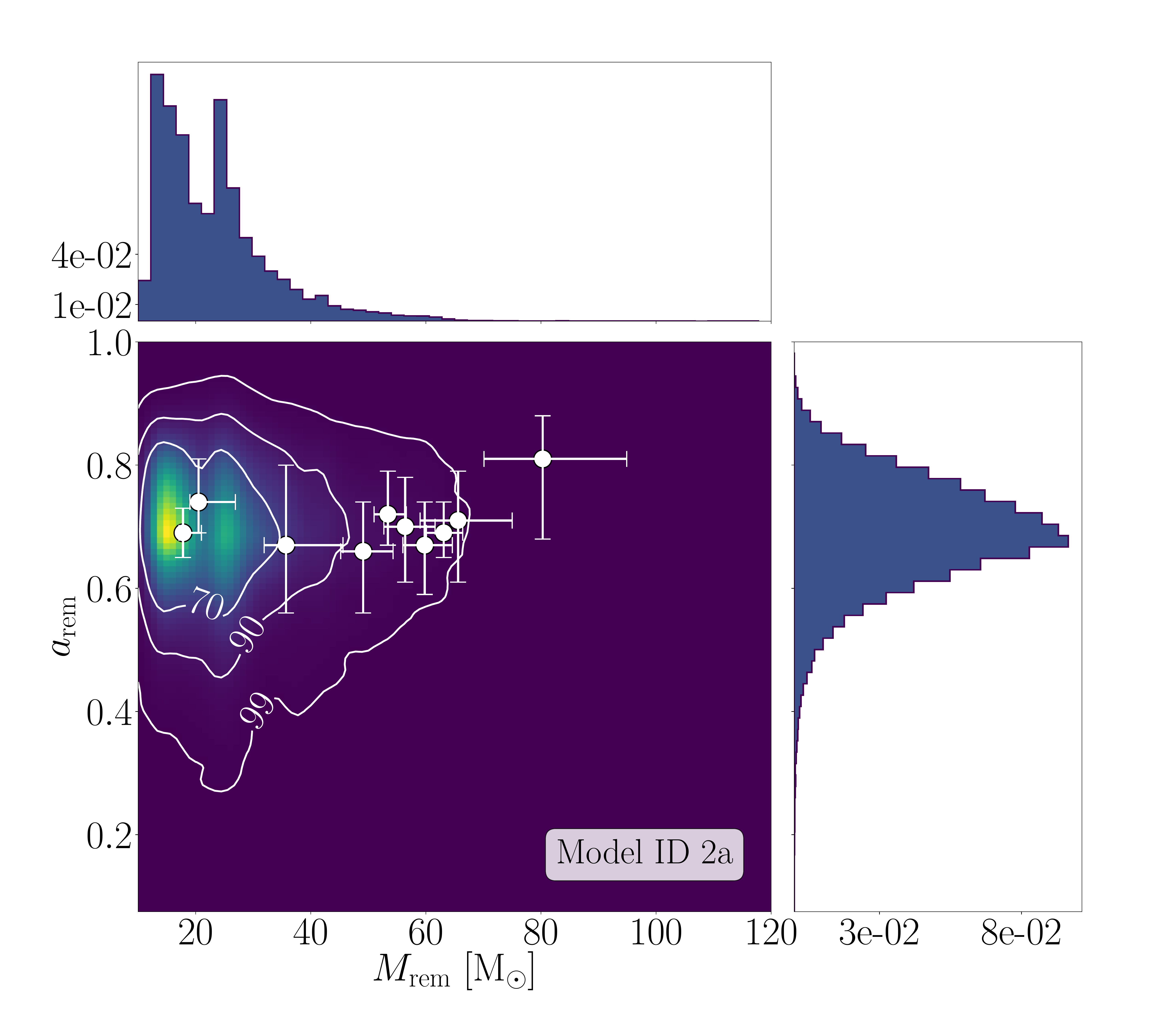}
    \includegraphics[width=7cm]{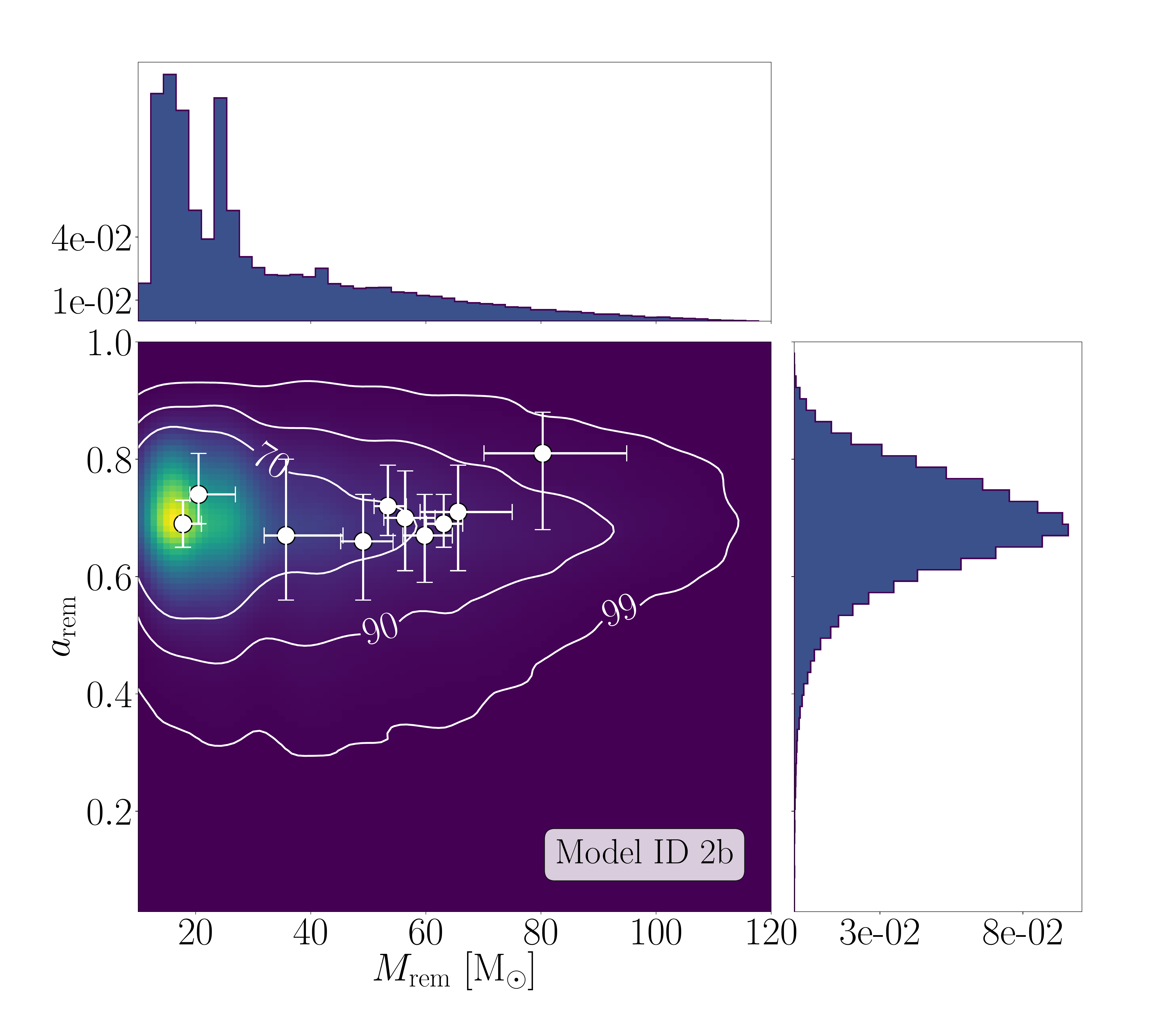}
    \includegraphics[width=7cm]{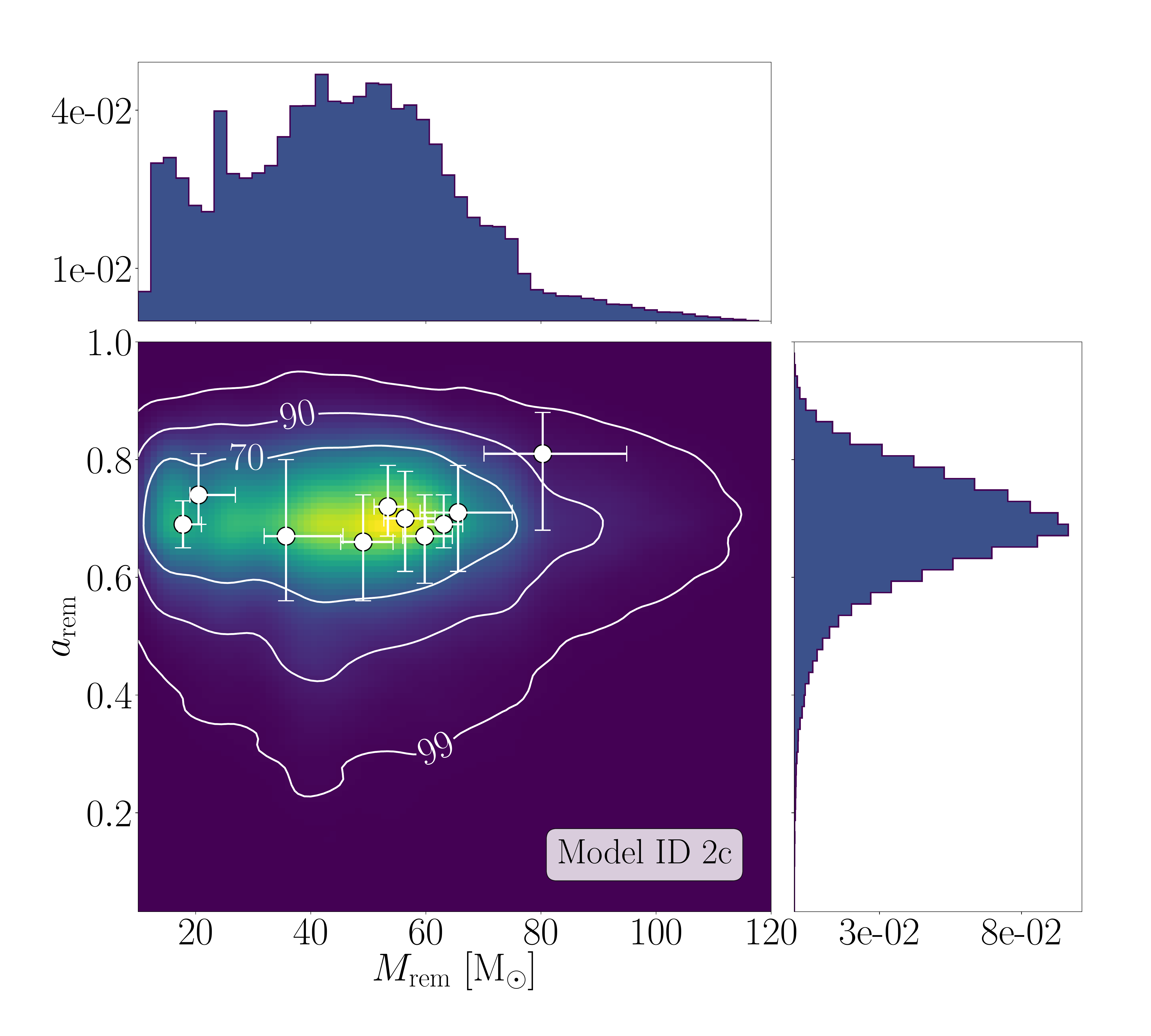}
    \caption{Remnant mass and spin plane in the assumption that the merger probability is independent of host metallicity. Top panel: galaxies and star clusters follow the observed $Z$ distribution, regardless of clusters' type (set ID 2a). Central panel: galaxies and open clusters follow the observed $Z$ distribution, while nuclear and globular clusters follow a logarithmically flat distribution (set ID 2b). Bottom panel: galaxies and star clusters follow a $Z$ distribution flat in logarithms, regardless of clusters' type (set ID2c). }
    \label{fig:nocorr}
\end{figure}

Under this assumption, we explore three different cases. In the first case, we assume that star clusters and host galaxies are characterised by the same metallicity distribution, regardless of clusters' type (set ID 2a). In the second case, we assume that metallicity of globular and nuclear clusters is equally distributed in logarithmic bins, while the open clusters and the Galaxy have $Z$ distributed according to SDSS observations (set ID 2b). In the third case, we assume that both galaxies and star clusters of all types follow a distribution flat in logarithmic bins (set ID 2c). The latter distribution serves to show how the merged BH population would change if the population of metal-poor galaxies inhabiting the volume scanned by the LVC contribute as much as metal-rich systems.
Figure \ref{fig:nocorr} shows how the $M_\rem-a_\rem$ plane would change in consequence of such choices. 

As is apparent from Figure~\ref{fig:nocorr}, assuming that environments' metallicity does not impact the merger probability, namely $f(Z)=1$, has strong implications for the $M_\rem - a_\rem$ plane. Indeed, it seems hard to reconcile the LVC detections with isolated and dynamical formation channels if we assume that both galaxies and star clusters have a $Z$ distribution shifted toward solar values as shown in SDSS observations. In this case, indeed, only the lightest BHs detected fall in the maximum of the $M_\rem-a_\rem$ plane, with the mass distribution peaking at $\sim 20-30\Ms$, as shown in the top panel of the figure. Assuming that star clusters have a flatter metallicity distribution, as in the fiducial model, but still assuming no metallicity - merger probability relation leads to a slightly broader $M_\rem$ distribution (central panel), still quite incompatible with observations. A way to obtain an $M_\rem - a_\rem$ distribution that embraces detected sources is to assume that the metallicity distribution of both galaxies and star clusters is flat in logarithms. For instance, by assuming a flat distribution in logarithmic metallicity values for both galaxies and star clusters, as shown in the bottom panel of Figure \ref{fig:nocorr}, we find that 9 observed sources lie in the region containing half the total number of mock sources, thus implying a much better comparison with observations.

\subsection{Impact of the formation channel} \label{sec:channel}

\begin{figure}
    \centering
    \includegraphics[width=8cm]{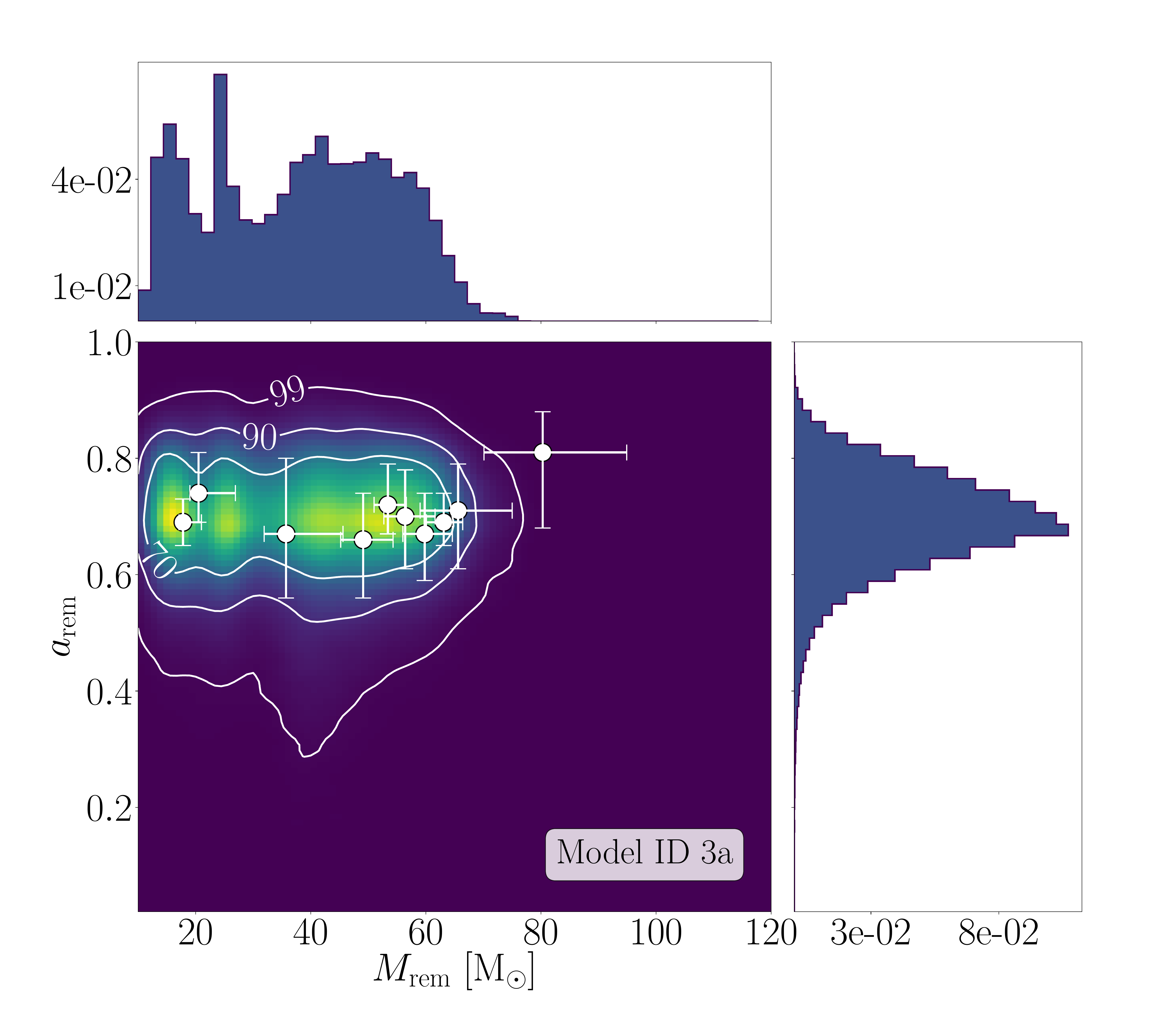}    \\
    \includegraphics[width=8cm]{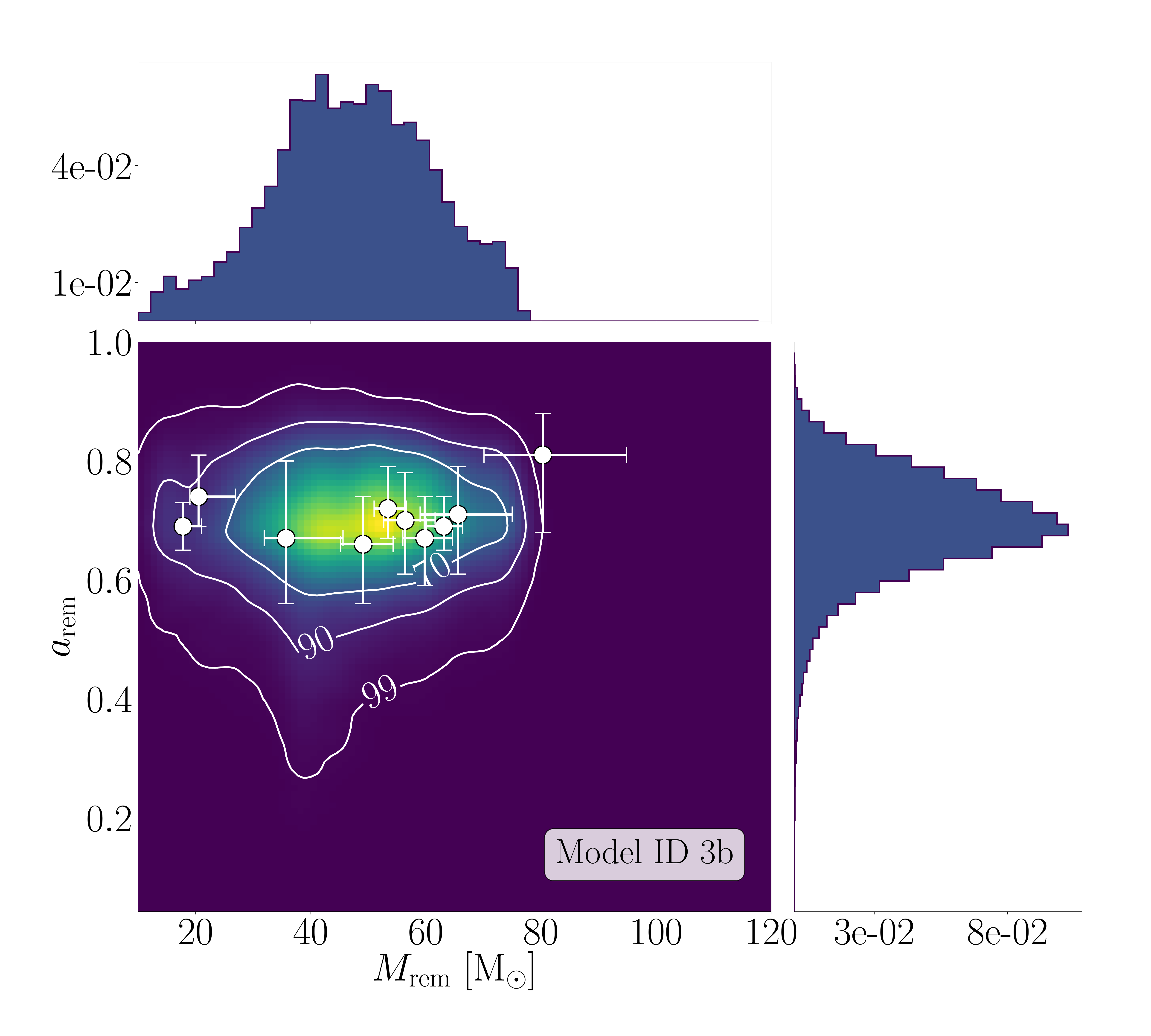}  \\
    \includegraphics[width=8cm]{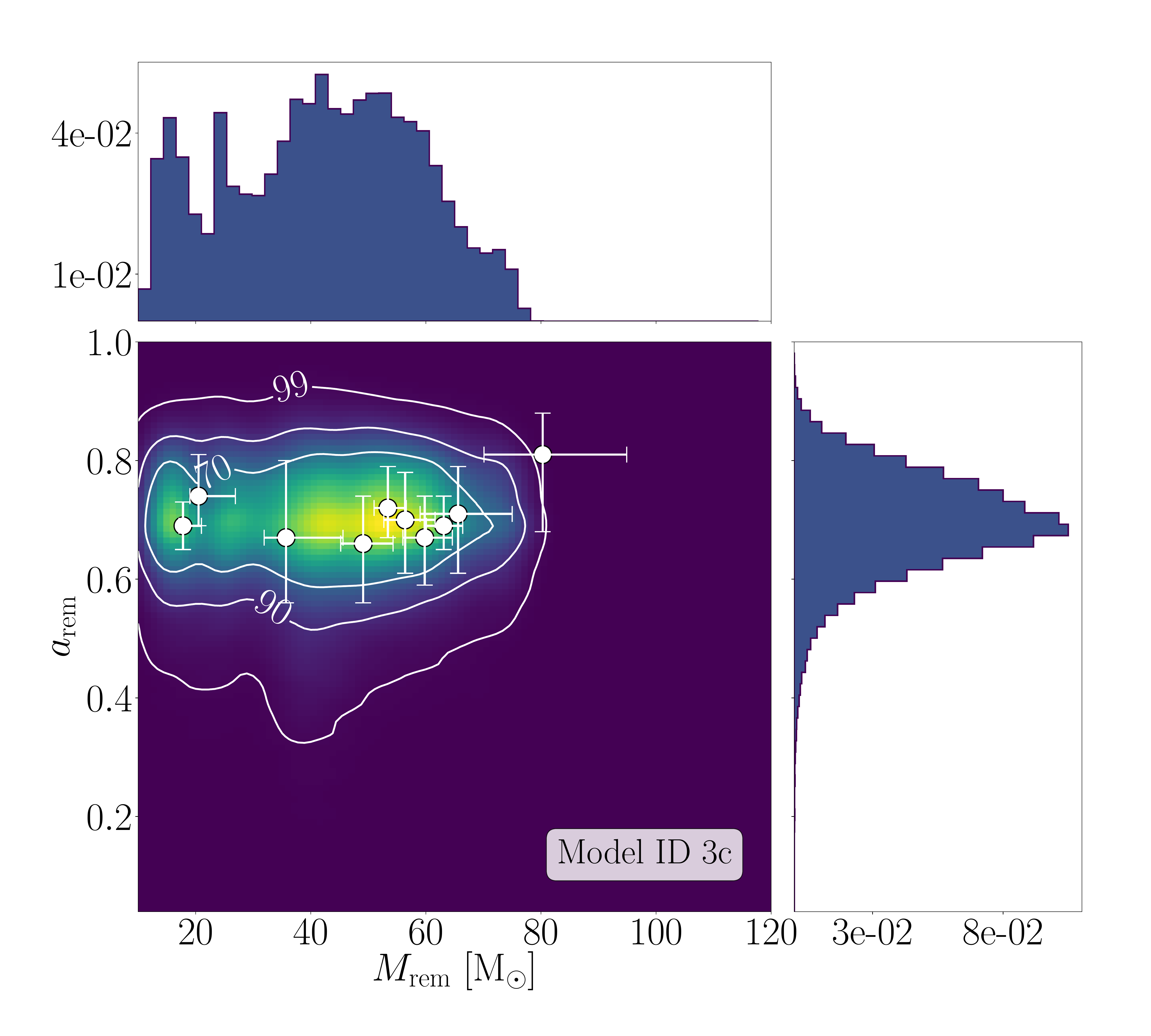}    \\
    \caption{Remnant mass and spin distribution for models ID 3a (top panel), 3b (central panel), 3c (bottom panel). }
    \label{fig:new}
\end{figure}

In this section we discuss the impact of the formation channel in determining the $M_\rem-a_\rem$ plane. We assume either that all the mergers originate in the field ($f_\iso=1$, set ID 3) or in star clusters ($f_\dyn=1$, set ID 4a), leaving all the other parameters unchanged with respect to the fiducial model.
As shown in Figure \ref{fig:new}, the two channels produce significantly different patterns in the plane. This is due to two main factors: i) the metallicity distribution, which is assumed to be different for galaxies and globular/nuclear clusters; ii) the assumed correlation between  observation probability and BBH primary mass.

\begin{figure*}
    \centering
    \includegraphics[width=0.4\textwidth]{fiducial_isolated}        \includegraphics[width=0.4\textwidth]{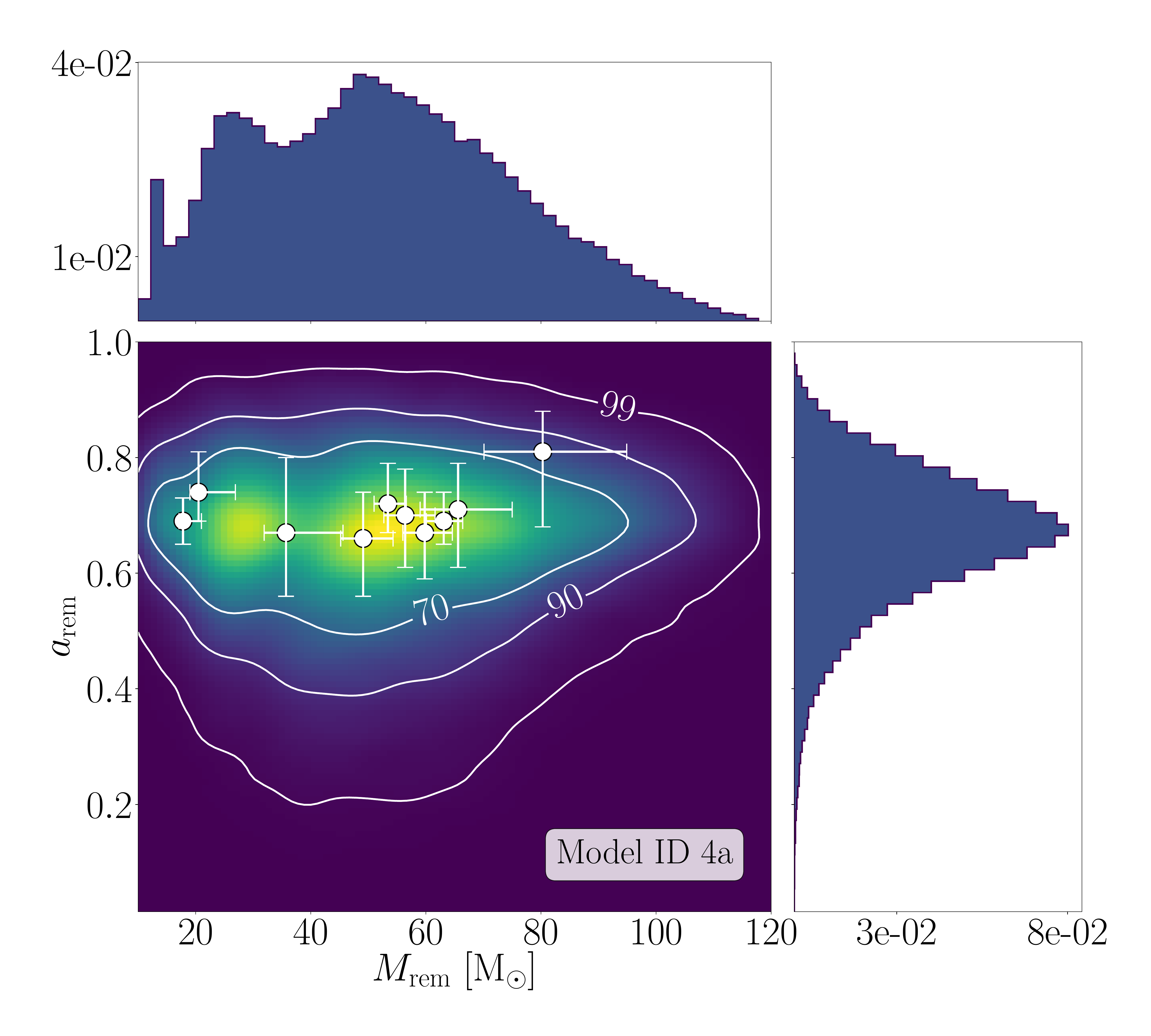}\\
    \includegraphics[width=0.3\textwidth]{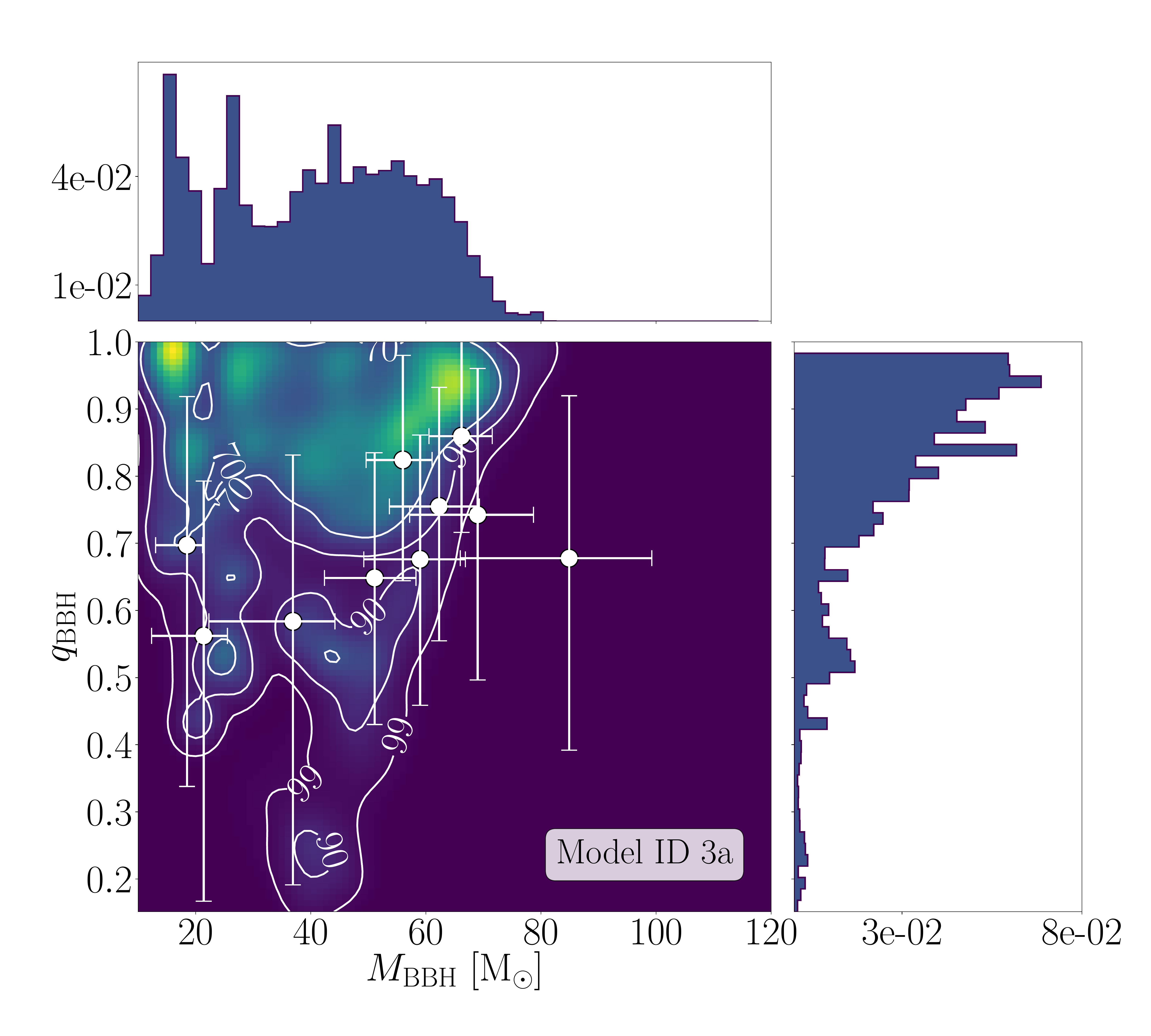}      \includegraphics[width=0.3\textwidth]{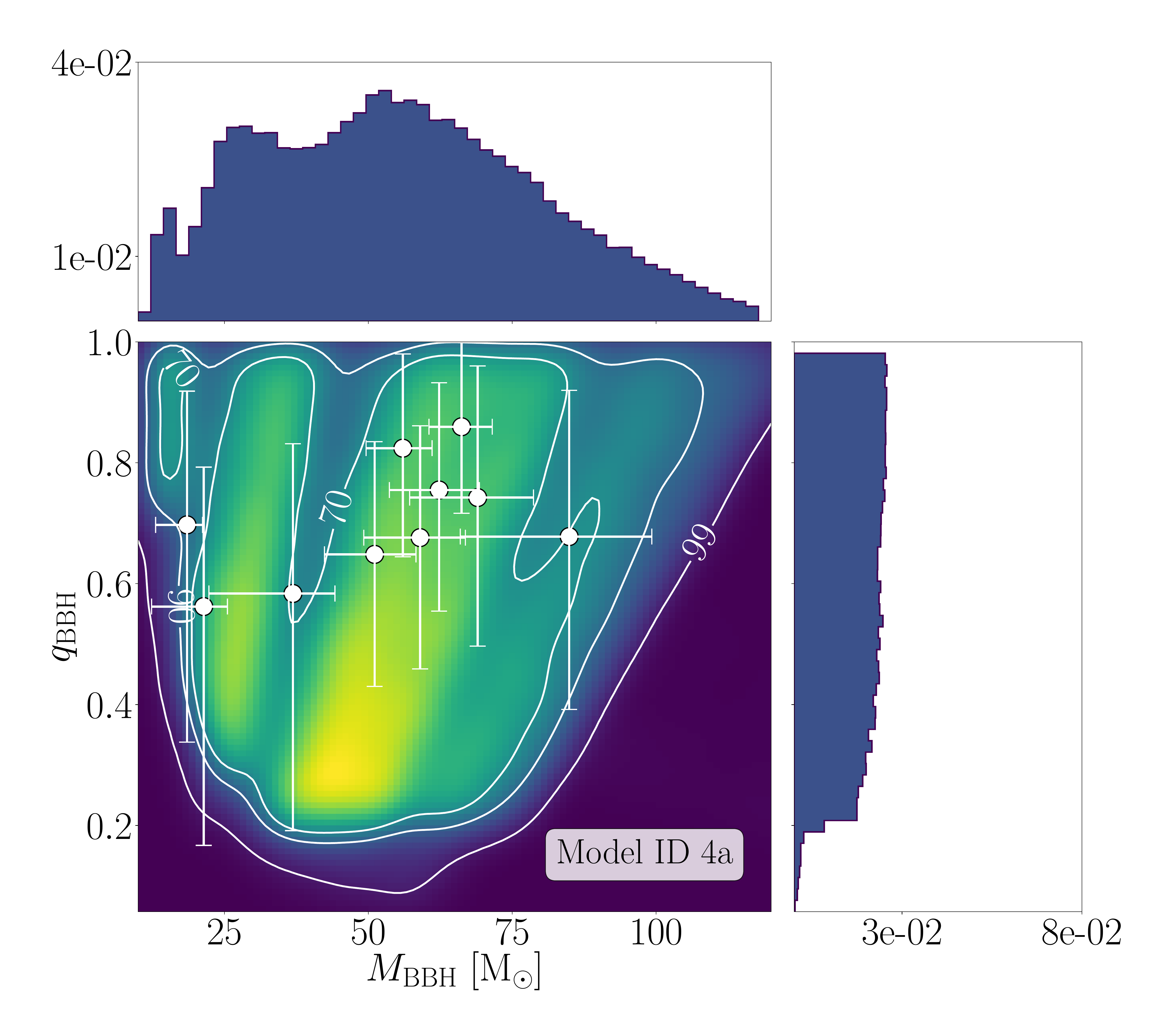}
    \includegraphics[width=0.3\textwidth]{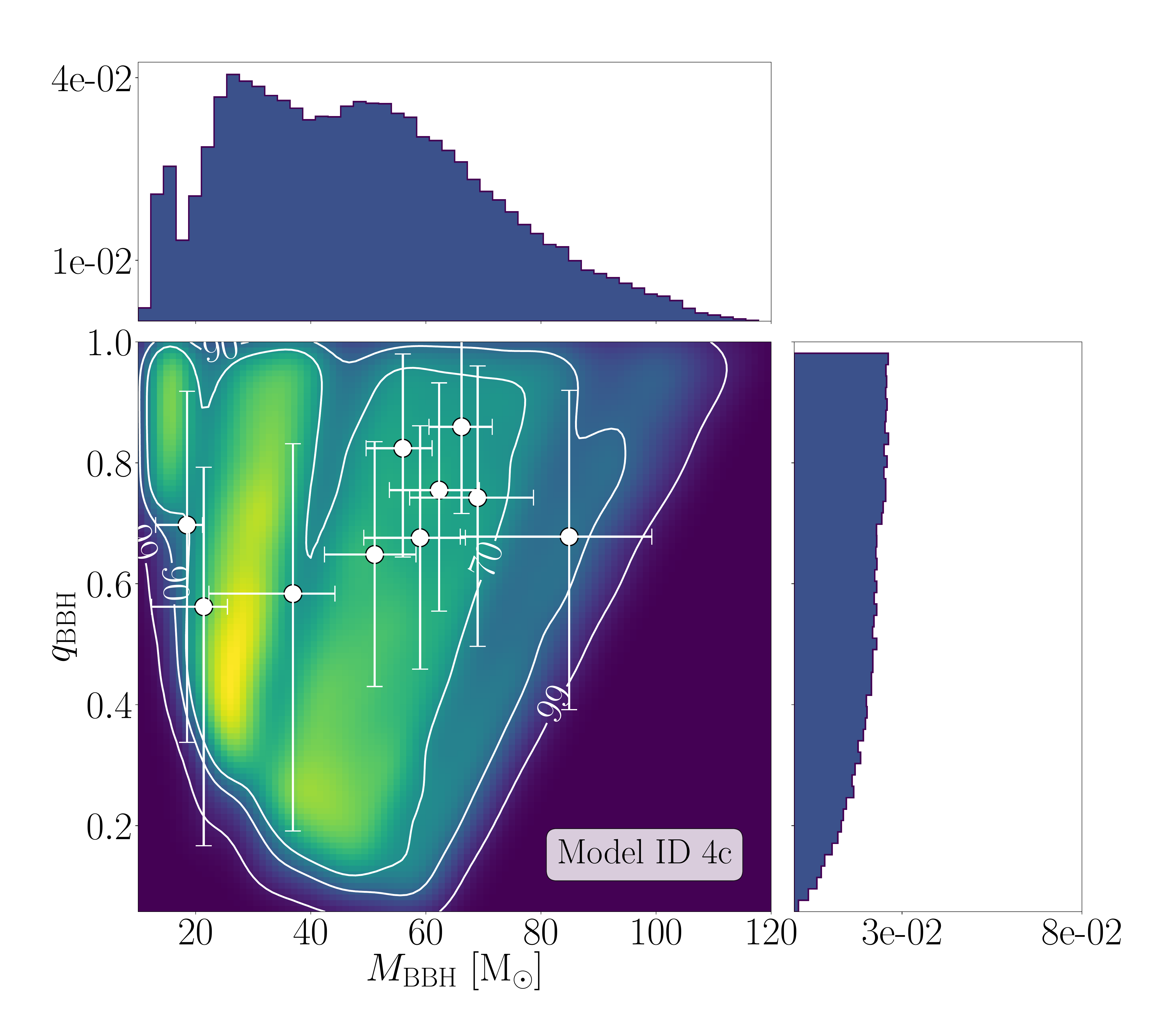}
    \caption{Top row: $M_\rem - a_\rem$ plane assuming $100\%$ of isolated mergers (left-hand panel) or $100\%$ of dynamical mergers (right-hand panel). Bottom row: total BBH mass as a function of its mass ratio for $100\%$ of isolated mergers (left-hand panel), $100\%$ of dynamical mergers (central panel), $100\%$ of mergers coming from young clusters (right-hand panel).} 
    \label{fig:formation}
\end{figure*}

The $M_\rem$ distribution for isolated binaries shows three peaks at $15,~25$ and $\sim 50\Ms$, and an abrupt decrease at values $\gtrsim 65\Ms$. Due to the sharp truncation at high mass, it seems quite hard to explain heavy remnants with an isolated origin, unless we assume that most observed BBHs formed several Gyr ago from metal-poor progenitors. Previous results exploring the cosmic evolution of merging BBHs pointed out that at least half the total mergers in the local Universe formed from metal-poor progenitors at high-redshift \citep{mapelli18,mapelli19}. 
We stress that our work likely underestimates the contribution of BBHs that formed at high redshift and merge in the local Universe, because our methodology cannot model the delay time self-consistently.
Nonetheless, in our models we can capture the contribution of high-redshift, metal-poor progenitors by modifying the metallicity distribution of galaxies. To this end, we  explore three different possibilities, namely i) that the distribution of galaxy metallicity equals that of local Universe galaxies as observed in the SDSS (model ID 3a), ii) that the distribution of galaxy metallicity is shifted to values three times smaller than observed in the local Universe (model ID 3c)\footnote{This corresponds to the case in which the metallicity distribution of BBH progenitors peaks at around $Z=0.001$ as discussed in \cite[][see their Figure 4]{mapelli19}}, and iii) that the metallicity distribution is flat in logarithmic values (model ID 3b). In all three cases, we include the corrective function $f(Z)$ to the merger probability $P(Z)$. Figure \ref{fig:formation} shows the remnant mass and spin distribution for all these models.

If all the observed sources have a dynamical origin, we find that 7 out of 10 detections fall in the maximum of the distribution. The $M_\rem$ distribution in this case is very broad, with a single peak at $\sim 60\Ms$ and a tail extending up to $\lesssim 200\Ms$. A dynamical origin provides a suitable explanation for the heaviest merger product observed so far, GW170729, as it falls in a region encompassing $70\%$ of all the modelled sources. The spin distribution is broader compared to isolated binaries, being characterised by a FWHM $\sim 0.2-0.25$ and a peak at 0.7. 

Clear differences among different channels and assumptions emerge also comparing the distribution of binary masses $M_\bbh$ and mass ratios $q_\bbh$, as shown in bottom row panels in Figure \ref{fig:formation}. For comparison, we also show the binary combined mass--mass ratio distribution for binaries forming only in young clusters (set ID 4c).

Isolated binaries (left bottom panel) seem to preferentially form nearly equal mass mergers, being the mass ratio distribution characterized by a nearly flat distribution in the $q_\bbh \simeq 0.2-0.7$ range, a steep rise at larger values and a peak at $q_\bbh = 0.9$ and BBH masses $M_\bbh = 63\Ms$. 

Dynamical binaries (central bottom panel), instead, show a broad distribution in mass ratio values that embraces heavy detections. It must be noted that in our model, such large distribution is obtained by construction, as we assume that the BBH mass ratio is randomly distributed between 1 and a minimum value $q_\Min$. Also, we note that the mass distribution, compared to isolated binaries, extends to values larger than $100\Ms$. This is due to the combined effect of two assumptions: i) isolated BBH masses are calculated via binary stellar evolution, whereas dynamical ones are calculated with single stellar evolution; and ii) star clusters' metallicity distributions depend on their type, with young clusters having the same distribution as galaxies.

The latter panel in the bottom row of the figure shows dynamical BBHs formed only in young clusters. We recall that in our models this correspond to the assumption that the metallicity distribution is the same as for galaxies, that there is no limit on the minimum mass ratio allowed, and that recycling depends only on the GW recoil kick after the previous merger.
In this case, the BBH mass distribution broadens toward lower values compared to a more heterogeneous population of dynamical binaries, shown in the central panel. This is due to the adopted metallicity distribution. This, combined with the looser assumption on the mass ratio, leads to a predominance of low-mass sources i.e. $M_\bbh<40\Ms$.

\begin{figure}
    \centering
    \includegraphics[width=\columnwidth]{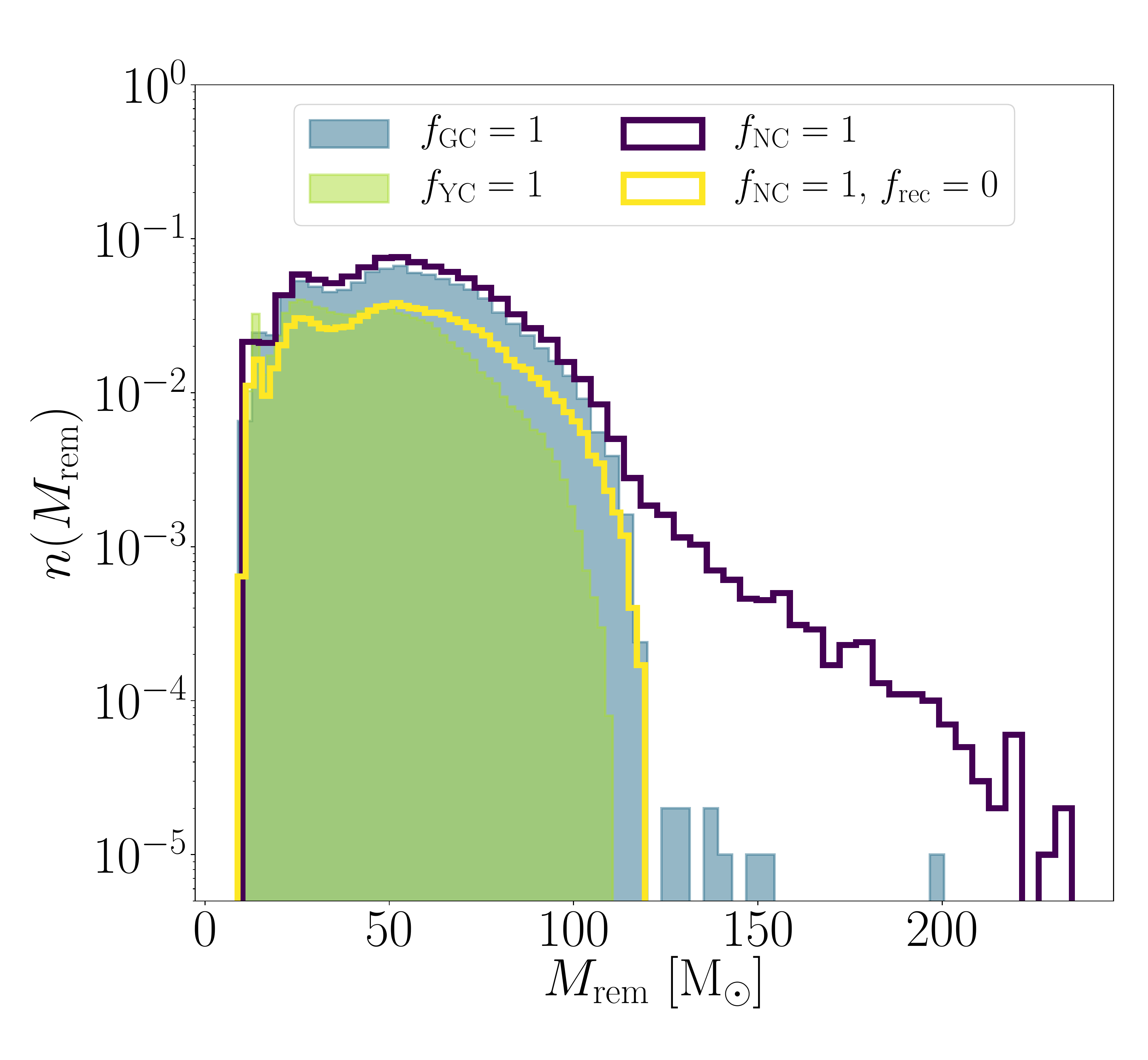}\\
    \includegraphics[width=\columnwidth]{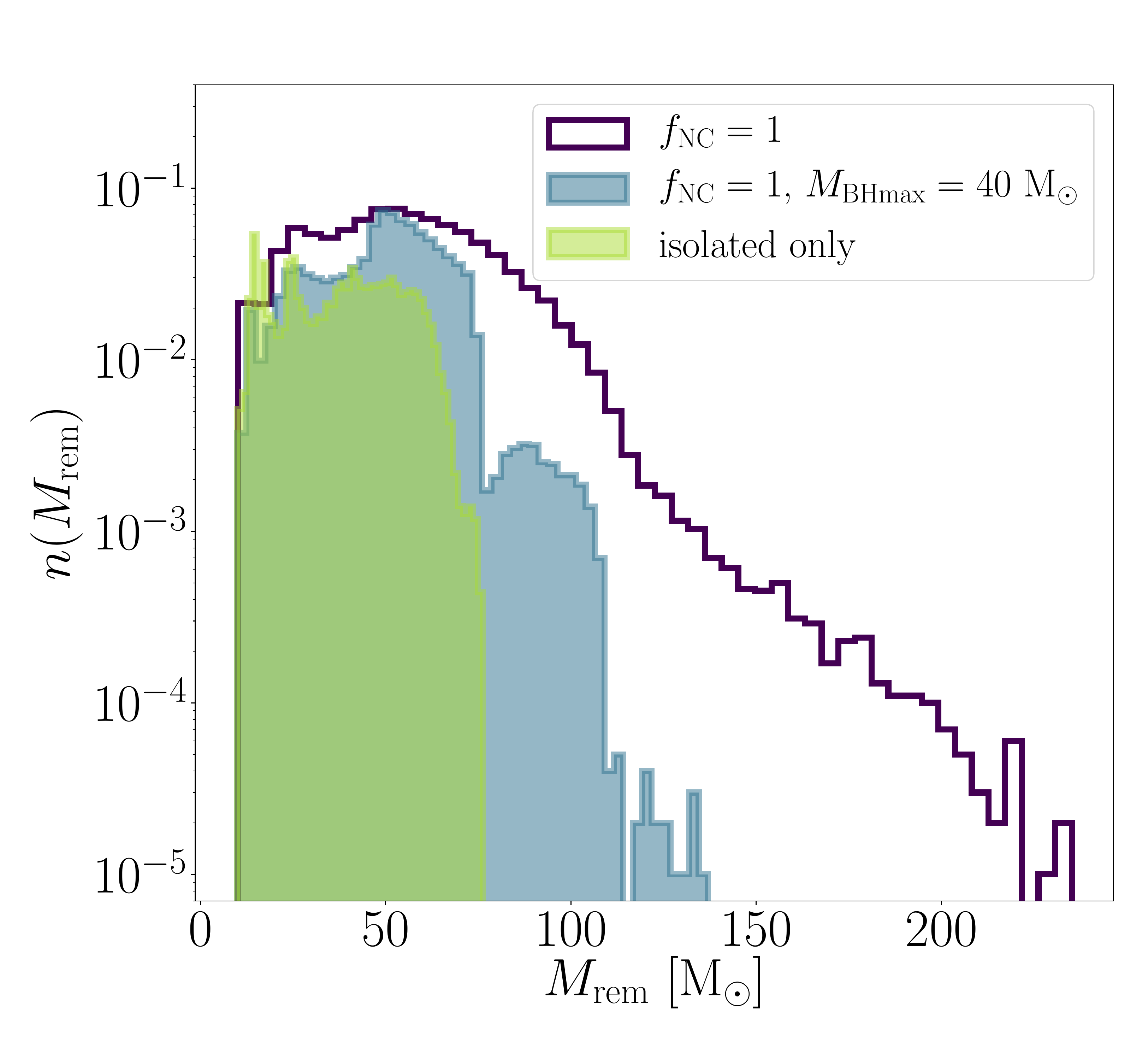}
    \caption{Top panel: $M_\rem$ distribution for globular (blue steps, model 4b), young (green steps, model 4c), and nuclear clusters (purple steps, model 4d) even in the case with no recycling allowed (yellow steps, model 4e). Bottom panel: $M_\rem$ distribution for nuclear clusters assuming \mobse{} BH mass spectrum (model 4d) and limiting this mass spectrum to a maximum value $M_{\rm BHmax} = 40\Ms$ (model 4d$\dagger$).}
    \label{fig:histo}
\end{figure}

In our treatment, distinguishing between different dynamical environments (i.e. globular clusters, nuclear star clusters and young star clusters) corresponds to varying metallicity distribution, minimum mass ratio  $q_\Min$, and the multiple merger probability via $v_{\rm max}$. Figure \ref{fig:histo} compares the $M_\rem - a_\rem$ plane for BBHs forming either in young, globular, and nuclear clusters. 

When comparing young and globular clusters, it is quite evident that the latter are characterised by a broader $M_\rem$ distribution. This is due to the different assumption on the $Z$ distribution, which for young clusters is double peaked at $Z\simeq 0.1\Zs$ and $Z = \Zs$, while for nuclear and globular clusters is equally distributed across logarithmic bins from $Z = 0.01\Zs$ up to solar values. In globular clusters, the higher escape velocity enables the formation of BHs with masses in the $120-200\Ms$ mass range, which are hard to explain under the assumptions made for young clusters. Hence, the potential detection of such massive BHs would allow us to place constraints on the metallicity distribution of the dynamical environments in which their progenitors developed.

Figure \ref{fig:histo} also quantifies the importance of hierarchical BH mergers in the case of nuclear clusters. These are the dynamical environment in which multiple mergers are most likely to happen. In model ID 4d, we allow BH mergers to undergo a further merger depending on the GW recoil kick, while in model ID 4e we forbid recycling ($f_{\rm rec}=0$) for BHs in nuclear clusters. Repeated mergers are responsible for the long tail at the high-end of the mass distribution and allow the formation of BHs as massive as $200 \Ms$. In model 4d (nuclear clusters with recycling), out of $10^5$ simulated BBHs, $\sim 5000$ BHs undergo 2 mergers , 115 undergo 3 mergers, and 4 undergo 4 mergers. BHs undergoing two or three subsequent mergers can reach masses up to $250\Ms$. Finding a number of BHs with such large masses would provide crucial insights on a) the probability of multiple mergers, and b) the merger rate from dense and massive clusters compared to other formation channels. We note that this result does not account for the possible formation of massive black holes through (multiple) mergers of massive stars \citep{zwart04,mapelli16}. This alternative channel might lead to the formation of BBHs with mass $>>100$ M$_\odot$ even in dense young star clusters, as described in \cite{dicarlo19,dicarlo20}.

\begin{figure}
    \centering
    \includegraphics[width=\columnwidth]{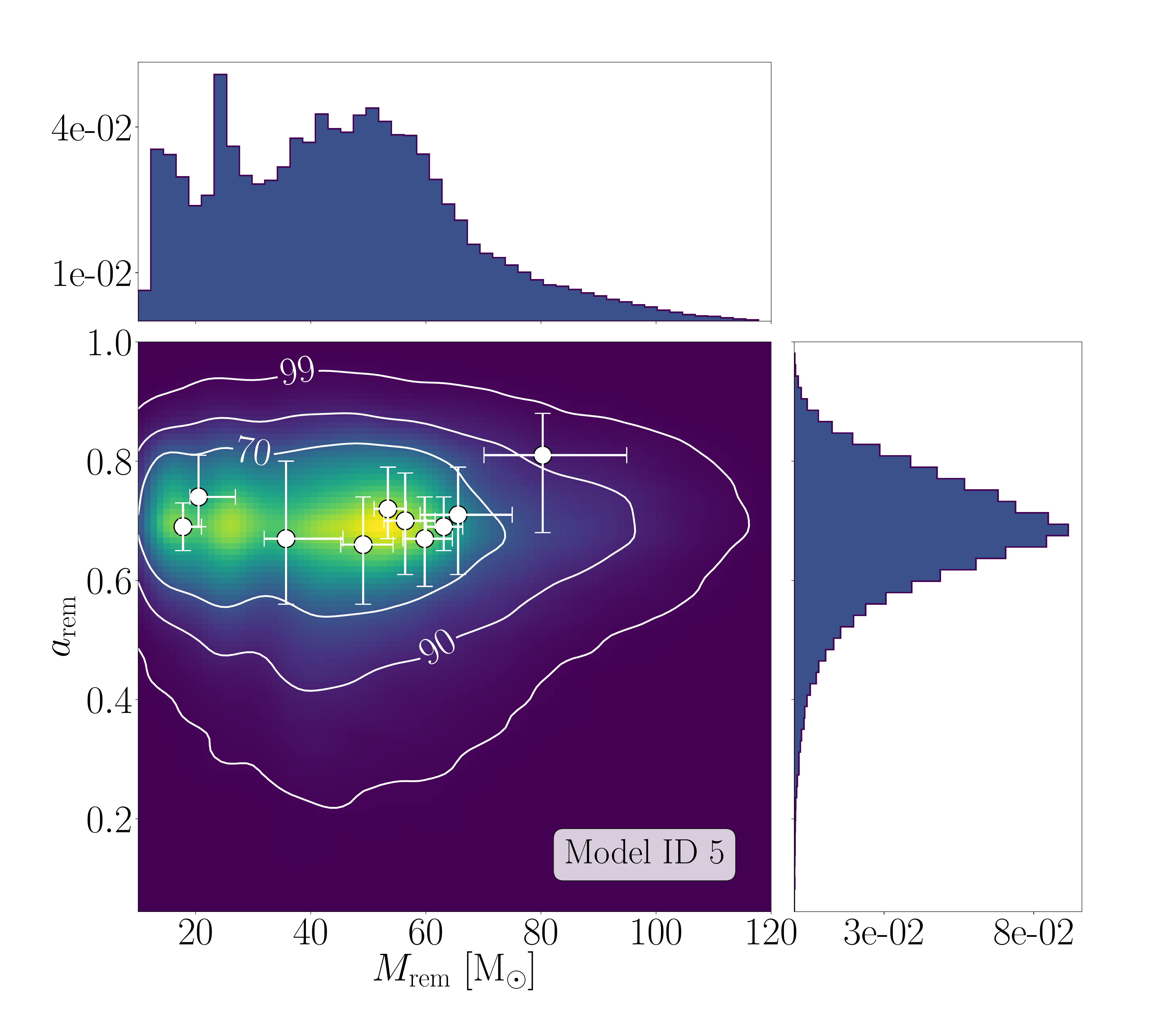}\\
    \includegraphics[width=\columnwidth]{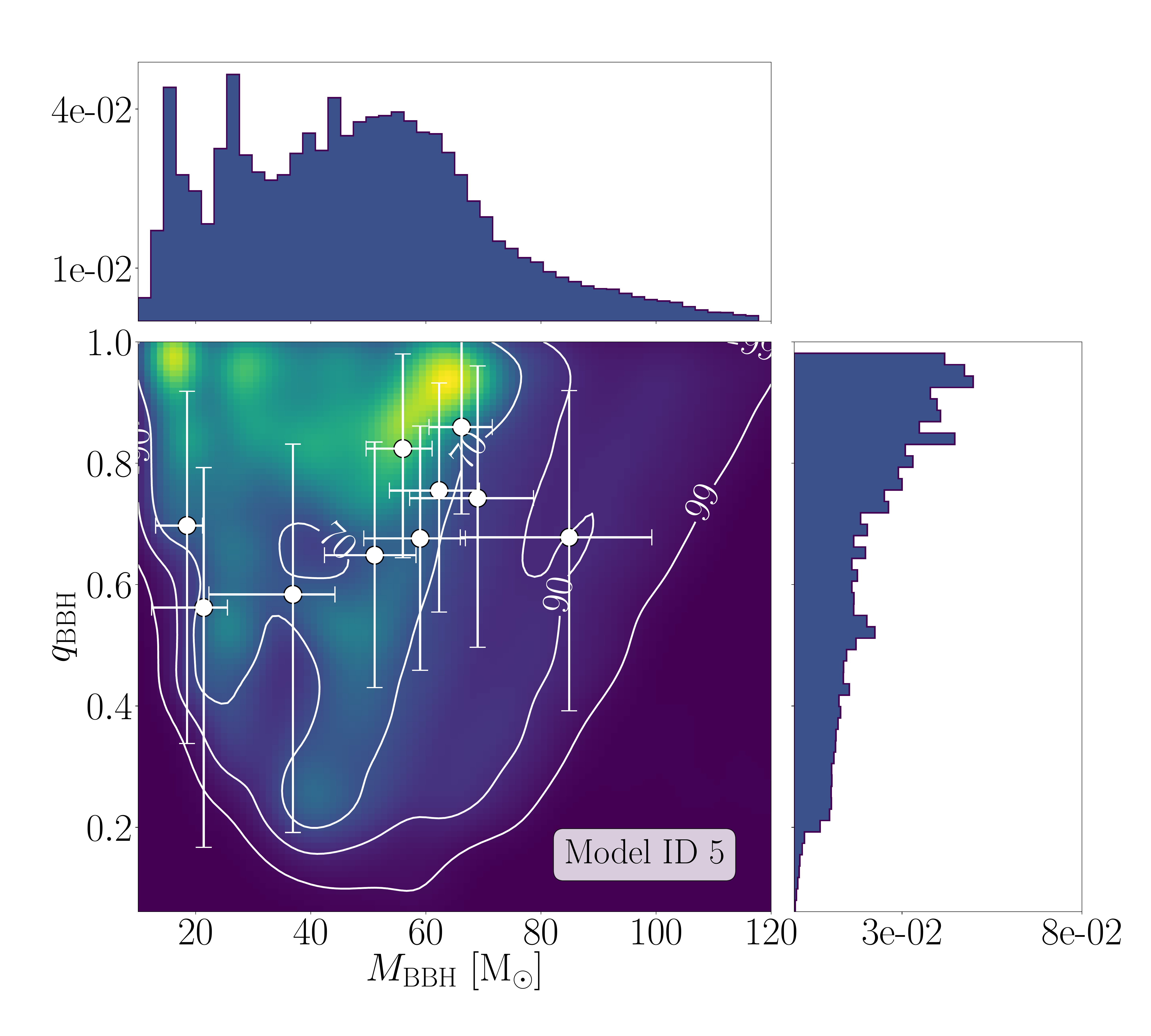}
    \caption{Top panel: $M_\rem-a_\rem$ plane assuming that BBH mergers are equally distributed among isolated and dynamical channels ($f_\iso=f_\dyn$), and among different cluster types ($f_\yc=f_\gc=f_\nc$), i.e. set ID 5. Bottom panel: same as above, but here the BBH mass-mass ratio plane is shown.}
    \label{fig:equal}
\end{figure}

A further model worthy of investigation relies upon the assumption that BBH mergers are equally contributed by isolated and dynamical binaries ($f_\iso = f_\dyn = 0.5$) and that all cluster types contribute equally to dynamical mergers ($f_\gc = f_\nc = f_\yc$, set ID 5). The $M_\rem-a_\rem$ and $q_\bbh-M_\bbh$ planes corresponding to such model are shown in Figure~\ref{fig:equal}. Having an equal contribution from dynamical and isolated binaries widens the BBH mass ratio distribution. Since $q_\bbh$ distribution is narrow and peaked around unity for isolated binaries, whereas it is flat for our dynamical ones by construction, increasing the percentage of dynamical binaries leads to a larger amount of unequal mass binaries, thus increasing the match between observations and models. At the same time, a larger number of dynamical mergers reduces the number of light remnant BHs, $M_\rem\simeq 20-40\Ms$, making harder to match observations and models in the $M_\rem - a_\rem$ plane.  

\subsection{Impact of the maximum BH mass}

One of the main open questions about BH formation is the maximum mass ($M_{\rm BHmax}$) of a BH born from a single star with zero-age main sequence mass $m_{\rm ZAMS}\leq 230$ M$_\odot$ . This is strongly affected by (pulsational) pair instability \citep{woosley17,woosley19,belczynski16a,spera17,giacobbo18,marchant19,stevenson19,renzo20}, by stellar rotation \citep{mapelli20}, by uncertainty on nuclear reaction rates \citep{farmer19} and by the collapse of a residual hydrogen envelope \citep{mapelli20}. As a result, $M_{\rm BHmax}$ might be as low as $\sim 40-45$ M$_\odot$ \citep{belczynski16} or as high as $\sim 65$ M$_\odot$ \citep{giacobbo18}.

To explore the role of $M_{\rm BHmax}$ in shaping the remnant mass distribution we run a further model, named 4d$\dagger$, which assumes $M_{\rm BHmax} = 40\Ms$ in our \mobse{} models. The bottom panel of Figure \ref{fig:histo} shows the comparison between models 4d (single BH masses from \mobse{}) and 4d$\dagger$ (single BH masses capped at $40\Ms$). As shown in the plot, limiting the maximum mass of zeroth generation BHs leads to a sharp drop of the number of systems with $M_\rem \geq 120-130\Ms$. The distribution of model 4d$\dagger$ with $f_{\rm NC}=1$ shows two peaks, one at $\sim{}50\Ms$ and the other at $\sim 100\Ms$.

\subsection{Caveats}
In this work, we developed a self-consistent statistical approach to construct catalogues of BBH mergers forming via different channels and in various environments. Our tool allowed us to explore a wide portion of the phase space and to place constraints on the role played by different parameters in determining the properties of BBH merger populations. Although quite fast and based on a set of physically motivated assumptions, our method comes with a number of caveats, which we discuss below. Overcoming these limitations requires a more in-depth study that faces the computational challenge of modelling at the same time star clusters and isolated binary populations altogether with their host galaxy, taking into account the chemo-dynamical evolution of the whole system. Our tool can serve as a basis to understand which parameters are the most effective in influencing the properties of BBH populations, and can be used to compare with observations but also to understand what kind of follow-up, more detailed, models are needed to improve our knowledge of BBH physics and evolution. 

Despite its versatility and wide applicability to study both dynamical and isolated binaries, our tool does not include any treatment for stellar collisions, which might affect the BH mass spectrum. We excluded this feature as its understanding is currently rather poor, since the number of studies addressing this aspect of dynamics is rather low \citep[see for instance]{dicarlo19}. The tool does not take into account the metallicity gradient in modelling BBH galaxy hosts, although this issue is partly addressed via defining different metallicity distribution choices (see for instance models ID 2a, b, and c). We excluded from our analysis BBH mergers forming via alternative processes, like BBH formation around a SMBH or in an AGN disc \citep{mckernan12,mckernan2014,mckernan2018,bartos2017,yang19}, or in triples \citep{antonini12,hoang18,fragione19,arca20}. Moreover, our tool does not account for chemically-homogeneous binary evolution \citep{marchant16}. This formation channel can produce remnants with mass $>130\Ms$, much higher than the limit posed by other models for isolated BBH formation.  
The fraction of chemically homogeneous stars over the total population might be larger at low metallicity \citep{ramachandran19}, but the 
delay time for BBHs from chemically-homogeneous is short \citep[$<0.4$ Gyr][]{marchant16}, thus limiting the possibility that this type of mergers
take place at $z<1$. 

In our method, we do not consider explicitly delay times in creating our mock BBHs. Since BBHs forming in metal poor galaxies at high redshift can constitute almost half of the mergers taking place at redshift zero \citep{mapelli19}, we encoded the information about the delay time in the choice of metallicity distribution as discussed in Section \ref{sec:met}. Regarding dynamical binaries, we do not make any assumption on the possible relation between the delay time and the BBH total mass and spin in star clusters. 
In a future versions of our tool, we will incorporate properly the time delay to quantify any possible correlation between this quantity and BBH population properties self-consistently.

Finally, we do not account for different stellar evolution recipes or different spin distributions. A discussion about the role played by these ingredients is already given in our previous paper \citep{ASBEN19}.

\section{Conclusions} \label{sec:concl}
We use a statistical approach to build BBH samples to be compared with observations. Our model takes into account the effects arising from BBH birth places, formation channels, initial metallicity,  progenitor's natal mass, kicks, and spins. Our results can be summarized as follows.
\begin{itemize}
\item We presented a versatile and self-consistent approach to construct catalogs of BBH mergers forming through different channels, which allows a fast exploration of a wide parameter space (Figure \ref{fig:sketch}). Our approach is an effective alternative to the computational challenge of dynamical simulations, but contains several approximations (e.g. the mass function of BBHs in star clusters and the time delays).
\item We show that the metallicity distribution of parent galaxies is a crucial parameter to assess the distribution of BBH masses, and can severely affect theoretical models. Moreover, we discuss the impact of post-merger GW recoil, which can limit the probability for a merged BH to undergo another merger event (Figure \ref{fig:Zdist}-\ref{fig:kick}).
\item Using our technique, we create samples of 100,000 BBH mergers forming either in isolation or via dynamical interactions in star clusters. For each merger, we calculate the remnant mass and spin and show how the global distribution compares with O1+O2 LVC BBHs. The results for our fiducial model are shown in Figure \ref{fig:fiducial}.
\item By comparing simulated and observed BBHs, we show that the fiducial model matches pretty well the properties of mergers observed during O1 and O2 (Figure \ref{fig:further}). 
\item  Based on the assumptions of our model, if the number of sources with remnant mass $\geq 70\Ms$ is significantly larger than that of sources with remnant mass $\leq 30$ M$_\odot$, dynamical BBHs dominate the population, while the absence of sources with $\geq 70\Ms$ implies that the isolated channel contributes to most of the BBH population in the local Universe (Table \ref{tab:mremo3}).
\item According to our fiducial model, we predict that at least one BBH out of 100 will leave a remnant with a mass $90 < M_\rem / $M$_\odot < 110$ and one out of 1000 will have $110 < M_\rem / $M$_\odot < 250\Ms$, thus in the range of intermediate-mass black holes (Figure \ref{fig:O3}).
\item We investigate a possible formation route for the heaviest BBH reported to date, GW170729 \citep{abbott19}, as the product of a series of mergers taking place in a dense star cluster. We find that the observed mass and spin of GW170729 can be explained by 1--3 subsequent mergers occurring in a dense cluster with velocity dispersion O($100$ km s$^{-1}$) and metallicity in the range $Z = 0.0002-0.002$.
\item We explore how different parameters affect the results. We demonstrate that the metallicity distribution and the relative amount of dynamical and isolated channels are the most important parameters determining the remnant mass distribution. Our results suggest that both the dynamical and isolated channel contribute to the overall population of BBH mergers.
\end{itemize}

\section*{Acknowledgements}
The authors are grateful to the referee for their comments and suggestions that helped to improve an earlier version of this manuscript.
MAS acknowledges financial support from the Alexander von Humboldt Foundation and the Federal Ministry for Education and Research for the research project "The evolution of black holes from stellar to galactic scales". 
MM acknowledges financial support by the European
Research Council for the ERC Consolidator grant DEMOBLACK, under contract no. 770017.
MS acknowledges funding from the European Union's Horizon 2020 research and innovation programme under the Marie-Sklodowska-Curie grant agreement No. 794393.
MB acknowledges the support while serving at the National Science Foundation through award 1755085. Any opinion, findings, and conclusions or recommendations expressed in this material are those of the authors and do not necessarily reflect the views of the National Science Foundation.
This work benefited from support by the International Space Science In-stitute (ISSI), Bern, Switzerland, through its International Team programme ref. no. 393 The Evolution of Rich Stellar Populations \& BH Binaries (2017-18), by the Sonderforschungsbereich SFB 881 ``The Milky Way System'' of the German Research Foundation (DFG) and by the COST Action CA16104 ``GWverse''.

\clearpage
\newpage
\footnotesize{
\bibliography{ASetal2015}

\newcommand{\noop}[1]{}
\begin{thebibliography}{}
\expandafter\ifx\csname natexlab\endcsname\relax\def\natexlab#1{#1}\fi
\providecommand{\url}[1]{\href{#1}{#1}}

\bibitem[{{Abbott} {et~al.}(2019{\natexlab{a}}){Abbott}, {Abbott}, {Abbott},
  {Abraham}, {Acernese}, {LIGO Scientific Collaboration}, \& {Virgo
  Collaboration}}]{LIGO19}
{Abbott}, B.~P., {Abbott}, R., {Abbott}, T.~D., {et~al.} 2019{\natexlab{a}},
  Physical Review X, 9, 031040

\bibitem[{{Abbott} {et~al.}(2019{\natexlab{b}}){Abbott}, {Abbott}, {Abbott},
  {Abraham}, {LIGO Scientific Collaboration}, \& {the Virgo
  Collaboration}}]{abbott19}
---. 2019{\natexlab{b}}, \apjl, 882, L24

\bibitem[{{Abbott} {et~al.}(2016{\natexlab{a}}){Abbott}, {Abbott}, {Abbott},
  {Abernathy}, {Acernese}, {Ackley}, {Adams}, {Adams}, {Addesso}, {Adhikari},
  \& {et al.}}]{abbott16a}
---. 2016{\natexlab{a}}, \prd, 93, 122003

\bibitem[{{Abbott} {et~al.}(2016{\natexlab{b}}){Abbott}, {Abbott}, {Abbott},
  {Abernathy}, {Acernese}, {Ackley}, {Adams}, {Adams}, {Addesso}, {Adhikari},
  \& et~al.}]{abbott16b}
---. 2016{\natexlab{b}}, Physical Review Letters, 116, 061102

\bibitem[{{Abbott} {et~al.}(2016{\natexlab{c}}){Abbott}, {Abbott}, {Abbott},
  {Abernathy}, {Acernese}, {Ackley}, {Adams}, {Adams}, {Addesso}, {Adhikari},
  \& et~al.}]{abbott16c}
---. 2016{\natexlab{c}}, Physical Review Letters, 116, 241103

\bibitem[{{Abbott} {et~al.}(2017{\natexlab{a}}){Abbott}, {Abbott}, {Abbott},
  {Acernese}, {Ackley}, {Adams}, {Adams}, {Addesso}, {Adhikari}, {Adya}, \&
  et~al.}]{abbott17a}
---. 2017{\natexlab{a}}, Physical Review Letters, 118, 221101

\bibitem[{{Abbott} {et~al.}(2017{\natexlab{b}}){Abbott}, {Abbott}, {Abbott},
  {Acernese}, {Ackley}, {Adams}, {Adams}, {Addesso}, {Adhikari}, {Adya}, \&
  et~al.}]{abbott17b}
---. 2017{\natexlab{b}}, Physical Review Letters, 119, 141101

\bibitem[{{Abbott} {et~al.}(2017{\natexlab{c}}){Abbott}, {Abbott}, {Abbott},
  {Acernese}, {Ackley}, {Adams}, {Adams}, {Addesso}, {Adhikari}, {Adya}, \&
  et~al.}]{abbott17c}
---. 2017{\natexlab{c}}, Physical Review Letters, 119, 161101

\bibitem[{{Abbott} {et~al.}(2017{\natexlab{d}}){Abbott}, {Abbott}, {Abbott},
  {Acernese}, {Ackley}, {Adams}, {Adams}, {Addesso}, {Adhikari}, {Adya}, \&
  et~al.}]{abbott17d}
---. 2017{\natexlab{d}}, \apjl, 848, L13

\bibitem[{{Amaro-Seoane} \& {Chen}(2016)}]{seoane16}
{Amaro-Seoane}, P., \& {Chen}, X. 2016, \mnras, 458, 3075

\bibitem[{{Antonini} {et~al.}(2016){Antonini}, {Chatterjee}, {Rodriguez},
  {Morscher}, {Pattabiraman}, {Kalogera}, \& {Rasio}}]{antonini16}
{Antonini}, F., {Chatterjee}, S., {Rodriguez}, C.~L., {et~al.} 2016, \apj, 816,
  65

\bibitem[{{Antonini} {et~al.}(2018){Antonini}, {Gieles}, \&
  {Gualandris}}]{antonini18c}
{Antonini}, F., {Gieles}, M., \& {Gualandris}, A. 2018, ArXiv e-prints,
  arXiv:1811.03640

\bibitem[{{Antonini} \& {Perets}(2012)}]{antonini12}
{Antonini}, F., \& {Perets}, H.~B. 2012, \apj, 757, 27

\bibitem[{{Antonini} \& {Rasio}(2016)}]{antonini16b}
{Antonini}, F., \& {Rasio}, F.~A. 2016, \apj, 831, 187

\bibitem[{{Antonini} {et~al.}(2017){Antonini}, {Toonen}, \&
  {Hamers}}]{antonini17}
{Antonini}, F., {Toonen}, S., \& {Hamers}, A.~S. 2017, \apj, 841, 77

\bibitem[{{Arca Sedda}(2020)}]{arca20}
{Arca Sedda}, M. 2020, arXiv e-prints, arXiv:2002.04037

\bibitem[{{Arca Sedda} \& {Benacquista}(2019)}]{ASBEN19}
{Arca Sedda}, M., \& {Benacquista}, M. 2019, \mnras, 482, 2991

\bibitem[{{Arca-Sedda} \& {Capuzzo-Dolcetta}(2019)}]{ASCD17b}
{Arca-Sedda}, M., \& {Capuzzo-Dolcetta}, R. 2019, \mnras, 483, 152

\bibitem[{{Arca-Sedda} \& {Gualandris}(2018)}]{ASG17}
{Arca-Sedda}, M., \& {Gualandris}, A. 2018, \mnras, 477, 4423

\bibitem[{{Arca-Sedda} {et~al.}(2018){Arca-Sedda}, {Li}, \& {Kocsis}}]{ASKLI18}
{Arca-Sedda}, M., {Li}, G., \& {Kocsis}, B. 2018, ArXiv e-prints,
  arXiv:1805.06458

\bibitem[{{Askar} {et~al.}(2017){Askar}, {Szkudlarek}, {Gondek-Rosi{\'n}ska},
  {Giersz}, \& {Bulik}}]{askar17}
{Askar}, A., {Szkudlarek}, M., {Gondek-Rosi{\'n}ska}, D., {Giersz}, M., \&
  {Bulik}, T. 2017, \mnras, 464, L36

\bibitem[{{Baker} {et~al.}(2006){Baker}, {Centrella}, {Choi}, {Koppitz}, \&
  {van Meter}}]{baker06}
{Baker}, J.~G., {Centrella}, J., {Choi}, D.-I., {Koppitz}, M., \& {van Meter},
  J. 2006, Physical Review Letters, 96, 111102

\bibitem[{{Banerjee}(2017)}]{banerjee16}
{Banerjee}, S. 2017, \mnras, 467, 524

\bibitem[{{Banerjee}(2018)}]{banerjee18}
---. 2018, \mnras, 473, 909

\bibitem[{{Banerjee} {et~al.}(2010){Banerjee}, {Baumgardt}, \&
  {Kroupa}}]{banerjee10}
{Banerjee}, S., {Baumgardt}, H., \& {Kroupa}, P. 2010, \mnras, 402, 371

\bibitem[{{Barrett} {et~al.}(2018){Barrett}, {Gaebel}, {Neijssel},
  {Vigna-G{\'o}mez}, {Stevenson}, {Berry}, {Farr}, \& {Mandel}}]{barrett18}
{Barrett}, J.~W., {Gaebel}, S.~M., {Neijssel}, C.~J., {et~al.} 2018, \mnras,
  477, 4685

\bibitem[{{Bartos} {et~al.}(2016){Bartos}, {Kocsis}, {Haiman}, \&
  {M{\'a}rka}}]{bartos16}
{Bartos}, I., {Kocsis}, B., {Haiman}, Z., \& {M{\'a}rka}, S. 2016, ArXiv
  e-prints, arXiv:1602.03831

\bibitem[{{Bartos} {et~al.}(2017){Bartos}, {Kocsis}, {Haiman}, \&
  {M{\'a}rka}}]{bartos2017}
---. 2017, \apj, 835, 165

\bibitem[{{Bavera} {et~al.}(2020){Bavera}, {Fragos}, {Qin}, {Zapartas},
  {Neijssel}, {Mandel}, {Batta}, {Gaebel}, {Kimball}, \&
  {Stevenson}}]{bavera20}
{Bavera}, S.~S., {Fragos}, T., {Qin}, Y., {et~al.} 2020, \aap, 635, A97

\bibitem[{{Belczynski} {et~al.}(2010){Belczynski}, {Bulik}, {Fryer}, {Ruiter},
  {Valsecchi}, {Vink}, \& {Hurley}}]{belczynski10}
{Belczynski}, K., {Bulik}, T., {Fryer}, C.~L., {et~al.} 2010, \apj, 714, 1217

\bibitem[{{Belczynski} {et~al.}(2016{\natexlab{a}}){Belczynski}, {Holz},
  {Bulik}, \& {O'Shaughnessy}}]{belczynski16}
{Belczynski}, K., {Holz}, D.~E., {Bulik}, T., \& {O'Shaughnessy}, R.
  2016{\natexlab{a}}, \nat, 534, 512

\bibitem[{{Belczynski} {et~al.}(2002){Belczynski}, {Kalogera}, \&
  {Bulik}}]{belczynski02}
{Belczynski}, K., {Kalogera}, V., \& {Bulik}, T. 2002, \apj, 572, 407

\bibitem[{{Belczynski} {et~al.}(2016{\natexlab{b}}){Belczynski}, {Heger},
  {Gladysz}, {Ruiter}, {Woosley}, {Wiktorowicz}, {Chen}, {Bulik},
  {O'Shaughnessy}, {Holz}, {Fryer}, \& {Berti}}]{belczynski16a}
{Belczynski}, K., {Heger}, A., {Gladysz}, W., {et~al.} 2016{\natexlab{b}},
  \aap, 594, A97

\bibitem[{{Belczynski} {et~al.}(2017){Belczynski}, {Klencki}, {Meynet},
  {Fryer}, {Brown}, {Chruslinska}, {Gladysz}, {O'Shaughnessy}, {Bulik},
  {Berti}, {Holz}, {Gerosa}, {Giersz}, {Ekstrom}, {Georgy}, {Askar}, {Lasota},
  \& {Wysocki}}]{Belczynski17}
{Belczynski}, K., {Klencki}, J., {Meynet}, G., {et~al.} 2017, ArXiv e-prints,
  arXiv:1706.07053

\bibitem[{{Bird} {et~al.}(2016){Bird}, {Cholis}, {Mu{\~n}oz}, {Ali-
  Ha{\"\i}moud}, {Kamionkowski}, {Kovetz}, {Raccanelli}, \& {Riess}}]{bird16}
{Bird}, S., {Cholis}, I., {Mu{\~n}oz}, J.~B., {et~al.} 2016, \prl, 116, 201301

\bibitem[{{Bouffanais} {et~al.}(2019){Bouffanais}, {Mapelli}, {Gerosa}, {Di
  Carlo}, {Giacobbo}, {Berti}, \& {Baibhav}}]{bouffanais2019}
{Bouffanais}, Y., {Mapelli}, M., {Gerosa}, D., {et~al.} 2019, arXiv e-prints,
  arXiv:1905.11054

\bibitem[{{Broadhurst} {et~al.}(2018){Broadhurst}, {Diego}, \&
  {Smoot}}]{broadhurst18}
{Broadhurst}, T., {Diego}, J.~M., \& {Smoot}, George, I. 2018, arXiv e-prints,
  arXiv:1802.05273

\bibitem[{{Campanelli} {et~al.}(2006){Campanelli}, {Lousto}, {Marronetti}, \&
  {Zlochower}}]{campanelli06}
{Campanelli}, M., {Lousto}, C.~O., {Marronetti}, P., \& {Zlochower}, Y. 2006,
  Physical Review Letters, 96, 111101

\bibitem[{{Campanelli} {et~al.}(2007){Campanelli}, {Lousto}, {Zlochower}, \&
  {Merritt}}]{campanelli07}
{Campanelli}, M., {Lousto}, C.~O., {Zlochower}, Y., \& {Merritt}, D. 2007,
  \prl, 98, 231102

\bibitem[{{Capano} {et~al.}(2016){Capano}, {Harry}, {Privitera}, \&
  {Buonanno}}]{capano16}
{Capano}, C., {Harry}, I., {Privitera}, S., \& {Buonanno}, A. 2016, \prd, 93,
  124007

\bibitem[{{Casares} {et~al.}(2017){Casares}, {Jonker}, \&
  {Israelian}}]{casares17}
{Casares}, J., {Jonker}, P.~G., \& {Israelian}, G. 2017, {X-Ray Binaries}, ed.
  A.~W. {Alsabti} \& P.~{Murdin}, 1499

\bibitem[{{Di Carlo} {et~al.}(2019{\natexlab{a}}){Di Carlo}, {Giacobbo},
  {Mapelli}, {Pasquato}, {Spera}, {Wang}, \& {Haardt}}]{dicarlo19}
{Di Carlo}, U.~N., {Giacobbo}, N., {Mapelli}, M., {et~al.} 2019{\natexlab{a}},
  arXiv e-prints, arXiv:1901.00863

\bibitem[{{Di Carlo} {et~al.}(2019{\natexlab{b}}){Di Carlo}, {Mapelli},
  {Bouffanais}, {Giacobbo}, {Bressan}, {Spera}, \& {Haardt}}]{dicarlo20}
{Di Carlo}, U.~N., {Mapelli}, M., {Bouffanais}, Y., {et~al.}
  2019{\natexlab{b}}, arXiv e-prints, arXiv:1911.01434

\bibitem[{{Do} {et~al.}(2015){Do}, {Kerzendorf}, {Winsor}, {St{\o}stad},
  {Morris}, {Lu}, \& {Ghez}}]{do15}
{Do}, T., {Kerzendorf}, W., {Winsor}, N., {et~al.} 2015, \apj, 809, 143

\bibitem[{{Doctor} {et~al.}(2019){Doctor}, {Wysocki}, {O'Shaughnessy}, {Holz},
  \& {Farr}}]{doctor19}
{Doctor}, Z., {Wysocki}, D., {O'Shaughnessy}, R., {Holz}, D.~E., \& {Farr}, B.
  2019, arXiv e-prints, arXiv:1911.04424

\bibitem[{{Dominik} {et~al.}(2013){Dominik}, {Belczynski}, {Fryer}, {Holz},
  {Berti}, {Bulik}, {Mandel}, \& {O'Shaughnessy}}]{dominik13}
{Dominik}, M., {Belczynski}, K., {Fryer}, C., {et~al.} 2013, \apj, 779, 72

\bibitem[{{Dominik} {et~al.}(2015){Dominik}, {Berti}, {O'Shaughnessy},
  {Mandel}, {Belczynski}, {Fryer}, {Holz}, {Bulik}, \& {Pannarale}}]{dominik15}
{Dominik}, M., {Berti}, E., {O'Shaughnessy}, R., {et~al.} 2015, \apj, 806, 263

\bibitem[{{Downing} {et~al.}(2010){Downing}, {Benacquista}, {Giersz}, \&
  {Spurzem}}]{downing10}
{Downing}, J.~M.~B., {Benacquista}, M.~J., {Giersz}, M., \& {Spurzem}, R. 2010,
  \mnras, 407, 1946

\bibitem[{{Eldridge} {et~al.}(2017){Eldridge}, {Stanway}, {Xiao}, {McClelland
  }, {Taylor}, {Ng}, {Greis}, \& {Bray}}]{eldridge17}
{Eldridge}, J.~J., {Stanway}, E.~R., {Xiao}, L., {et~al.} 2017, \pasa, 34, e058

\bibitem[{{Farmer} {et~al.}(2019){Farmer}, {Renzo}, {de Mink}, {Marchant}, \&
  {Justham}}]{farmer19}
{Farmer}, R., {Renzo}, M., {de Mink}, S.~E., {Marchant}, P., \& {Justham}, S.
  2019, \apj, 887, 53

\bibitem[{{Farr} {et~al.}(2017){Farr}, {Stevenson}, {Miller}, {Mandel}, {Farr},
  \& {Vecchio}}]{farr17}
{Farr}, W.~M., {Stevenson}, S., {Miller}, M.~C., {et~al.} 2017, \nat, 548, 426

\bibitem[{Fernández \& Kobayashi(2019)}]{fernandez19}
Fernández, J.~J., \& Kobayashi, S. 2019, Monthly Notices of the Royal
  Astronomical Society, 487, 1200.
\newblock \url{https://doi.org/10.1093/mnras/stz1353}

\bibitem[{{Fishbach} \& {Holz}(2017)}]{fishbach17b}
{Fishbach}, M., \& {Holz}, D.~E. 2017, \apjl, 851, L25

\bibitem[{{Fishbach} {et~al.}(2017){Fishbach}, {Holz}, \& {Farr}}]{fishbach17a}
{Fishbach}, M., {Holz}, D.~E., \& {Farr}, B. 2017, \apjl, 840, L24

\bibitem[{{Fragione} {et~al.}(2018){Fragione}, {Grishin}, {Leigh}, {Perets}, \&
  {Perna}}]{fragione19}
{Fragione}, G., {Grishin}, E., {Leigh}, N.~W.~C., {Perets}, H.~B., \& {Perna},
  R. 2018, ArXiv e-prints, arXiv:1811.10627

\bibitem[{{Gallazzi} {et~al.}(2005){Gallazzi}, {Charlot}, {Brinchmann},
  {White}, \& {Tremonti}}]{Gallazzi05}
{Gallazzi}, A., {Charlot}, S., {Brinchmann}, J., {White}, S.~D.~M., \&
  {Tremonti}, C.~A. 2005, \mnras, 362, 41

\bibitem[{Georgiev {et~al.}(2009)Georgiev, Hilker, Puzia, Goudfrooij, \&
  Baumgardt}]{georgiev09}
Georgiev, I.~Y., Hilker, M., Puzia, T.~H., Goudfrooij, P., \& Baumgardt, H.
  2009, Monthly Notices of the Royal Astronomical Society, 396, 1075.
\newblock \url{https://doi.org/10.1111/j.1365-2966.2009.14776.x}

\bibitem[{{Gerosa} \& {Berti}(2017)}]{gerosa17}
{Gerosa}, D., \& {Berti}, E. 2017, \prd, 95, 124046

\bibitem[{{Gerosa} {et~al.}(2018){Gerosa}, {Berti}, {O'Shaughnessy},
  {Belczynski}, {Kesden}, {Wysocki}, \& {Gladysz}}]{gerosa18}
{Gerosa}, D., {Berti}, E., {O'Shaughnessy}, R., {et~al.} 2018, \prd, 98, 084036

\bibitem[{{Giacobbo} \& {Mapelli}(2019)}]{giacobbo2019}
{Giacobbo}, N., \& {Mapelli}, M. 2019, arXiv e-prints, arXiv:1909.06385

\bibitem[{{Giacobbo} {et~al.}(2018{\natexlab{a}}){Giacobbo}, {Mapelli}, \&
  {Spera}}]{giacobbo18b}
{Giacobbo}, N., {Mapelli}, M., \& {Spera}, M. 2018{\natexlab{a}}, \mnras, 474,
  2959

\bibitem[{{Giacobbo} {et~al.}(2018{\natexlab{b}}){Giacobbo}, {Mapelli}, \&
  {Spera}}]{giacobbo18}
---. 2018{\natexlab{b}}, \mnras, 474, 2959

\bibitem[{{Gonz{\'a}lez} {et~al.}(2007){Gonz{\'a}lez}, {Sperhake},
  {Br{\"u}gmann}, {Hannam}, \& {Husa}}]{gonzalez07}
{Gonz{\'a}lez}, J.~A., {Sperhake}, U., {Br{\"u}gmann}, B., {Hannam}, M., \&
  {Husa}, S. 2007, \prl, 98, 091101

\bibitem[{{Harris} {et~al.}(2014){Harris}, {Morningstar}, {Gnedin},
  {O'Halloran}, {Blakeslee}, {Whitmore}, {C{\^o}t{\'e}}, {Geisler}, {Peng},
  {Bailin}, {Rothberg}, {Cockcroft}, \& {Barber DeGraaff}}]{harris14}
{Harris}, W.~E., {Morningstar}, W., {Gnedin}, O.~Y., {et~al.} 2014, \apj, 797,
  128

\bibitem[{{Hoang} {et~al.}(2019){Hoang}, {Naoz}, {Kocsis}, {Farr}, \&
  {McIver}}]{hoang19}
{Hoang}, B.-M., {Naoz}, S., {Kocsis}, B., {Farr}, W., \& {McIver}, J. 2019,
  arXiv e-prints, arXiv:1903.00134

\bibitem[{{Hoang} {et~al.}(2018){Hoang}, {Naoz}, {Kocsis}, {Rasio}, \&
  {Dosopoulou}}]{hoang18}
{Hoang}, B.-M., {Naoz}, S., {Kocsis}, B., {Rasio}, F.~A., \& {Dosopoulou}, F.
  2018, \apj, 856, 140

\bibitem[{{Hofmann} {et~al.}(2016){Hofmann}, {Barausse}, \&
  {Rezzolla}}]{hofmann16}
{Hofmann}, F., {Barausse}, E., \& {Rezzolla}, L. 2016, \apjl, 825, L19

\bibitem[{{Hong} {et~al.}(2018){Hong}, {Vesperini}, {Askar}, {Giersz},
  {Szkudlarek}, \& {Bulik}}]{hong18}
{Hong}, J., {Vesperini}, E., {Askar}, A., {et~al.} 2018, \mnras, 480, 5645

\bibitem[{{Hughes} \& {Blandford}(2003)}]{hughes03}
{Hughes}, S.~A., \& {Blandford}, R.~D. 2003, \apjl, 585, L101

\bibitem[{{Hurley} {et~al.}(2002){Hurley}, {Tout}, \& {Pols}}]{hurley02}
{Hurley}, J.~R., {Tout}, C.~A., \& {Pols}, O.~R. 2002, \mnras, 329, 897

\bibitem[{{Jim{\'e}nez-Forteza} {et~al.}(2017){Jim{\'e}nez-Forteza}, {Keitel},
  {Husa}, {Hannam}, {Khan}, \& {P{\"u}rrer}}]{jimenez17}
{Jim{\'e}nez-Forteza}, X., {Keitel}, D., {Husa}, S., {et~al.} 2017, \prd, 95,
  064024

\bibitem[{{Kimball} {et~al.}(2019){Kimball}, {Berry}, \&
  {Kalogera}}]{kimball2019}
{Kimball}, C., {Berry}, C. P.~L., \& {Kalogera}, V. 2019, arXiv e-prints,
  arXiv:1903.07813

\bibitem[{{Kroupa}(2001)}]{kroupa01}
{Kroupa}, P. 2001, \mnras, 322, 231

\bibitem[{{Kruckow} {et~al.}(2018){Kruckow}, {Tauris}, {Langer}, {Kramer}, \&
  {Izzard}}]{kruckow18}
{Kruckow}, M.~U., {Tauris}, T.~M., {Langer}, N., {Kramer}, M., \& {Izzard},
  R.~G. 2018, \mnras, 481, 1908

\bibitem[{{Kumamoto} {et~al.}(2019){Kumamoto}, {Fujii}, \&
  {Tanikawa}}]{kumamoto19}
{Kumamoto}, J., {Fujii}, M.~S., \& {Tanikawa}, A. 2019, \mnras, 486, 3942

\bibitem[{{Lamers} {et~al.}(2017){Lamers}, {Kruijssen}, {Bastian}, {Rejkuba},
  {Hilker}, \& {Kissler-Patig}}]{lamers17}
{Lamers}, H.~J.~G.~L.~M., {Kruijssen}, J.~M.~D., {Bastian}, N., {et~al.} 2017,
  \aap, 606, A85

\bibitem[{{Lee}(1995)}]{lee95}
{Lee}, H.~M. 1995, \mnras, 272, 605

\bibitem[{{Lousto} \& {Zlochower}(2008)}]{lousto08}
{Lousto}, C.~O., \& {Zlochower}, Y. 2008, \prd, 77, 044028

\bibitem[{{Lousto} {et~al.}(2012){Lousto}, {Zlochower}, {Dotti}, \&
  {Volonteri}}]{lousto12}
{Lousto}, C.~O., {Zlochower}, Y., {Dotti}, M., \& {Volonteri}, M. 2012, \prd,
  85, 084015

\bibitem[{{Mapelli}(2016)}]{mapelli16}
{Mapelli}, M. 2016, \mnras, 459, 3432

\bibitem[{{Mapelli} \& {Bressan}(2013)}]{mapelli13}
{Mapelli}, M., \& {Bressan}, A. 2013, \mnras, 430, 3120

\bibitem[{{Mapelli} {et~al.}(2009){Mapelli}, {Colpi}, \&
  {Zampieri}}]{mapelli09}
{Mapelli}, M., {Colpi}, M., \& {Zampieri}, L. 2009, \mnras, 395, L71

\bibitem[{{Mapelli} \& {Giacobbo}(2018)}]{mapelli18}
{Mapelli}, M., \& {Giacobbo}, N. 2018, \mnras, 479, 4391

\bibitem[{{Mapelli} {et~al.}(2017){Mapelli}, {Giacobbo}, {Ripamonti}, \&
  {Spera}}]{mapelli17}
{Mapelli}, M., {Giacobbo}, N., {Ripamonti}, E., \& {Spera}, M. 2017, \mnras,
  472, 2422

\bibitem[{{Mapelli} {et~al.}(2019){Mapelli}, {Giacobbo}, {Santoliquido}, \&
  {Artale}}]{mapelli19}
{Mapelli}, M., {Giacobbo}, N., {Santoliquido}, F., \& {Artale}, M.~C. 2019,
  \mnras, 487, 2

\bibitem[{{Mapelli} {et~al.}(2010){Mapelli}, {Huwyler}, {Mayer}, {Jetzer}, \&
  {Vecchio}}]{mapelli10}
{Mapelli}, M., {Huwyler}, C., {Mayer}, L., {Jetzer}, P., \& {Vecchio}, A. 2010,
  \apj, 719, 987

\bibitem[{{Mapelli} {et~al.}(2020){Mapelli}, {Spera}, {Montanari}, {Limongi},
  {Chieffi}, {Giacobbo}, {Bressan}, \& {Bouffanais}}]{mapelli20}
{Mapelli}, M., {Spera}, M., {Montanari}, E., {et~al.} 2020, \apj, 888, 76

\bibitem[{{Marchant} {et~al.}(2016){Marchant}, {Langer}, {Podsiadlowski},
  {Tauris}, \& {Moriya}}]{marchant16}
{Marchant}, P., {Langer}, N., {Podsiadlowski}, P., {Tauris}, T.~M., \&
  {Moriya}, T.~J. 2016, \aap, 588, A50

\bibitem[{{Marchant} {et~al.}(2019){Marchant}, {Renzo}, {Farmer}, {Pappas},
  {Taam}, {de Mink}, \& {Kalogera}}]{marchant19}
{Marchant}, P., {Renzo}, M., {Farmer}, R., {et~al.} 2019, \apj, 882, 36

\bibitem[{{Martynov} {et~al.}(2016){Martynov}, {Hall}, {Abbott}, {Abbott},
  {Abbott}, {Adams}, {Adhikari}, {Anderson}, {Anderson}, {Arai}, {Arain},
  {Aston}, {Austin}, {Ballmer}, {Barbet}, {Barker}, {Barr}, {Barsotti},
  {Bartlett}, {Barton}, {Bartos}, {Batch}, {Bell}, {Belopolski}, {Bergman},
  {Betzwieser}, {Billingsley}, {Birch}, {Biscans}, {Biwer}, {Black}, {Blair},
  {Bogan}, {Bork}, {Bridges}, {Brooks}, {Celerier}, {Ciani}, {Clara}, {Cook},
  {Countryman}, {Cowart}, {Coyne}, {Cumming}, {Cunningham}, {Damjanic},
  {Dannenberg}, {Danzmann}, {Costa}, {Daw}, {DeBra}, {DeRosa}, {DeSalvo},
  {Dooley}, {Doravari}, {Driggers}, {Dwyer}, {Effler}, {Etzel}, {Evans},
  {Evans}, {Factourovich}, {Fair}, {Feldbaum}, {Fisher}, {Foley}, {Frede},
  {Fritschel}, {Frolov}, {Fulda}, {Fyffe}, {Galdi}, {Giaime}, {Giardina},
  {Gleason}, {Goetz}, {Gras}, {Gray}, {Greenhalgh}, {Grote}, {Guido}, {Gushwa},
  {Gustafson}, {Gustafson}, {Hammond}, {Hanks}, {Hanson}, {Hardwick}, {Harry},
  {Heefner}, {Heintze}, {Heptonstall}, {Hoak}, {Hough}, {Ivanov}, {Izumi},
  {Jacobson}, {James}, {Jones}, {Kandhasamy}, {Karki}, {Kasprzack}, {Kaufer},
  {Kawabe}, {Kells}, {Kijbunchoo}, {King}, {King}, {Kinzel}, {Kissel},
  {Kokeyama}, {Korth}, {Kuehn}, {Kwee}, {Landry}, {Lantz}, {Le Roux}, {Levine},
  {Lewis}, {Lhuillier}, {Lockerbie}, {Lormand}, {Lubinski}, {Lundgren},
  {MacDonald}, {MacInnis}, {Macleod}, {Mageswaran}, {Mailand}, {M{\'a}rka},
  {M{\'a}rka}, {Markosyan}, {Maros}, {Martin}, {Martin}, {Marx}, {Mason},
  {Massinger}, {Matichard}, {Mavalvala}, {McCarthy}, {McClelland}, {McCormick},
  {McIntyre}, {McIver}, {Merilh}, {Meyer}, {Meyers}, {Miller}, {Mittleman},
  {Moreno}, {Mueller}, {Mueller}, {Mullavey}, {Munch}, {Nuttall}, {Oberling},
  {O'Dell}, {Oppermann}, {Oram}, {O'Reilly}, {Osthelder}, {Ottaway},
  {Overmier}, {Palamos}, {Paris}, {Parker}, {Patrick}, {Pele}, {Penn},
  {Phelps}, {Pickenpack}, {Pierro}, {Pinto}, {Poeld}, {Principe}, {Prokhorov},
  {Puncken}, {Quetschke}, {Quintero}, {Raab}, {Radkins}, {Raffai}, {Ramet},
  {Reed}, {Reid}, {Reitze}, {Robertson}, {Rollins}, {Roma}, {Romie}, {Rowan},
  {Ryan}, {Sadecki}, {Sanchez}, {Sandberg}, {Sannibale}, {Savage}, {Schofield},
  {Schultz}, {Schwinberg}, {Sellers}, {Sevigny}, {Shaddock}, {Shao}, {Shapiro},
  {Shawhan}, {Shoemaker}, {Sigg}, {Slagmolen}, {Smith}, {Smith},
  {Smith-Lefebvre}, {Sorazu}, {Staley}, {Stein}, {Stochino}, {Strain},
  {Taylor}, {Thomas}, {Thomas}, {Thorne}, {Thrane}, {Torrie}, {Traylor},
  {Vajente}, {Valdes}, {van Veggel}, {Vargas}, {Vecchio}, {Veitch},
  {Venkateswara}, {Vo}, {Vorvick}, {Waldman}, {Walker}, {Ward}, {Warner},
  {Weaver}, {Weiss}, {Welborn}, {We{\ss}els}, {Wilkinson}, {Willems},
  {Williams}, {Willke}, {Winkelmann}, {Wipf}, {Worden}, {Wu}, {Yamamoto},
  {Yancey}, {Yu}, {Zhang}, {Zucker}, \& {Zweizig}}]{martynov16}
{Martynov}, D.~V., {Hall}, E.~D., {Abbott}, B.~P., {et~al.} 2016, \prd, 93,
  112004

\bibitem[{{McKernan} {et~al.}(2014){McKernan}, {Ford}, {Kocsis}, {Lyra}, \&
  {Winter}}]{mckernan2014}
{McKernan}, B., {Ford}, K.~E.~S., {Kocsis}, B., {Lyra}, W., \& {Winter}, L.~M.
  2014, \mnras, 441, 900

\bibitem[{{McKernan} {et~al.}(2012){McKernan}, {Ford}, {Lyra}, \&
  {Perets}}]{mckernan12}
{McKernan}, B., {Ford}, K.~E.~S., {Lyra}, W., \& {Perets}, H.~B. 2012, \mnras,
  425, 460

\bibitem[{{McKernan} {et~al.}(2018){McKernan}, {Ford}, {Bellovary}, {Leigh},
  {Haiman}, {Kocsis}, {Lyra}, {Mac Low}, {Metzger}, {O'Dowd}, {Endlich}, \&
  {Rosen}}]{mckernan2018}
{McKernan}, B., {Ford}, K.~E.~S., {Bellovary}, J., {et~al.} 2018, \apj, 866, 66

\bibitem[{{Miller} \& {Hamilton}(2002)}]{miller02}
{Miller}, M.~C., \& {Hamilton}, D.~P. 2002, \apj, 576, 894

\bibitem[{{Miller} \& {Lauburg}(2009)}]{miller09}
{Miller}, M.~C., \& {Lauburg}, V.~M. 2009, \apj, 692, 917

\bibitem[{{Morawski} {et~al.}(2018){Morawski}, {Giersz}, {Askar}, \&
  {Belczynski}}]{morawski18}
{Morawski}, J., {Giersz}, M., {Askar}, A., \& {Belczynski}, K. 2018, \mnras,
  481, 2168

\bibitem[{{Neijssel} {et~al.}(2019){Neijssel}, {Vigna-G{\'o}mez}, {Stevenson},
  {Barrett}, {Gaebel}, {Broekgaarden}, {de Mink}, {Sz{\'e}csi}, {Vinciguerra},
  \& {Mandel}}]{neijssel2019}
{Neijssel}, C.~J., {Vigna-G{\'o}mez}, A., {Stevenson}, S., {et~al.} 2019,
  \mnras, 490, 3740

\bibitem[{{Netopil} {et~al.}(2016){Netopil}, {Paunzen}, {Heiter}, \&
  {Soubiran}}]{netopil15}
{Netopil}, M., {Paunzen}, E., {Heiter}, U., \& {Soubiran}, C. 2016, \aap, 585,
  A150

\bibitem[{{Neumayer} {et~al.}(2020){Neumayer}, {Seth}, \&
  {Boeker}}]{neumayer20}
{Neumayer}, N., {Seth}, A., \& {Boeker}, T. 2020, arXiv e-prints,
  arXiv:2001.03626

\bibitem[{{O'Leary} {et~al.}(2009){O'Leary}, {Kocsis}, \& {Loeb}}]{oleary09}
{O'Leary}, R.~M., {Kocsis}, B., \& {Loeb}, A. 2009, \mnras, 395, 2127

\bibitem[{{O'Leary} {et~al.}(2016){O'Leary}, {Meiron}, \& {Kocsis}}]{OLeary16}
{O'Leary}, R.~M., {Meiron}, Y., \& {Kocsis}, B. 2016, \apjl, 824, L12

\bibitem[{{Paudel} {et~al.}(2011){Paudel}, {Lisker}, \&
  {Kuntschner}}]{paudel11}
{Paudel}, S., {Lisker}, T., \& {Kuntschner}, H. 2011, \mnras, 413, 1764

\bibitem[{{Perna} {et~al.}(2019){Perna}, {Wang}, {Farr}, {Leigh}, \&
  {Cantiello}}]{perna19}
{Perna}, R., {Wang}, Y.-H., {Farr}, W.~M., {Leigh}, N., \& {Cantiello}, M.
  2019, \apjl, 878, L1

\bibitem[{{Pilyugin} {et~al.}(2014){Pilyugin}, {Grebel}, \&
  {Kniazev}}]{Pilyugin14}
{Pilyugin}, L.~S., {Grebel}, E.~K., \& {Kniazev}, A.~Y. 2014, \aj, 147, 131

\bibitem[{{Portegies Zwart} {et~al.}(2004){Portegies Zwart}, {Baumgardt},
  {Hut}, {Makino}, \& {McMillan}}]{zwart04}
{Portegies Zwart}, S.~F., {Baumgardt}, H., {Hut}, P., {Makino}, J., \&
  {McMillan}, S.~L.~W. 2004, \nat, 428, 724

\bibitem[{{Portegies Zwart} \& {McMillan}(2000)}]{zwart00b}
{Portegies Zwart}, S.~F., \& {McMillan}, S.~L.~W. 2000, \apjl, 528, L17

\bibitem[{{Portegies Zwart} {et~al.}(2010){Portegies Zwart}, {McMillan}, \&
  {Gieles}}]{portegieszwart2010}
{Portegies Zwart}, S.~F., {McMillan}, S. L.~W., \& {Gieles}, M. 2010, \araa,
  48, 431

\bibitem[{{Pretorius}(2005)}]{pretorius05}
{Pretorius}, F. 2005, Classical and Quantum Gravity, 22, 425

\bibitem[{{Qin} {et~al.}(2018){Qin}, {Fragos}, {Meynet}, {Andrews},
  {S{\o}rensen}, \& {Song}}]{qin18}
{Qin}, Y., {Fragos}, T., {Meynet}, G., {et~al.} 2018, \aap, 616, A28

\bibitem[{{Ramachandran} {et~al.}(2019){Ramachandran}, {Hamann}, {Oskinova},
  {Gallagher}, {Hainich}, {Shenar}, {Sand er}, {Todt}, \&
  {Fulmer}}]{ramachandran19}
{Ramachandran}, V., {Hamann}, W.~R., {Oskinova}, L.~M., {et~al.} 2019, \aap,
  625, A104

\bibitem[{{Rasskazov} \& {Kocsis}(2019)}]{rasskazov19}
{Rasskazov}, A., \& {Kocsis}, B. 2019, arXiv e-prints, arXiv:1902.03242

\bibitem[{{Rastello} {et~al.}(2019){Rastello}, {Amaro-Seoane}, {Arca-Sedda},
  {Capuzzo-Dolcetta}, {Fragione}, \& {Tosta e Melo}}]{rastello19}
{Rastello}, S., {Amaro-Seoane}, P., {Arca-Sedda}, M., {et~al.} 2019, \mnras,
  483, 1233

\bibitem[{{Remillard} \& {McClintock}(2006)}]{xbinrev}
{Remillard}, R.~A., \& {McClintock}, J.~E. 2006, \araa, 44, 49

\bibitem[{{Renzo} {et~al.}(2020){Renzo}, {Farmer}, {Justham}, {G{\"o}tberg},
  {de Mink}, {Zapartas}, {Marchant}, \& {Smith}}]{renzo20}
{Renzo}, M., {Farmer}, R., {Justham}, S., {et~al.} 2020, arXiv e-prints,
  arXiv:2002.05077

\bibitem[{{Rezzolla} {et~al.}(2008){Rezzolla}, {Barausse}, {Dorband},
  {Pollney}, {Reisswig}, {Seiler}, \& {Husa}}]{rezzolla08}
{Rezzolla}, L., {Barausse}, E., {Dorband}, E.~N., {et~al.} 2008, \prd, 78,
  044002

\bibitem[{{Rodriguez} {et~al.}(2018){Rodriguez}, {Amaro-Seoane}, {Chatterjee},
  \& {Rasio}}]{rodriguez18}
{Rodriguez}, C.~L., {Amaro-Seoane}, P., {Chatterjee}, S., \& {Rasio}, F.~A.
  2018, Physical Review Letters, 120, 151101

\bibitem[{{Rodriguez} {et~al.}(2016){Rodriguez}, {Chatterjee}, \&
  {Rasio}}]{rodriguez16}
{Rodriguez}, C.~L., {Chatterjee}, S., \& {Rasio}, F.~A. 2016, \prd, 93, 084029

\bibitem[{{Rodriguez} \& {Loeb}(2018)}]{rodriguez2018}
{Rodriguez}, C.~L., \& {Loeb}, A. 2018, \apjl, 866, L5

\bibitem[{{Rodriguez} {et~al.}(2015){Rodriguez}, {Morscher}, {Pattabiraman},
  {Chatterjee}, {Haster}, \& {Rasio}}]{rodriguez15}
{Rodriguez}, C.~L., {Morscher}, M., {Pattabiraman}, B., {et~al.} 2015, Physical
  Review Letters, 115, 051101

\bibitem[{{Rodriguez} {et~al.}(2019){Rodriguez}, {Zevin}, {Amaro-Seoane},
  {Chatterjee}, {Kremer}, {Rasio}, \& {Ye}}]{rodriguez2019}
{Rodriguez}, C.~L., {Zevin}, M., {Amaro-Seoane}, P., {et~al.} 2019, \prd, 100,
  043027

\bibitem[{{Rossa} {et~al.}(2006){Rossa}, {van der Marel}, {B{\"o}ker},
  {Gerssen}, {Ho}, {Rix}, {Shields}, \& {Walcher}}]{rossa}
{Rossa}, J., {van der Marel}, R.~P., {B{\"o}ker}, T., {et~al.} 2006, \aj, 132,
  1074

\bibitem[{{Samsing}(2018)}]{samsing18}
{Samsing}, J. 2018, \prd, 97, 103014

\bibitem[{{Sana} {et~al.}(2012){Sana}, {de Mink}, {de Koter}, {Langer},
  {Evans}, {Gieles}, {Gosset}, {Izzard}, {Le Bouquin}, \& {Schneider}}]{sana12}
{Sana}, H., {de Mink}, S.~E., {de Koter}, A., {et~al.} 2012, Science, 337, 444

\bibitem[{{Sigurdsson} \& {Phinney}(1993)}]{sigurdsson93}
{Sigurdsson}, S., \& {Phinney}, E.~S. 1993, \apj, 415, 631

\bibitem[{{Spera} \& {Mapelli}(2017)}]{spera17}
{Spera}, M., \& {Mapelli}, M. 2017, \mnras, 470, 4739

\bibitem[{{Spera} {et~al.}(2015){Spera}, {Mapelli}, \& {Bressan}}]{spera15}
{Spera}, M., {Mapelli}, M., \& {Bressan}, A. 2015, \mnras, 451, 4086

\bibitem[{{Spera} {et~al.}(2019){Spera}, {Mapelli}, {Giacobbo}, {Trani},
  {Bressan}, \& {Costa}}]{spera18}
{Spera}, M., {Mapelli}, M., {Giacobbo}, N., {et~al.} 2019, \mnras, 485, 889

\bibitem[{{Sperhake}(2015)}]{sperhake2015}
{Sperhake}, U. 2015, Classical and Quantum Gravity, 32, 124011

\bibitem[{{Stephan} {et~al.}(2016){Stephan}, {Naoz}, {Ghez}, {Witzel},
  {Sitarski}, {Do}, \& {Kocsis}}]{stephan16}
{Stephan}, A.~P., {Naoz}, S., {Ghez}, A.~M., {et~al.} 2016, \mnras, 460, 3494

\bibitem[{{Stevenson} {et~al.}(2017){Stevenson}, {Berry}, \&
  {Mandel}}]{stevenson17}
{Stevenson}, S., {Berry}, C.~P.~L., \& {Mandel}, I. 2017, \mnras, 471, 2801

\bibitem[{{Stevenson} {et~al.}(2019){Stevenson}, {Sampson}, {Powell},
  {Vigna-G{\'o}mez}, {Neijssel}, {Sz{\'e}csi}, \& {Mandel}}]{stevenson19}
{Stevenson}, S., {Sampson}, M., {Powell}, J., {et~al.} 2019, \apj, 882, 121

\bibitem[{{Talbot} \& {Thrane}(2017)}]{talbot17}
{Talbot}, C., \& {Thrane}, E. 2017, \prd, 96, 023012

\bibitem[{{Tutukov} \& {Yungelson}(1973)}]{tutukov73}
{Tutukov}, A., \& {Yungelson}, L. 1973, Nauchnye Informatsii, 27, 70

\bibitem[{{VanLandingham} {et~al.}(2016){VanLandingham}, {Miller}, {Hamilton},
  \& {Richardson}}]{vanL16}
{VanLandingham}, J.~H., {Miller}, M.~C., {Hamilton}, D.~P., \& {Richardson},
  D.~C. 2016, \apj, 828, 77

\bibitem[{{Wang} {et~al.}(2016){Wang}, {Spurzem}, {Aarseth}, {Giersz}, {Askar},
  {Berczik}, {Naab}, {Schadow}, \& {Kouwenhoven}}]{wang16}
{Wang}, L., {Spurzem}, R., {Aarseth}, S., {et~al.} 2016, \mnras, 458, 1450

\bibitem[{{Wen}(2003)}]{wen03}
{Wen}, L. 2003, \apj, 598, 419

\bibitem[{{Woosley}(2017)}]{woosley17}
{Woosley}, S.~E. 2017, \apj, 836, 244

\bibitem[{{Woosley}(2019)}]{woosley19}
---. 2019, \apj, 878, 49

\bibitem[{{Yang} {et~al.}(2019){Yang}, {Bartos}, {Haiman}, {Kocsis}, {Marka},
  {Stone}, \& {Marka}}]{yang19}
{Yang}, Y., {Bartos}, {Haiman}, Z., {et~al.} 2019, arXiv e-prints,
  arXiv:1903.01405

\bibitem[{{Zevin} {et~al.}(2017){Zevin}, {Pankow}, {Rodriguez}, {Sampson},
  {Chase}, {Kalogera}, \& {Rasio}}]{zevin17}
{Zevin}, M., {Pankow}, C., {Rodriguez}, C.~L., {et~al.} 2017, \apj, 846, 82

\bibitem[{{Zevin} {et~al.}(2018){Zevin}, {Samsing}, {Rodriguez}, {Haster}, \&
  {Ramirez-Ruiz}}]{zevin19}
{Zevin}, M., {Samsing}, J., {Rodriguez}, C., {Haster}, C.-J., \&
  {Ramirez-Ruiz}, E. 2018, arXiv e-prints, arXiv:1810.00901

\bibitem[{{Ziosi} {et~al.}(2014){Ziosi}, {Mapelli}, {Branchesi}, \&
  {Tormen}}]{ziosi14}
{Ziosi}, B.~M., {Mapelli}, M., {Branchesi}, M., \& {Tormen}, G. 2014, \mnras,
  441, 3703

\end{thebibliography}
}

\end{document}